\documentclass[aps,prx,twocolumn,superscriptaddress,dvipsnames,longbibliography]{revtex4-2}

\usepackage{amssymb,amsfonts,amsmath} 
\usepackage{graphicx,epsfig,psfrag}
\usepackage{color}
\usepackage{natbib}
\usepackage{url}
\usepackage[breaklinks=true]{hyperref}
\usepackage{mathtools}
\usepackage{subfigure}
\hypersetup{
        colorlinks = true,
        urlcolor = blue,
        citecolor = blue
}
\usepackage{bookmark}
\usepackage{epic}
\usepackage{longtable}
\usepackage{bbm}
\usepackage{ulem}
\usepackage{braket}
\usepackage{xcolor}
\usepackage{csquotes}
\usepackage{tikz}
\usepackage{extarrows}
\usetikzlibrary{trees}
\usetikzlibrary{decorations.pathmorphing}
\usetikzlibrary{decorations.markings}
\usetikzlibrary{positioning,arrows}
\usetikzlibrary{shapes.arrows}
\usepackage{leftidx}
\usepackage{framed}
\usepackage{tcolorbox}
\usepackage{bm}
\usepackage{physics}
\usepackage{lipsum}
\usepackage{outlines}

\tikzset{->-/.style={decoration={
  markings,
  mark=at position #1 with {\arrow{>}}},postaction={decorate}}} 

\definecolor{lime}{HTML}{A6CE39}
\DeclareRobustCommand{\orcidicon}{%
    \begin{tikzpicture}
    \draw[lime, fill=lime] (0,0) 
    circle [radius=0.16] 
    node[white] {{\fontfamily{qag}\selectfont \tiny ID}};
    \draw[white, fill=white] (-0.0625,0.095) 
    circle [radius=0.007];
    \end{tikzpicture}
    \hspace{-2mm}
}

\foreach \x in {A, ..., Z}{%
    \expandafter\xdef\csname orcid\x\endcsname{\noexpand\href{https://orcid.org/\csname orcidauthor\x\endcsname}{\noexpand\orcidicon}}
}
\newcommand{\orcid}[1]{\href{https://orcid.org/#1}{\textcolor[HTML]{A6CE39}{\orcidicon}}}

\begin{document} 

\title{Coexistence of localization and transport in many-body two-dimensional Aubry-Andr\'{e} models}

\author{Antonio \v{S}trkalj \orcid{0000-0002-9062-6001}}
\email{as3157@cam.ac.uk}
\affiliation{\mbox{T.C.M. Group, Cavendish Laboratory,
University of Cambridge, Cambridge, CB3 0HE, United Kingdom}}

\author{Elmer V. H. Doggen \orcid{0000-0002-0644-8610}}
\affiliation{\mbox{Institute for Quantum Materials and Technologies, Karlsruhe Institute of Technology, 76021 Karlsruhe, Germany}}
\affiliation{\mbox{Institut f\"{u}r Theorie der Kondensierten Materie, Karlsruhe Institute of Technology, 76128 Karlsruhe, Germany}}

\author{Claudio Castelnovo}
\affiliation{\mbox{T.C.M. Group, Cavendish Laboratory,
University of Cambridge, Cambridge, CB3 0HE, United Kingdom}}

%==========================================================
\begin{abstract} 
Whether disordered and quasiperiodic many-body quantum systems host a long-lived localized phase in the thermodynamic limit has been the subject of intense recent debate. While in one dimension substantial evidence for such a many-body localized (MBL) phase exists, the behavior in higher dimensions remains an open puzzle.
In disordered systems, for instance, it has been argued that rare regions may lead to thermalization of the whole system through a mechanism dubbed the avalanche instability. 
In quasiperiodic systems, however, rare regions are altogether absent and the fate of a putative many-body localized phase has hitherto remained largely unexplored. 
In this work, we investigate the localization properties of two many-body quasiperiodic models, which are two-dimensional generalizations of the Aubry-Andr\'{e} model.
By studying numerically the out-of-equilibrium dynamics of large systems, we find very long-lived localization on experimentally relevant time scales. Surprisingly, we also observe large-scale transport along deterministic lines of weak potential, which appear in the investigated quasiperiodic models. 
Our results demonstrate that quasiperiodic many-body systems have the remarkable and counter-intuitive capability of exhibiting coexisting localization and transport properties -- a phenomenon reminiscent of the behavior of supersolids. 
Our findings are of direct experimental relevance and can be tested, for instance, using state-of-the-art cold atomic systems. 
\end{abstract}
%==========================================================
\pacs{} 
\date{\today} 
\maketitle
%==========================================================

%%%%%%%%%%%%%%%%%%%%%%%%%%%%%%%%%%%%%%%%%%%%%%%%%%%
\section{Introduction}
%%%%%%%%%%%%%%%%%%%%%%%%%%%%%%%%%%%%%%%%%%%%%%%%%%%
%
Conventional equilibrium statistical mechanics relies on the notion of thermalization, which requires that all parts of the system exchange particles and energy. Such systems are ergodic and obey the eigenstate thermalization hypothesis~\cite{Deutsch1991,Srednicki1994,Srednicki1999,Polkovnikov2011,Nandkishore2015,Abanin2019}.
However, there exist a class of systems that defy thermalization due to localization of their eigenstates. In his pivotal work, Anderson~\cite{Anderson1958} showed that random disorder present in non-interacting quantum systems can cause the coherent localization of particle wave functions, a phenomenon that became known as \textit{Anderson localization}~\cite{Abrahams2010}. 

One naively expects that adding interactions to a system leads to dephasing of the coherently localized wave functions, and to their delocalization. In other words, it leads to the thermalization of the system.
However, both analytical~\cite{Gornyi2005, Basko2006, Altshuler1997, Imbrie2016} and numerical works~\cite{Oganesyan2007, Znidaric2008, Pal2010, Nandkishore2015, Abanin2019, Gopalakrishnan2020}, as well as experiments~\cite{Schreiber2015, Abanin2019}, have shown that localization can persist to appreciable length and timescales even in interacting systems, thanks to a phenomenon known as \textit{many-body localization} (MBL)~\cite{Gornyi2005, Basko2006, Nandkishore2015, Alet2018, Abanin2019}. 
This phenomenon has been investigated in disordered spin chains~\cite{Oganesyan2007, Pal2010, Luitz2015}, systems with spinful fermions~\cite{Mondaini2015,Prelovsek2016,Zakrzewski2018} and bosons~\cite{Sierant2018,Orell2019,Hopjan2020}.
MBL has since attracted significant interest in the literature as a generic mechanism that may prevent thermalization in closed interacting systems and is robust with respect to small local perturbations (e.g., changing the disorder or interaction strength)~\cite{Abanin2019}. 

The existence and stability of the MBL phase is now considered to be established in one-dimensional disordered systems, where it is proven~\cite{Imbrie2016} -- under certain plausible assumptions -- that, once the disorder strength exceeds some critical value, a system is localized and its dynamics remains frozen indefinitely. 
Whether the same is true in higher dimensions is still under debate. Recent experiments on two-dimensional (2D) random systems indicate that MBL can survive on intermediate timescales~\cite{Choi2016}, and several theoretical works support this scenario~\cite{Wahl2018,Kennes2018,Theveniaut2020,Kshetrimayum2020,Li2021,Foo2022}, while others dispute it~\cite{DeRoeck2017a,DeRoeck2017b,Gopalakrishnan2019,Potirniche2019,Doggen2020,Suntajs2020}. The MBL phase has been argued to be destabilized through the appearance of thermal bubbles mediated by rare regions of weak disorder that appear in random systems.
Such thermal bubbles act as local thermal baths, which thermalize their vicinity and grow according to a mechanism dubbed the \textit{avalanche mechanism}~\cite{DeRoeck2017a,DeRoeck2017b,Thiery2018,Potirniche2019,Leonard2020,Gopalakrishnan2020,Morningstar2021,Sels2022}. The growth of thermal bubbles is unbounded for dimensions larger than one~\cite{DeRoeck2017a,DeRoeck2017b}, meaning that the MBL phase cannot survive in the thermodynamic limit. 

Decades after Anderson's work~\cite{Anderson1958}, it was shown that quasiperiodic potentials can also lead to similar localization phenomena in noninteracting systems~\cite{Aubry1980, Senechal1995, Devakul2017}. 
In contrast to random systems, however, quasiperiodic potentials are deterministic and lack stochastic rare regions. 
They are therefore not directly susceptible to avalanche instabilities and one may wonder whether, in the presence of interactions, quasiperiodic systems may host a thermodynamically stable MBL phase in higher dimensions~\cite{Bordia2017}. 

One of the most famous quasiperiodic models studied in both 1D and 2D is the Aubry-Andr\'{e} (AA) model~\cite{Aubry1980}. The quasiperiodicity in the AA model comes from the cosine modulation of onsite energies that is incommensurate with the underlying lattice. The AA model in two and higher dimensions does not host rare regions, but, depending on its realization, it can have deterministic lines of weak potential (WPLs)~\cite{Szabo2020,Johnstone2021,Johnstone2022}. Such lines are delocalized in the noninteracting case~\cite{Szabo2020} and are expected to be thermal once interactions are present~\cite{Johnstone2021,Johnstone2022}. A natural question then arises: can WPLs destabilize the MBL phase in 2D quasiperiodic models, in analogy to the avalanche mechanism for random systems? 

In this work, we explore the localization properties of the 2D many-body Aubry-Andr\'{e} model by studying its out-of-equilibrium dynamics. 
The main question we answer is whether this model supports a stable long-lived MBL phase.
More precisely, we investigate the influence of the deterministic weak potential lines on the MBL phase.
By analyzing the decay of the particle imbalance, we obtain evidence for a critical point $W_{\rm C}/J$ of an MBL transition, and we show that $W_{\rm C}/J$ is not sensitive to changes in the system size. 
This suggest that the MBL phase is stable over experimentally relevant lengths and timescales. Indeed, the same system size scaling that evidenced the \textit{instability} of MBL in random 2D system~\cite{Doggen2020} shows instead \textit{stability} in quasiperiodic 2D systems, raising the intriguing question of its possible stability in the thermodynamic limit.
Furthermore, by analyzing specific samples, we explicitly demonstrate how the WPLs fail to thermalize the system in the MBL phase, in contrast to the avalanche picture for disordered systems, at least on our accessible length and time scales. Even if the coexistence of ergodic WPLs and localized bulk were to be a finite time effect, it is at the very least remarkable, and possibly of fundamental and practical importance, because it highlights an exceptional separation of time and length scales that appears to be unique to quasiperiodic systems.

Although ergodicity is broken in the MBL phase, the WPLs are capable of supporting transport of particles in certain situations. 
The coexistence of ergodicity breaking without the complete suppression of transport is another main result of our work, intuitively reminiscent of the behaviour of supersolids. 
To the best of our knowledge, it has not been hitherto predicted in many-body out-of-equilibrium quasiperiodic systems. 

The outline of this paper is as follows. In Sec.~\ref{sec:model}, we introduce the Hamiltonian of the 2D quasiperiodic model and discuss the two types of Aubry-Andr\'{e} potentials used in our analysis. Sec.~\ref{sec:single_particle} presents a summary of the single-particle properties for both realizations of the potential. In Sec.~\ref{sec:method_int_sys}, we introduce the methods we use to study these many-body systems, together with the observables that serve as a measure of localization. Secs.~\ref{sec:MB_properties_NS&S},~\ref{sec:stability} and~\ref{sec:WPLs} present the main results of this paper. There we show evidence of localization in our finite-size many-body interacting systems, we demonstrate its stability at timescales of at least $\mathcal{O}(100)$ hopping times, and we discuss the role of weak potential lines on localization and transport properties. Our findings are summarized and an outlook is presented in Sec.~\ref{sec:discussion_and_conclusion}. 

%
%%%%%%%%%%%%%%%%%%%%%%%%%%%%%%%%%%%%%%%%%%%%%%%%%%%
\section{Model  \label{sec:model}}
\begin{figure}[t!]
	\centering
    \includegraphics[scale=1]{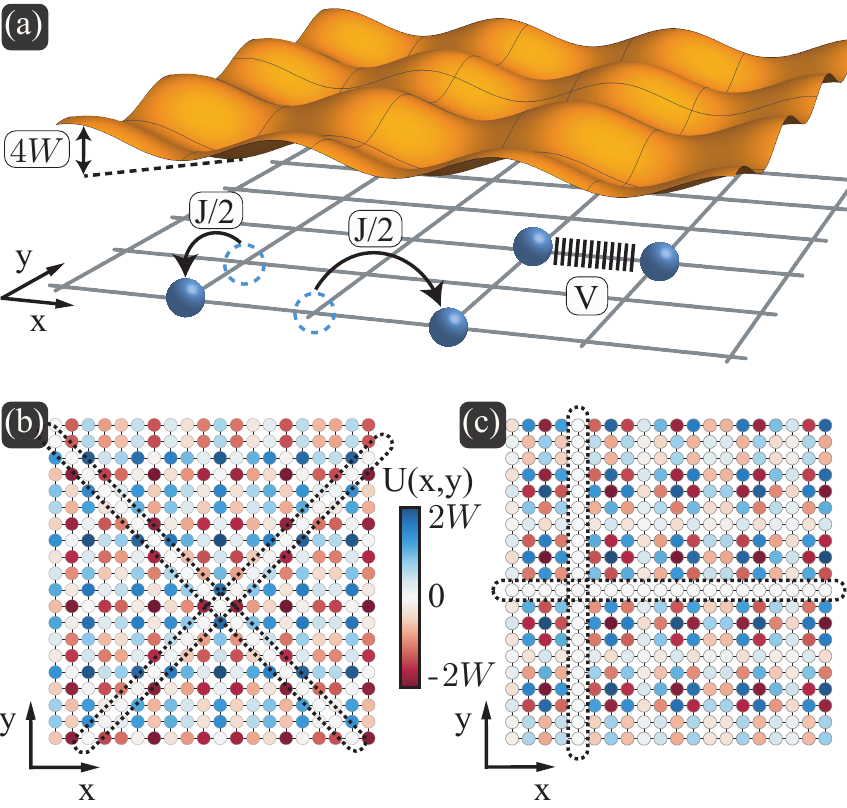}
    \caption{(a) Sketch of the systems studied in this work. 
    The kinetic energy of the particles is determined by a uniform hopping parameter $J$. The particles are subjected to an external quasiperiodic potential $U_{ij}$ of strength $W$ and interact via nearest neighbor coupling $V$. (b) Illustration of the potential $U^{\rm S}$ for the separable AA model considered in this work, with diagonal weak potential lines. (c) Illustration of the potential $U^{\rm NS}$ for the nonseparable AA model, with horizontal and vertical weak potential lines. Selected weak potential lines are highlighted by dashed ovals.
    }
    \label{fig:intro}
\end{figure}

%%%%%%%%%%%%%%%%%%%%%%%%%%%%%%%%%%%%%%%%%%%%%%%%%%%
%
We consider a system of hard-core bosons with nearest-neighbor interactions on a 2D square lattice with open boundary conditions. Particles are subjected to an external potential that is incommensurate with the underlying lattice, i.e., quasiperiodic. The Hamiltonian is given by
\begin{align}
	H =& \sum_{\expval{ij; i'j'}} \Bigg[ -\frac{J}{2}(b^{\dagger}_{ij} b_{i'j'} + \mathrm{h.c.}) + V \hat{n}_{ij}\hat{n}_{i'j'} \Bigg] \nonumber\\ 
	&+ \sum_{ij} U_{ij} \hat{n}_{ij} 
	\, .
	\label{eq:TB_hamiltonian}
\end{align}
where the pair of indices $ij$ labels the sites $(i,j)$ of a square lattice; $b^{\dagger}_{ij}$/$b_{ij}$ are bosonic creation/annihilation operators that act on site $(i,j)$; $\hat{n}_{ij}=b^{\dagger}_{ij} b_{ij}$ is a bosonic density restricted to be $n_{ij} \leq 1$; the summation over $\expval{ij; i'j'}$ couples only adjacent sites; and $U_{ij}$ is the onsite potential. Accordingly, particles move on the lattice with constant hopping $J$, and interact with nearest-neighbor interaction strength $V$, see Fig.~\ref{fig:intro}(a). 
In this work, we set $V=J$ and work in units where $\hbar=1$.  

The discrete quasiperiodic potential $U_{ij}$ is taken from a continuous function $U(x,y)$ given by the sum of two perpendicular cosine functions that can be (i) aligned with the underlying lattice (which we denote $U^{\rm S}$), or (ii) form a $45^{\circ}$ angle with the lattice (which we denote $U^{\rm NS}$):
\begin{align}
    U^{\rm S}(x,y) &= W \bigg\{ \cos(2 \pi b \,  x + \phi_x) + \cos(2 \pi b \, y + \phi_y) \bigg\} \nonumber\\
    U^{\rm NS}(x,y) &= W \bigg\{ \cos\left[ 2 \pi b \,  (x+y) + \phi_+ \right] \nonumber\\
    & \qquad + \cos\left[ 2 \pi b \, (x-y) + \phi_- \right] \bigg\} \, .
	\label{eq:QP_potentials}
\end{align} 
The frequency is taken to be the inverse of the golden mean, $b=2/(1+\sqrt{5})$. 
Examples of the potentials in Eq.~\eqref{eq:QP_potentials} are shown in Fig.~\ref{fig:intro}(b) and (c). 
In the noninteracting case, Hamiltonian~\eqref{eq:TB_hamiltonian} with potential $U^{\rm S}$ is separable into $x$- and $y$-components, while with potential $U^{\rm NS}$ it becomes nonseparable; hence, we refer to the two models as the \textit{separable} and \textit{nonseparable} 2D Aubry-Andr\'{e} (AA) models. 

Note that the phases $\phi_{x,y,+,-}$ translate the potential with respect to the underlying lattice~\cite{footnote_phases}, and they do not play a role in the localization properties when the system is in the thermodynamic limit. However, since we can numerically study only finite-size systems, the phases serve us as generators of samples with different realizations of the potential over which we average the observables we calculate. In this way, together with using large enough samples such that finite-size effects are mitigated, we aim to obtain information about the behavior of the system in the thermodynamic limit. 

%
%%%%%%%%%%%%%%%%%%%%%%%%%%%%%%%%%%%%%%%%%%%%%%%%%%%
\section{Single-particle properties \label{sec:single_particle}}
%%%%%%%%%%%%%%%%%%%%%%%%%%%%%%%%%%%%%%%%%%%%%%%%%%%
%
Before discussing the many-body physics of the separable and nonseparable AA models, let us recap the known localization properties of their noninteracting ($V=0$) limit. 

\begin{figure}[t!]
	\centering
    \includegraphics[scale=1]{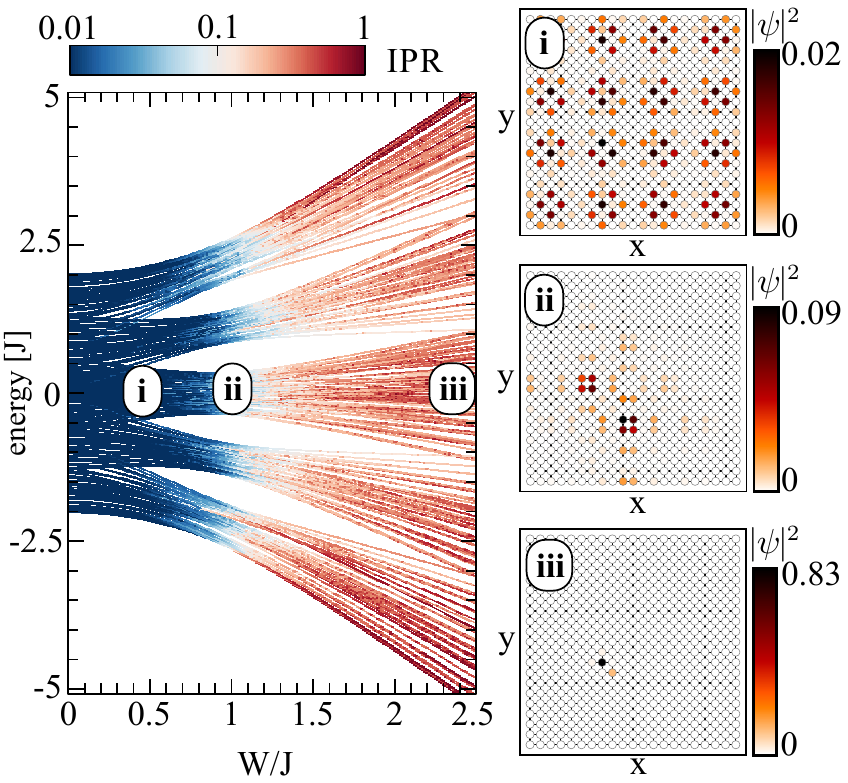}
    \caption{Single-particle localization properties of the 2D separable AA model. The left panel shows the IPR of each eigenstate. The three panels on the right show the spatial dependence of the density of three eigenstates at $E \approx 0$ taken from the (i) extended, (ii) critical, and (iii) localized phase. Note that different colorbar intensity scales are used in (i-iii). The system size used in all the plots is $L_x \times L_y = 21 \times 21$.
    }
    \label{fig:noninteracting_separable}
\end{figure}

We start with the separable model~\cite{Bordia2017,Rossignolo2019}. 
The 2D system in this case inherits the localization properties of the 1D AA model, namely a metal-to-insulator transition that occurs uniformly throughout the spectrum at the critical point $W/J=1$~\cite{Aubry1980,Bordia2017,Huang2019,Rossignolo2019,Strkalj2020}. This can be seen from the left panel of Fig.~\ref{fig:noninteracting_separable}, where we show the inverse participation ratio (IPR) of each eigenstate. 
The IPR of an eigenstate $\psi(E_n)$ with eigenenergy $E_n$ is defined as
$\mathrm{IPR}(E_n) = \sum_{m=1}^{L_x \times L_y} |\psi_m(E_n)|^4 / \sum_{m=1}^{L_x \times L_y} |\psi_m(E_n)|^2$, where the sum is over all states in the spectrum, and it serves as a measure of localization~\cite{Bell1970,Edwards1972}. 
In the thermodynamic limit it tends to $0$ for extended states while it remains equal to $1$ for states localized on a single site. In the right three panels of Fig.~\ref{fig:noninteracting_separable}, we show spatial density profiles of three representative states taken from the (i) extended, (ii) critical, and (iii) localized phase. 

\begin{figure}[t!]
	\centering
    \includegraphics[scale=1]{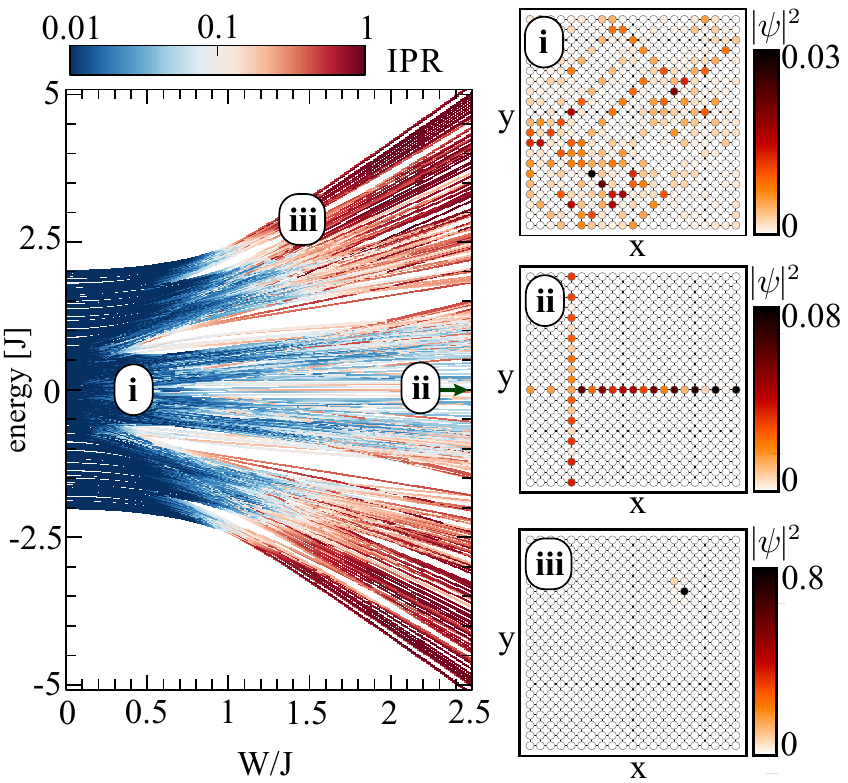}
    \caption{Single-particle localization properties of the 2D nonseparable AA model. The left panel shows the IPR of each eigenstate. The three panels on the right show the spatial dependence of the density of three eigenstates that are (i) extended over the whole system, (ii) extended only over WPLs, and (iii) localized. The potential strength used for (ii) is $W/J=50$. Note that different colorbar intensity scales are used in (i-iii). The system size used in all plots is $L_x \times L_y = 21 \times 21$.
    }
    \label{fig:noninteracting_nonseparable}
\end{figure}

Conversely, in the nonseparable AA model there is no uniform metal-to-insulator phase transition. A large fraction of the states remain extended even above the $W/J=1$ point~\cite{Szabo2020,Johnstone2021,Johnstone2022}, see the left panel of Fig.~\ref{fig:noninteracting_nonseparable} where the IPR of all eigenstates is shown. 
It is easy to understand the existence of the aforementioned extended states by rewriting the potential in a multiplicative form: $U^{\rm NS}(x,y) = 2 W \cos(2 \pi b \,  x ) \cdot \cos(2 \pi b \, y)$, where we have set without loss of generality $\phi_+=\phi_-=0$~\cite{footnote_phases}. 
For a fixed $x$ ($y$) line, the potential in the $y$-~\mbox{($x$-)} direction is a cosine with strength $W(x)$ ($W(y)$) that can be arbitrary small. This means that the system will contain \textit{weak potential lines} (WPLs) that lie in the $x$- and $y$-direction and are characterised by having a potential strength much smaller than $W$; this is illustrated in Fig.~\ref{fig:intro}(c), where two lines with vanishing potential are shown. 
Generally, throughout the text we refer to WPLs as lines that encompass sites with an effective potential strength that is smaller than a critical point $W_{\rm C}$ where a 1D AA chain localizes~\cite{Doggen2019}.
In the thermodynamic limit, it is always possible to find WPLs with arbitrarily small values of the potential even when $W$ is extremely large. Therefore, there will be some eigenstates in the spectrum that are extended over such WPLs; see the panel marked (ii) in Fig.~\ref{fig:noninteracting_nonseparable}. 

Note that $U^{\rm S}$ also hosts WPLs, but they always connect next-to-nearest neighbor sites that lie along a lattice diagonal, see Fig.~\ref{fig:intro}(b). The matrix element that connects two closest sites in such WPLs is proportional to $J^2/U_{ij}$, and particle hopping along the diagonal is again quasiperiodic, thus leading to localization for large enough $W/J$. Further details of the analysis of noninteracting WPLs in the separable model can be found in App.~\ref{app:details_separable_model}.

%
%%%%%%%%%%%%%%%%%%%%%%%%%%%%%%%%%%%%%%%%%%%%%%%%%%%
\section{Observables and methods for interacting systems    \label{sec:method_int_sys}}
%%%%%%%%%%%%%%%%%%%%%%%%%%%%%%%%%%%%%%%%%%%%%%%%%%%
%
Let us now turn to study the behavior of the interacting models. To probe the localization properties, we investigate the quench dynamics from an initial state where the occupied sites always neighbor unoccupied sites, forming a checkerboard pattern. Such a state corresponds to half-filling and, in the absence of the external potential, it is generally expected to thermalize quickly, usually within a few hopping times. 
For all our numerical calculations, we use finite systems with open boundary conditions. 
The observable we concentrate on is the particle imbalance, defined as the memory of the initial checkered state: 
\begin{align}
    \mathcal{I}(\tau) = \frac{2}{L_x L_y}\sum_{i,j} (-1)^{i+j} \bra{\psi(\tau)}\hat{n}_{ij}\ket{\psi(\tau)} \, .
    \label{eq:total_imbalance}
\end{align}
The choice of a prefactor is appropriate for a half-filled system to ensure that, at the initial time $\tau = 0$, the value of the imbalance is $\mathcal{I}(0) = 1$. In localized systems, the imbalance is expected to saturate to a constant value for $\tau \gg 1$, while in ergodic systems it decays towards zero in the long-time and thermodynamic limit. 
As we will see in the next section, the imbalance averaged over different samples with randomly chosen phases $\phi_{x,y,+,-}$, see Eq.~\eqref{eq:QP_potentials}, approximately follows an inverse power law behavior for the timescales studied here (cf. Fig.~\ref{fig:imbalance}), 
\begin{align}
    \overline{\mathcal{I}}(\tau) \propto \tau^{-\gamma} \, ,
    \label{eq:averaged_imbalance}
\end{align}
where the bar denotes averaging, and the exponent $\gamma$ depends on $W/J$. The vanishing of $\gamma$ inside the MBL phase reflects the persistence of the memory of the initial state and serves as a quantitative measure of localization in interacting systems. 

The numerical method we use for our simulations is time-dependent variational principle (TDVP)~\cite{Haegeman2016a} applied to matrix product states (MPS)~\cite{Schollwock2011a}. Geometrically, this method can be viewed as the projection of the time evolution onto the MPS manifold:
\begin{equation}
\frac{d}{dt} |\psi \rangle = -i \mathcal{P}_\mathrm{MPS} H |\psi\rangle, \label{eq:tdvp}  
\end{equation}
where $\mathcal{P}_\mathrm{MPS}$ is the projection operator, and the size of the variational MPS subspace scales polynomially with the bond dimension $\chi$. MPS algorithms probe the low-entanglement subspace of the entire Hilbert space (exponentially large in system size), which makes these methods especially suitable for simulating disordered systems~\cite{Doggen2021a}. Convergence of the algorithm can be checked by increasing $\chi$. Compared to other MPS approaches, TDVP has the advantage that time evolution respects the conservation of global quantities -- in particular, for the case of the unitary evolution of a closed system, the energy~\cite{Paeckel2019a}. Like all MPS-based methods, however, TDVP is restricted to the low- to moderately entangled subspace of the Hilbert space, a drawback compared to exact diagonalization (ED) methods that probe the entire Hilbert space. This implies that dynamics can be simulated only up to relatively short times for ergodic systems -- meaning in our case the limit of small $W/J$. This is because the entanglement grows rapidly in ergodic systems. As the strength of the quasiperiodic potential is increased, the available timescales likewise increase as the system becomes only weakly ergodic or even nonergodic. In this case, timescales of $O(10^2 J^{-1})$ are accessible, with system sizes much larger than those accessible to ED: $O(10^2)$ compared to around $20$ sites only for ED. Further details and benchmarks are presented in App.~\ref{app:numdetails}, and an extensive discussion of the benefits and drawbacks of the method is presented in Ref.~\cite{Doggen2021a}.

Notwithstanding the caveats mentioned above, TDVP is a powerful numerical method that allows us to study the dynamics up to a hundred hopping times -- a timescale comparable to state-of-art experiments~\cite{Schreiber2015, Bordia2016, Bordia2017}. 
In our simulations, of course, we have full control of the closed system and insight into its microscopical behavior. For instance, we can choose the initial state without any errors; there is no decoherence and loss of particles caused by coupling to the environment; the system can be scaled from small to relatively large sizes; and a high tunability of parameters in the Hamiltonian allows for simulating a wide range of physical phenomena. TDVP is one of the few methods suitable for treating large interacting systems in two dimensions, and is especially suited for simulating nonergodic systems~\cite{Doggen2020, Doggen2021a, Strkalj2021}.

%
%%%%%%%%%%%%%%%%%%%%%%%%%%%%%%%%%%%%%%%%%%%%%%%%%%%
\section{Many-body properties   \label{sec:MB_properties_NS&S}}
\begin{figure}[h!]
	\centering
    \includegraphics[scale=1]{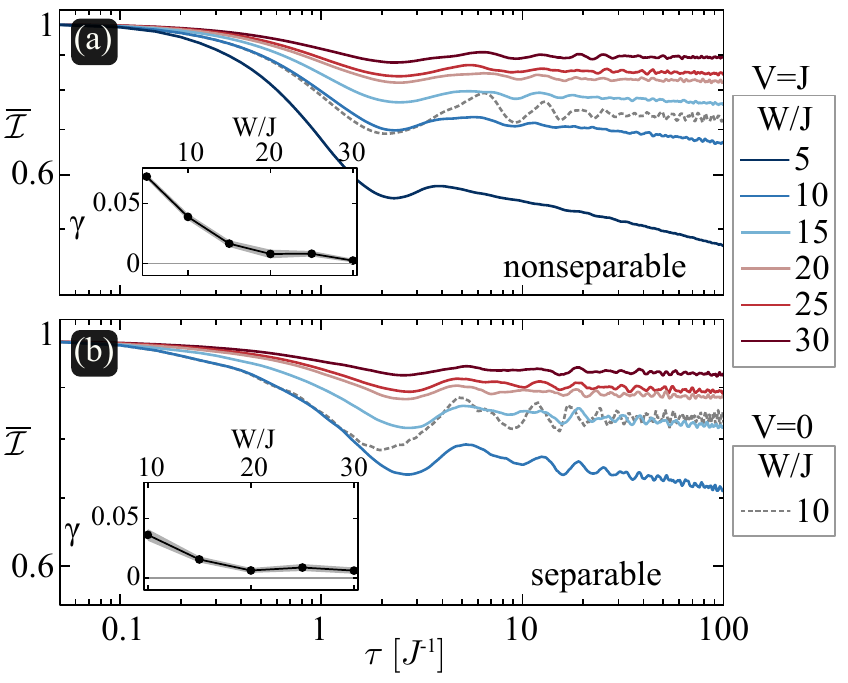}
    \caption{Log-log plot of time evolution of the average imbalance for the (a) nonseparable and (b) separable interacting AA models at different values of $W/J$. Dashed grey line is the noninteracting case with $W/J=10$ in both models. 
    The insets show the decay coefficient $\gamma$, Eq.~\eqref{eq:averaged_imbalance}, extracted from fits to the averaged imbalances of the interacting systems. The fitting window we used is $\tau \in [50, 100]$.
    In all plots we used systems of size $L_x \times L_y = 16 \times 5$, averaging over 64 different samples. All results converged with a bond dimension $\chi=128$ and a time step $\delta \tau = 0.05$. Error bars are 1$\sigma$ intervals based on a bootstrapping procedure~\cite{Efron1979}.
    }
    \label{fig:imbalance}
\end{figure}

%%%%%%%%%%%%%%%%%%%%%%%%%%%%%%%%%%%%%%%%%%%%%%%%%%%
%
We numerically calculate the averaged imbalance~\eqref{eq:averaged_imbalance} at different values of $W/J$ for finite systems of size $L_x \times L_y = 16 \times 5$. The results for both the nonseparable and the separable AA model are shown in Figs.~\ref{fig:imbalance}(a) and (b), respectively. 
Over the accessible timescales, and for potential values close to the MBL transition ($W/J \gtrsim 10$), we find that the dynamics of the averaged imbalance is well approximated by an inverse power law decay. The decay of the imbalance slows down with increasing $W/J$ until it saturates to a constant value for $W/J \gtrsim 20-30$.
The insets in Figs.~\ref{fig:imbalance}(a) and (b) show the decay coefficient $\gamma$ obtained from the curves in the main panel. 
With increasing $W/J$, $\gamma$ decreases rapidly. It shows a drastic reduction around $W/J = 20$ and essentially vanishes for $W/J \gtrsim 30$ (within $2$ and $3$ standard deviations, for the S and NS model, respectively). 
These finite-time simulations suggest that there is a dynamical transition from an ergodic to a localized phase in both models occurring somewhere between $W/J=20$ and $W/J=30$.

%
%%%%%%%%%%%%%%%%%%%%%%%%%%%%%%%%%%%%%%%%%%%%%%%%%%%
\section{Stability of the transition \label{sec:stability}}
%%%%%%%%%%%%%%%%%%%%%%%%%%%%%%%%%%%%%%%%%%%%%%%%%%%
%
In this section, we discuss whether the transition discussed above is stable upon increasing the system size, which by extrapolation indicates possible stability in the thermodynamic limit.
In a recent paper~\cite{Doggen2020}, some of us showed that for a 2D model with a random onsite potential, the critical disorder strength $W_{\rm C}/J$ diverges with increasing system size due to the avalanche instability. This suggests that the MBL phase is unstable in higher dimensional random systems in the thermodynamic limit~\cite{DeRoeck2017a,DeRoeck2017b,Thiery2018,Doggen2020}. 

\begin{figure}[t!]
	\centering
    \includegraphics[width=\columnwidth]{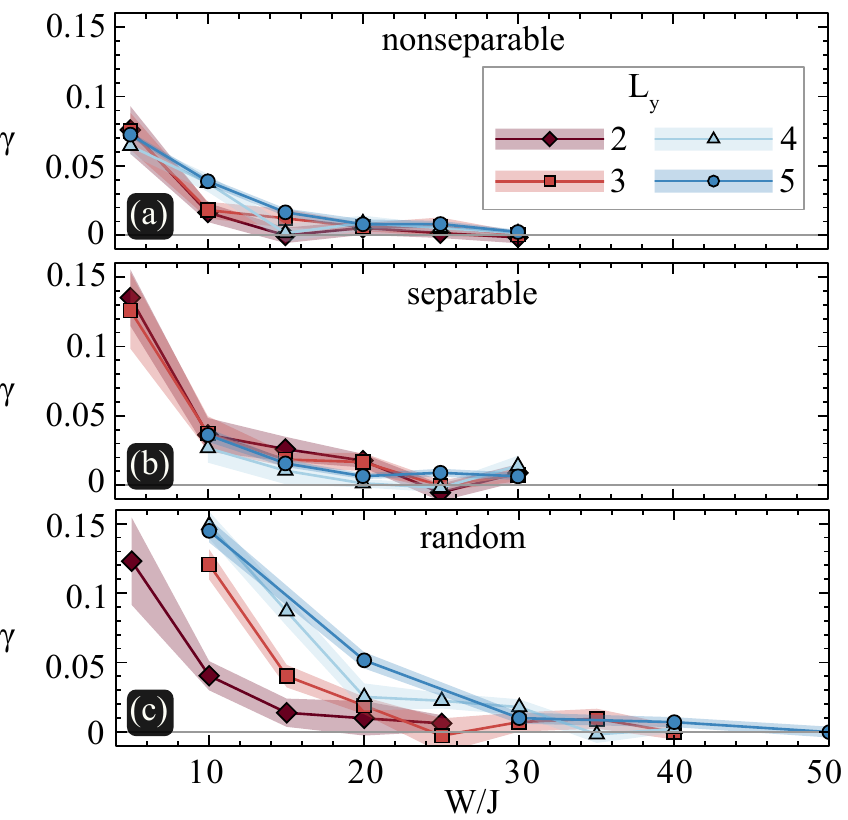}
    \caption{Decay coefficient $\gamma$ of the average imbalance, Eq.~\eqref{eq:averaged_imbalance}, as a function of potential strength $W/J$ and width of the system $L_y$ for the (a) nonseparable, (b) separable, and (c) random models. In all plots, we used $L_x=16$ and do the inverse power law fit over a time window $\tau \in [50, 100]$. The averaging was performed over $32$ samples for $L_y=2,3,4$ and $64$ samples for $L_y=5$ in the quasiperiodic cases, while for random disorder, we used $32$ samples for every $L_y$. All results converged with a bond dimension $\chi=128$ and a time step $\delta \tau = 0.05$. Error bars are 1$\sigma$ intervals based on a bootstrapping procedure~\cite{Efron1979}.
    }
    \label{fig:decay_coefs}
\end{figure}

To test if the same is true for separable and nonseparable interacting AA models, we numerically calculate the averaged imbalance for several systems with fixed $L_x=16$ and varying $L_y$. We then extract the decay coefficient $\gamma$ from Eq.~\eqref{eq:averaged_imbalance} and show its dependence on $W/J$ in Fig.~\ref{fig:decay_coefs}.  
In both quasiperiodic models, $\gamma(W/J)$ behaves similarly with increasing $L_y$, see Figs.~\ref{fig:decay_coefs}(a) and (b), and such behavior is markedly different from the case of random disorder, see Fig.~\ref{fig:decay_coefs}(c). In the latter, curves for larger $L_y$ shift towards higher values of $W/J$, in agreement with $W_{\rm C}/J$ diverging with $L_y$~\cite{Doggen2020}. In the former, on the other hand, curves for different $L_y$ lie on top of each other suggesting that $W_{\rm C}/J$ does not depend on $L_y$. 
These results provide evidence that the MBL phase in quasiperiodic AA models is more stable than in random systems.

Surprisingly, the transition points in the interacting separable and nonseparable AA models occur at similar values, even though their single-particle properties are different. Furthermore, in the averaged imbalance and its decay coefficient we do not see any signatures of the WPLs, which may have been a priori expected to play a similar role to rare regions in random systems in destabilizing the MBL phase. 

\begin{figure}[t!]
	\centering
    \includegraphics[width=\columnwidth]{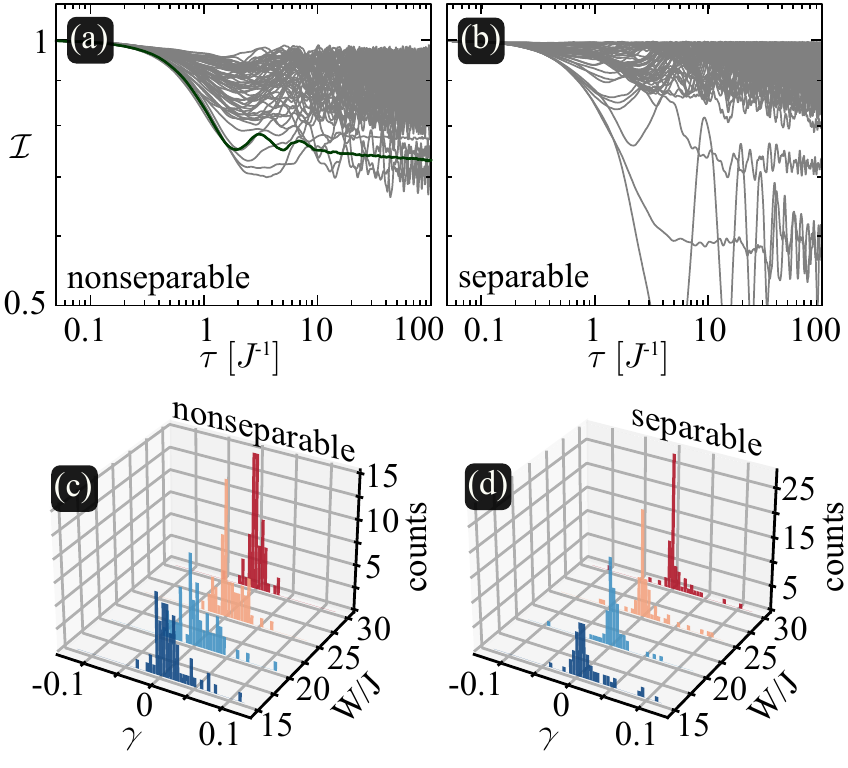}
    \caption{Imbalance $\mathcal{I}$ for $64$ samples with different randomly chosen phases $\phi$ in the case of (a) nonseparable and (b) separable AA model. We set $W/J=30$, well inside the localized phase. The dark green line in (a) shows one of the imbalance curves that decays even though the system is in the localized phase. In (c) and (d), distributions of $\gamma$ across different samples and as a function of $W/J$ for the nonseparable and the separable AA model are shown. The system size used in all plots is $L_x \times L_y = 16 \times 5$, and the fitting widow used in (c) and (d) is $\tau \in [50,100]$. In all plots, system is initialized in a checkerboard state. All results converged with a bond dimension $\chi=128$ and a time step $\delta \tau = 0.05$
    }
    \label{fig:histograms}
\end{figure}

To gain deeper insight into the differences between the separable and nonseparable AA models, we analyze each sample separately. In Figs.~\ref{fig:histograms}(a) and (b) we show the imbalance of each sample inside the MBL phase, for $W/J=30$. 
In the nonseparable model, Fig.~\ref{fig:histograms}(a), there are some samples in which the imbalance slowly decays, (see e.g., the one highlighted by a dark green line). Such decay occurs in samples that contain WPLs (cf. Fig.~\ref{fig:NS_density_evolution_WPL}), where the potential only weakly perturbs the particle motion.  
However, most of the samples do not contain such WPLs when the phases of the potential, Eq.~\eqref{eq:QP_potentials}, are chosen randomly. Therefore, the decay due to the WPLs becomes negligible after averaging over many samples.    
In the separable model, Fig.~\ref{fig:histograms}(b), no imbalance decay is observed at long times. This is a feature of the checkerboard initial state, which leaves the diagonal WPLs either completely fully or empty, and therefore particles cannot move along such WPLs. In Sec.~\ref{subsub:transport_WPL_S}, we shall analyze the diagonal WPLs starting from different initial states and observe nonvanishing transport even when the rest of the system is strongly localized. Note that some samples in Fig.~\ref{fig:histograms}(b) saturate at significantly lower values due to the short time dynamics where particles hop between pairs of sites with similar values of onsite potential -- e.g., see the top two rows of sites in Fig.~\ref{fig:intro}(b).

In Figs.~\ref{fig:histograms}(c) and (d), histograms of the decay coefficient for the nonseparable and the separable model, respectively, are shown as a function of $W/J$ close to the MBL transition point. In the separable model, Fig.~\ref{fig:histograms}(d), the width of the distribution of $\gamma$ rapidly reduces with increasing $W/J$, resulting in a growing number of counts around $\gamma=0$. 
In the nonseparable model, Figs.~\ref{fig:histograms}(c), the width of the distribution also reduces with increasing $W/J$, but much more slowly than in the case of the separable model. Comparing the distributions at $W/J=30$ from Figs.~\ref{fig:histograms}(c) and (d), we observe that in the latter, the peak around $\gamma=0$ is twice as high as in the former. This difference comes from the ergodic WPLs in the nonseparable model, which increase the value of $\gamma$ in samples that contain them. Even for extremely large $W/J$, where the distribution of $\gamma$ in the separable model becomes a delta peak located at $\gamma = 0$, in the nonseparable model the distribution is still expected to have a finite width. %

%
%%%%%%%%%%%%%%%%%%%%%%%%%%%%%%%%%%%%%%%%%%%%%%%%%%%
\section{Weak potential lines (WPLs)    \label{sec:WPLs}}
%%%%%%%%%%%%%%%%%%%%%%%%%%%%%%%%%%%%%%%%%%%%%%%%%%%
%
In this section, we focus on the WPLs and we investigate their impact on the behavior of the system. For simplicity, we tune the phases in the potential so that it vanishes exactly along the WPLs. Furthermore, we set $W/J=50$ which is, according to the analysis from previous sections, deeply in the localized phase.
The main goal of this section is (i) to confirm that thermal WPLs do not spread out to the localized parts of the system, and (ii) to study transport along the WPLs.

%
%%%
\subsection{WPLs in the nonseparable AA model      \label{subsec:WPL_nonseparable}}
\begin{figure}[t!]
	\centering
    \includegraphics[width=\columnwidth]{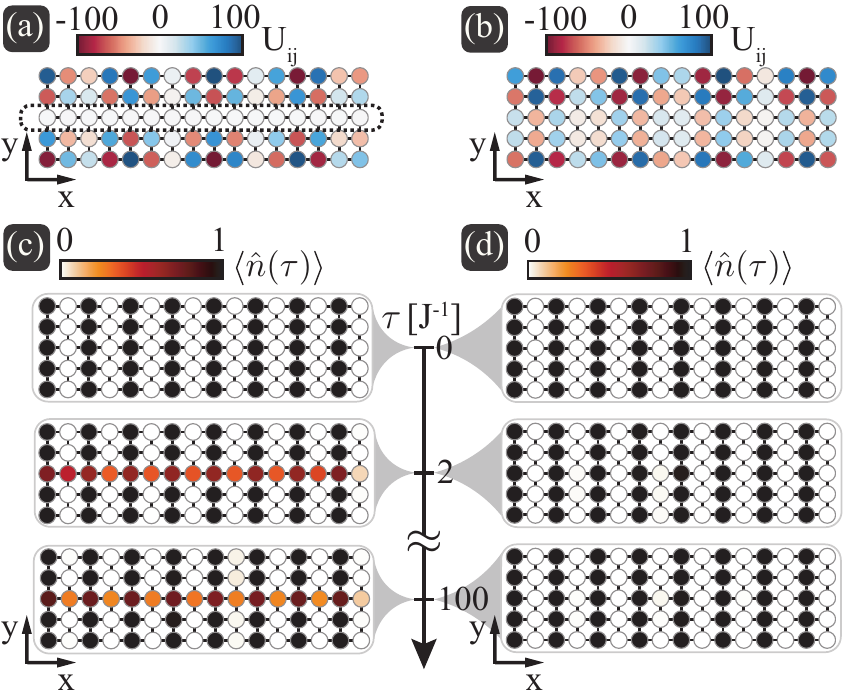}
    \caption{Time evolution of the density in the nonseparable model. The two top panels show, respectively, a sample with (a) and without (b) a horizontal WPL. A dashed oval marks the WPL in (a) where the potential is tuned to vanish. The bottom panels (c) and (d) show snapshots of the density $\expval{\hat{n}(\tau)}$ at three different times, calculated using the potentials shown in (a) and (b), respectively. 
    The system size is $L_x \times L_y = 16 \times 5$ and we set $W/J=50$, deeply in the localized phase. All results converged with a bond dimension $\chi=128$ and a time step $\delta \tau = 0.05$. Note that we use open-boundary conditions in both the $x$- and $y$-direction.
    }
    \label{fig:NS_density_evolution_WPL}
\end{figure}

%%%
%
Let us start with the interacting nonseparable AA model. First, we investigate the behavior of the particle density in two samples, one containing a WPL in the x-direction with vanishing potential, Fig.~\ref{fig:NS_density_evolution_WPL} (a), and the other with no WPLs, Fig.~\ref{fig:NS_density_evolution_WPL} (b).
To create the aforementioned samples, we first set the phases of the potential to $\phi_+ = \phi_0 + \delta \phi_+$ and $\phi_- = -\phi_0 + \delta \phi_+$, so that we obtain 
\begin{align}
    U^{\rm NS}_{ij} = 2 W \cos(2 \pi b \,  i + \delta \phi_+) \cos(2 \pi b \,  j + \phi_0) 
    \, ,
\end{align}
where $\delta \phi_+$ and $\phi_0$ shift the potential along the $x$- and $y$-directions, respectively. Now, by tuning $\phi_0$, we set one sample to have a horizontal WPL in the middle ($j=3$), i.e., $\cos(2 \pi b \cdot 3 + \phi_0)=0$, whereas we set the second configuration to not contain any horizontal WPLs. The residual phase parameter, $\delta \phi_+$, is chosen randomly. 
Furthermore, we set $W/J=50$ for both samples, which is in the localized phase, see Figs.~\ref{fig:imbalance} and~\ref{fig:decay_coefs}(a). We numerically calculate the time evolution of the particle density $\expval{\hat{n}_{ij}(\tau)}$ starting from an initial state that resembles a columnar density wave with a period of two sites, see the $\tau=0$ snapshot in Figs.~\ref{fig:NS_density_evolution_WPL}(c) and (d). Such an initial state reduces the transport signature of particles hopping between two neighboring sites with similar values of the potential, see Fig.~\ref{fig:NS_density_evolution_WPL}(b) where pairs of sites that are located at $j=2$ and $j=3$ (counted from the bottom) have almost degenerate values of the potential, i.e., $V_{i,j=2} \approx V_{i,j=3}$. 

The evolution of the density is shown in Figs.~\ref{fig:NS_density_evolution_WPL}(c) and (d).
In the former, the density quickly spreads over the WPL while remaining localized away from it. In the latter, the density profile does not change for at least one hundred hopping times, which is the longest time that we simulated. 

To quantitatively show the difference between samples with and without WPLs, we calculate the averaged imbalance $\overline{\mathcal{I}}(\tau)$, Eq.~\eqref{eq:averaged_imbalance}, as well as the $y$-dependent averaged imbalance 
\begin{align}
    \mathcal{I}_{j}=\frac{2}{L_x} \sum_{i=1}^{L_x} (-1)^{i+1}\bra{\psi(\tau)}\hat{n}_{ij}\ket{\psi(\tau)} \, .
\end{align}
The former is the same observable we used in the previous sections, but modified for a columnar density wave as the initial state, while the latter can capture the impact of the WPL on its localized neighborhood. 
For the two categories of samples, we use the same value of $\phi_0$ as in Fig.~\ref{fig:NS_density_evolution_WPL}(a) and (b), and average over different randomly chosen $\delta \phi_+$. Note that the number of horizontal WPLs is conserved when changing $\delta \phi_+$. 

\begin{figure}[t!]
	\centering
    \includegraphics[width=\columnwidth]{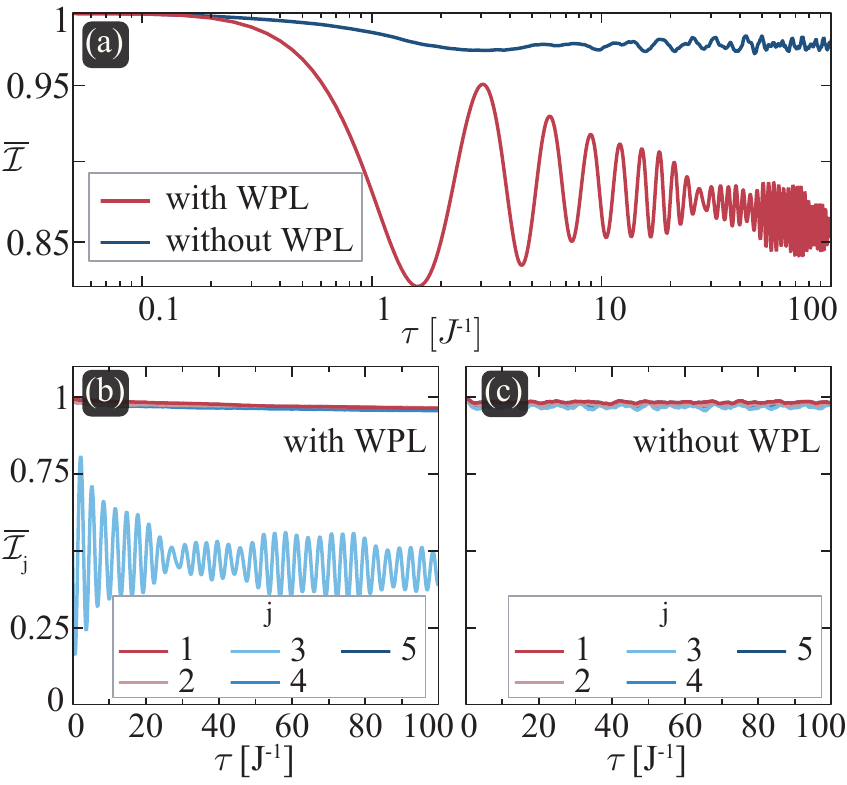}
    \caption{Quantitative effect of the WPLs. (a) Imbalance defined in Eq.~\eqref{eq:total_imbalance}, and modified for a columnar density wave as the initial state, averaged over samples with (red line) and without (blue line) horizontal WPL. Panels (b) and (c) show the $y$-dependent imbalance defined in the main text, averaged over samples with and without horizontal WPL, respectively. The index $j$ labels the rows of the system starting from the bottom. 
    In all plots we use $L_x \times L_y = 16 \times 5$, $W/J=50$, and we average over $10$ different samples all having the same number of horizontal WPLs. All results converged with a bond dimension $\chi=128$ and a time step $\delta \tau = 0.05$
    }
    \label{fig:NS_imbalance_WPL}
\end{figure}

The results for $\overline{\mathcal{I}}(\tau)$ and $\overline{\mathcal{I}_{j}}(\tau)$ are shown in Fig.~\ref{fig:NS_imbalance_WPL}. 
The total imbalance averaged over samples without horizontal WPLs has a value close to 1 at all times, while in the samples that contain a horizontal WPL the imbalance decays over time, see Fig.~\ref{fig:NS_imbalance_WPL}(a). The decay in the latter case is due to the ergodic WPL, which can be seen from Fig.~\ref{fig:NS_imbalance_WPL}(b), where the $y$-dependent imbalance $\overline{\mathcal{I}_{j}}(\tau)$ is seen to decay only when $j$ coincides with the position of the horizontal WPL, i.e., when $j=3$. For other values of $j$, the $y$-dependent imbalance remains close to unity with a small deviation from it due to the presence of vertical WPLs, see App.~\ref{app:details_nonseparable_model}.
For samples without horizontal WPLs, $\overline{\mathcal{I}_{j}}(\tau)$ does not decay for any $j$. 

We can safely conclude that a thermal WPL in the nonseparable AA model \emph{does not} thermalize its closest localized neighborhood, at least on the simulated timescale of one hundred hopping times~\cite{footnote:weakWPL}. 
Therefore, we argue that the system remains nonergodic for $W \gtrsim W_{\rm C}$, where $W_{\rm C}/J \approx 30$ as follows from Figs.~\ref{fig:imbalance}(a) and~\ref{fig:decay_coefs}(a), but transport is not zero for any $W/J$ and also independent the initial state. 

%
%%%
\subsection{WPLs in the separable model and state-dependent transport     \label{subsec:WPL_separable}}
%%%
%
\begin{figure*}[t!]
	\centering
    \includegraphics[width=\textwidth]{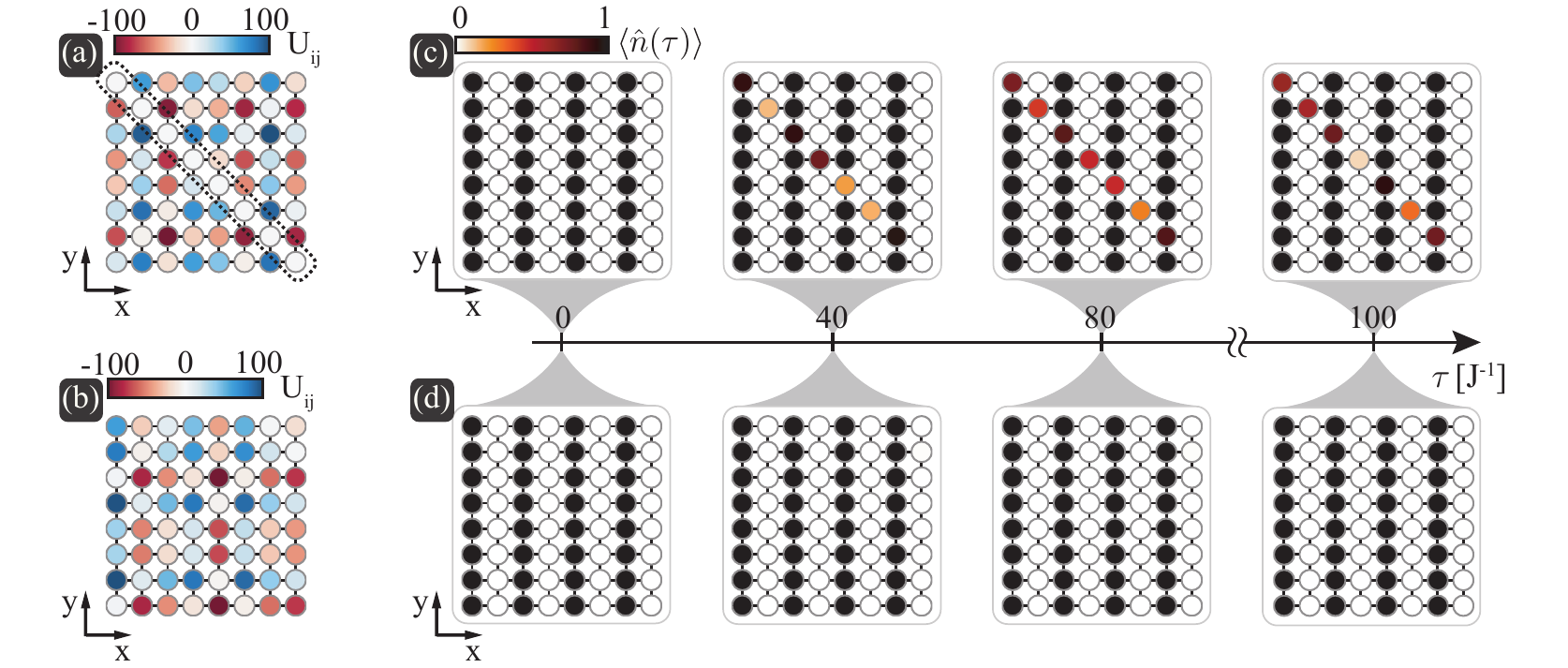}
    \caption{Time evolution of the density of particles in the separable model. In the leftmost panels, a single sample with (a) and without (b) diagonal WPL is shown. A dashed oval marks the WPL in (a) where the potential is tuned to vanish. Panels (c) and (d) show snapshots of the density $\expval{\hat{n}(\tau)}$ at four times calculated using the potential landscape shown in (a) and (b), respectively. 
    The system size is $L_x \times L_y = 8 \times 8$ and we set $W/J=50$, deeply in the localized phase. We used a bond dimension $\chi=128$ and a time step $\delta \tau = 0.05$.
    }
    \label{fig:S_density_evolution_WPL}
\end{figure*}
As discussed in Sec.~\ref{sec:single_particle}, the separable AA model exhibits diagonal WPLs, with no direct hopping between their sites. Therefore, to move along the WPL, particles need to hop to sites outside the WPL which have finite onsite potential, and consequently make the transport properties more complex. 

In the rectangular geometry discussed thus far, the diagonal WPLs span only a few sites, resulting in a relatively weak impact on the averaged observables. To maximize the effect that we want to study, we use a square geometry, $L_x=L_y$, and choose the phases $\phi_x$ and $\phi_y$ in Eq.~\eqref{eq:QP_potentials} such that one WPL (with potential equal to zero) always lies on the longest diagonal, see the dashed oval in Fig.~\ref{fig:S_density_evolution_WPL}(a).
We then compare the time evolution of the density of particles $\expval{\hat{n}(\tau)}$ in two samples that are deeply in the localized phase.
One of the samples contains a WPL with zero potential, Fig.~\ref{fig:S_density_evolution_WPL}(a), and the other one does not, Fig.~\ref{fig:S_density_evolution_WPL}(b). 
While in the latter the initial density profile stays constant for all times, in the former we notice some changes of the density on sites that belong to the WPL. More precisely, we see once again that the initial density profile smears out on the WPL and stays unaffected away from it.

To quantify the behavior observed in Fig.~\ref{fig:S_density_evolution_WPL}, we calculate the averaged imbalance $\overline{\mathcal{I}}(\tau)$, for samples with and without diagonal WPL. For the averaging, we adopt a similar procedure as for the nonseparable model discussed in Sec.~\ref{subsec:WPL_nonseparable}. We first write the phases in Eq.~\eqref{eq:QP_potentials} as $\phi_{x/y} \equiv \pm \phi_{x'} + \phi_{y'}$, so that we obtain 
\begin{align}
    U^{\rm S}_{ij} = 2 W \cos(\pi b \, (i+j) + \phi_{y'}) \cos(\pi b \,  (i-j) + \phi_{x'}) 
    \, ,
\end{align}
where $\phi_{y'}$ controls the $y'$ position (cf. Fig.~\ref{fig:diamond_chain}(a)) of the WPL marked with a dashed oval in Fig.~\ref{fig:S_density_evolution_WPL}(a), and $\phi_{x'}$ shifts the potential along the $x'$-direction. The averaging is then done over samples with different $\phi_{x'}$. 
Note again that such averaging preserves the number of WPLs along the $x'$-direction, and therefore the WPL marked with a dashed oval in Fig.~\ref{fig:S_density_evolution_WPL}(a). The number of WPLs along the perpendicular direction, $y'$, can change from sample to sample, but due to sample averaging, their impact on the final averaged observable is small.

\begin{figure}[t!]
	\centering
    \includegraphics[width=\columnwidth]{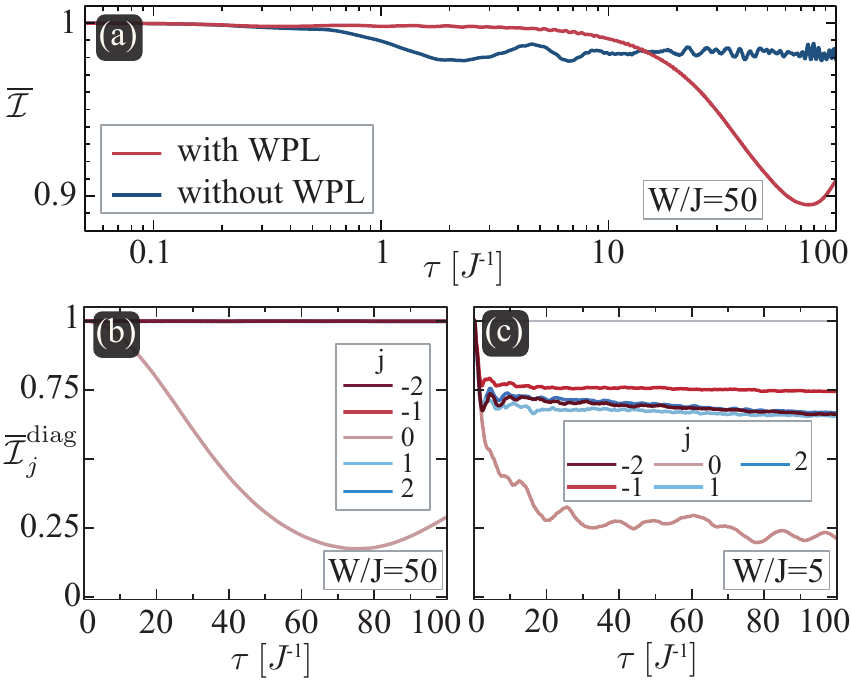}
    \caption{Quantitative effects of the diagonal WPL in the separable model. (a) Imbalance of the whole system defined in Eq.~\eqref{eq:total_imbalance} averaged over samples with (red line) and without (blue line) diagonal WPL. The system is deeply in the localized phase with potential strength set to $W/J=50$.
    Panels (b) and (c) show the diagonal imbalance defined in the main text for the localized and thermal phase, respectively. A negative/positive $j$ denotes the diagonals below/above the main diagonal marked with a dashed oval in Fig.~\ref{fig:S_density_evolution_WPL}(a). For (a) and (b) we used bond dimension $\chi=128$, while for (c) we had to use a higher bond dimension, $\chi=256$, to reach convergence.
    In all plots we use a square system of size $L_x \times L_y = 8 \times 8$ and we average over $10$ different samples, as described in the main text. A time step used in all plots is $\delta \tau = 0.05$.
    }
    \label{fig:S_imbalance_WPL}
\end{figure}

In Fig.~\ref{fig:S_imbalance_WPL}(a) we show the averaged imbalance, defined in Eq.~\eqref{eq:averaged_imbalance} and modified for a columnar density wave as the initial state, and compare the cases with and without zero-potential WPL. 
For the latter case, the averaged imbalance quickly saturates to a value close to 1. In the former, the imbalance is expected to oscillate around a value slightly lower then $1$ and, assuming that a complete delocalization occurs only on sites along the WPL, never go below $1 - L^{\rm diag}/L^2 = 1 - 8/64 = 0.875$. Furthermore, since the particles that move on the diagonal WPL need to tunnel through high potential barriers given by the values of the onsite potential of the sites adjacent to the WPL, their hopping amplitudes are $t_i^{\rm eff} \propto t^2/U_{i,i \pm 1}$, $t^2/U_{i \pm 1,i}$. In some places throughout the WPL, the effective hopping is $t_i^{\rm eff} \ll t$ and the oscillations in the imbalance caused by the finite size of the WPL have a large period. Our numerical results in Fig.~\ref{fig:S_imbalance_WPL}(a) are indeed consistent with these expectations. 

After analyzing the averaged imbalance of the whole system, we turn to the microscopic behavior of particles on the WPL and its surroundings. To check whether the ergodic WPL hybridizes with neighboring sites, therefore seeding a thermal phase, we concentrate on the partial imbalance calculated separately for several diagonals that lie along the $x'$-direction. Such imbalance is defined as 
\begin{align}
    \overline{\mathcal{I}}^{\rm diag}_j = \frac{1}{\lfloor L_j^{\rm diag}/2 \rfloor} \sum_{\vec{r} \in j-{\rm th \, diag}} (-1)^{\alpha_{\vec{r}}} \bra{\psi(\tau)} \hat{n}_{\vec{r}} \ket{\psi(\tau)}
    \label{eq:averaged_diagonal_imbalance}
\end{align}
where $\vec{r}$ marks the position of sites that are part of the diagonal, $j$ indicates the position of the diagonal, with $j=0$ being the main diagonal, explicitly marked in Fig.~\ref{fig:S_density_evolution_WPL}(a), and $\pm 1$ are the neighboring diagonals above/below it, respectively. $L_j^{\rm diag}$ is the number of sites on the $j$-th diagonal and $\alpha_{\vec{r}} \equiv \bra{\psi(0)} \hat{n}_{\vec{r}} \ket{\psi(0)}+1$.
In Fig.~\ref{fig:S_imbalance_WPL}(b), we show the numerically obtained $\overline{\mathcal{I}}^{\rm diag}_j$ for the main diagonal and two of its neighboring diagonals in each perpendicular direction. While $\overline{\mathcal{I}}^{\rm diag}_0$ shows the same oscillatory behavior as the imbalance of the whole system $\overline{\mathcal{I}}$, the other $\overline{\mathcal{I}}^{\rm diag}_{j \neq 0}$ remain constant and never go below 1. In contrast, when the system is in the ergodic phase, all diagonal imbalances $\overline{\mathcal{I}}^{\rm diag}_{j}$ decay, as it can be seen in Fig.~\ref{fig:S_imbalance_WPL}(c) where the potential strength is set to $W/J=5$. We therefore conclude that WPL does not destabilize localized parts of a 2D system, at least on the simulated timescales. Furthermore, if the system is initialised in a columnar density wave, a finite particle transport occurs along the diagonal WPLs. 

%
%%%%%
\subsubsection{Transport through the WPL \label{subsub:transport_WPL_S}}
\begin{figure}[t!]
	\centering
    \includegraphics[scale=1]{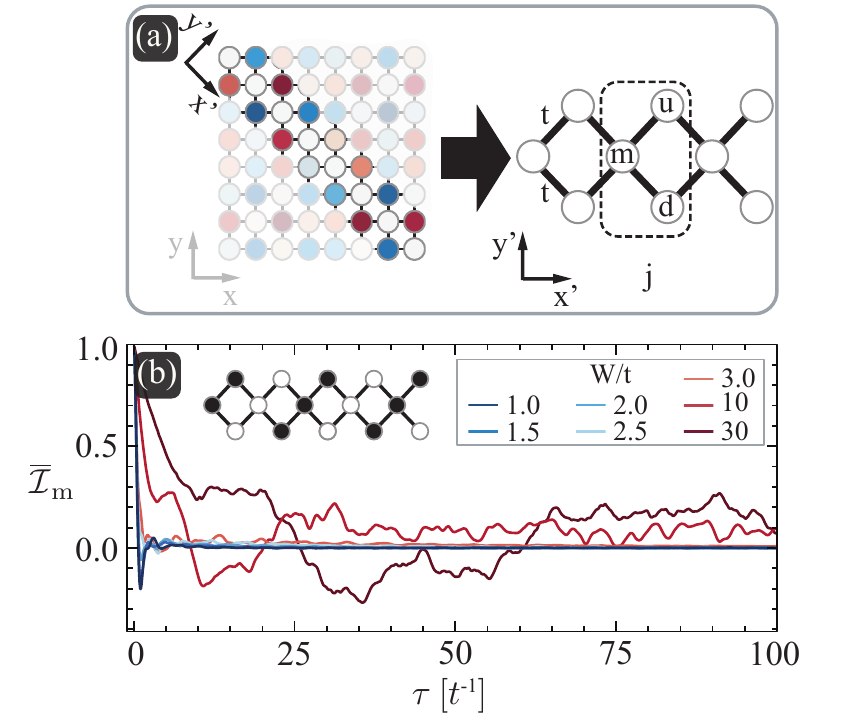}
    \caption{Many-body diamond chain and its localization properties.
    (a) By removing sites from a 2D separable AA model (left) that do not belong to the WPL or to its neighboring diagonals, one arrives at the diamond chain (right). The sites that were part of the WPL in the 2D system now belong to middle sites of the diamond chain and are denoted with $m$. Sites from the upper/lower diagonal in the 2D model lie on the upper/lower sites in the diamond chain and are denoted with $u$/$d$.
    (b) The average particle imbalance defined on the middle chain, see the main text, and calculated for an initial state that resembles the density wave used in Figs.~\ref{fig:S_density_evolution_WPL} and~\ref{fig:S_imbalance_WPL}, reproduced in the inset. The imbalance decays for all values of $W/t$ and there is no sign of a localization transition.
    We used systems with $L=20$ unit cells and the averaging was performed over $10$ different samples. We also used a time step $\delta \tau = 0.1$ and a bond dimension $\chi=128$. 
    \label{fig:diamond_chain}}
\end{figure}

%%%%%
%
To investigate transport of particles along the diagonal WPL in more detail, we turn to a simpler system shown in Fig.~\ref{fig:diamond_chain}(a), known as a \textit{diamond chain}~\cite{Vidal_2000,Danieli2015,Roy2020}.
This quasi-one-dimensional geometry corresponds to three neighboring diagonals of a 2D square lattice. We set the hopping $t$ to be constant (i.e., $t=J/2$, where $J=1$) and the onsite potentials to be 
\begin{align}
    U_j^u &= 2 W \cos(-\pi b + \phi_{y'}) \, \cos(2 \pi b j + \pi b + \phi_{x'})  \nonumber\\
    U_j^m &= 2 W \cos( \phi_{y'}) \, \cos(2 \pi b j + \phi_{x'})  \nonumber\\
    U_j^d &= 2 W \cos(\pi b + \phi_{y'}) \, \cos(2 \pi b j + \pi b + \phi_{x'}) \, ,
    \label{eq:WPL_potential}
\end{align}
corresponding to any three adjacent diagonals of the separable AA model that extend in $\vec{x'}$ direction. To obtain a WPL, we need to set $\phi_{y'}=\pi/2 + \delta$, where $\delta$ is small enough, such that the on-site potential on the middle chain ($m$) is much smaller than on the upper ($u$) and lower ($d$) chains. The case $\delta=0$ describes a WPL with zero potential. With a diamond chain geometry, we can simulate the dynamics of substantially longer WPLs than the ones studied in square and rectangular systems.

Single-particle properties of a diamond chain are discussed in detail in App.~\ref{app:details_separable_model}. In the case of a zero-potential WPL ($\delta = 0$), the states in the spectrum with $E\neq 0$ undergo a localization transition at the critical point $W_{\rm C}/t = \sqrt{2}/|\cos(\pi b + \frac{\pi}{2})|$, see Eq.~\eqref{eq:diamond_chain_Wc}. At $E=0$, there exist an extensive number of states that remain extended for any $W$. The wave function of these states spreads mostly over the middle chain $m$ with excursions into $u$ and $d$ chains at positions $j$ where the onsite potential $U_j^{u,d}$ is weak. 
On the other hand, for a WPL with finite onsite potentials, i.e., when $\delta \neq 0$, a mobility edge develops and the localization of $E \neq 0$ is no longer uniform throughout the spectrum. Furthermore, there are no extended zero energy states, but all states localize above some $W$. 
By adding the interactions to a many-body diamond chain with $\delta \neq 0$, we expect a many-body localization transition to occur, see App.~\ref{app:details_separable_model} and Fig.~\ref{fig:MB_diamond_chain}, because all the states in the spectrum are localized above a certain potential strength. For $\delta = 0$ case, the interactions will lift the degeneracy of $E=0$ states and localize them, therefore, again leading to a many-body localization.

To understand the dynamics along the WPL illustrated in Figs.~\ref{fig:S_density_evolution_WPL} and~\ref{fig:S_imbalance_WPL}, one ought to consider the peculiarity of the corresponding initial state shown in the inset of Fig.~\ref{fig:diamond_chain}(b). For a strong potential, particles in the upper and lower sites are localized and therefore restrict the hopping of particles from the middle sites (recall that multiple occupancy of a site is not allowed). 
As a consequence, particles initialized on the middle sites are able to move only along a one-dimensional zigzag pattern which is obtained by excluding the occupied upper and lower sites, see Fig.~\ref{fig:single_particle_mosaic_chain}(a). The geometry that follows from the aforementioned zigzag pattern is known as a quasiperiodic \textit{mosaic lattice}~\cite{Wang2020}. 
The single-particle spectrum of a quasiperiodic mosaic lattice has a mobility edge given by 
$E_{\rm C} = 2V \pm t^2/(W |\cos(\pi b + \frac{\pi}{2})|)$ when $\delta=0$, Eq.~\eqref{eq:E_c_mosaic_lattice} (a full derivation is given in App.~\ref{app:details_separable_model_mosaic} for completeness). This means that for any value of $W$ there exist extended states with energies in the interval $-|E_{\rm C}| < E < |E_{\rm C}|$. 
These single-particle states can then make the interacting system thermal even above the transition point of a diamond chain given by Eq.~\eqref{eq:diamond_chain_Wc}.
For the case of a WPL that has non-zero onsite potential, i.e., when $\delta \neq 0$, the single-particle mobility edge described above breaks down and a localization of all states occurs above a certain $W/t$ that depends on $\delta$, see App.~\ref{app:details_separable_model_mosaic} and Fig.~\ref{fig:single_particle_mosaic_chain}(c). The many-body system is then also expected to be localized for strong enough potentials. However, if $\delta$ is very small, we expect the localization in a many-body mosaic lattice to occur at values that are much larger than $W_{\rm C}$ discussed in Sec.~\ref{sec:MB_properties_NS&S}. 
Therefore, the transport through WPLs with small enough $\delta$ can survive even if the rest of the 2D system described by the separable AA model is deeply inside the localized phase.

Let us now study the effect of nearest neighbor interactions in a diamond chain, using the same TDVP method as before. We initialize the system in the state shown in Fig.~\ref{fig:diamond_chain}(b), evolve it with the TDVP algorithm, and calculate the imbalance $\overline{\mathcal{I}}_m$ on the middle chain only. This corresponds to the quantity $\overline{\mathcal{I}}^{\rm diag}_0$ defined in Eq.~\eqref{eq:averaged_diagonal_imbalance}.
Note that we do not restrict the particles to move in a zig-zag pattern, but the simulations are preformed on the full diamond chain lattice with the described initial state. 
The result for an interacting model with $V=t$ is shown in Fig.~\ref{fig:diamond_chain}(b).
The imbalance $\overline{\mathcal{I}}_m$ decays for all values of $W/t$ meaning that the system does not localize even well above the single-particle transition $W_{\rm C}$ obtained in Eq.~\eqref{eq:diamond_chain_Wc}. Therefore, the chain supports transport irrespective of the potential strength, as predicted in agreement with our previous considerations. 

%
%%%%%%%%%%%%%%%%%%%%%%%%%%%%%%%%%%%%%%%%%%%%%%%%%%%
\section{Discussion and conclusions  \label{sec:discussion_and_conclusion}}
%%%%%%%%%%%%%%%%%%%%%%%%%%%%%%%%%%%%%%%%%%%%%%%%%%%
%
In this work, we studied the out-of-equilibrium dynamics of interacting hard-core bosons in two realizations of the 2D Aubry-Andr\'{e} model, namely a separable (S) and a nonseparable (NS) model. The time-dependent variational principle (TDVP) was employed to numerically study the time evolution of large systems, up to $\sim 100$ sites.

We find that a stable many-body localized (MBL) phase exists in 2D quasiperiodic systems, at least as far as we can infer from the system sizes and timescales that we were able to access -- in stark contrast to analogous two-dimensional random models, where over comparable system sizes and time scales a clear instability is seen, in the form of a drift with system size~\cite{Doggen2020}.
The occurrence of MBL is inferred from the saturation of the averaged particle imbalance in large systems. More precisely, by fitting a power law decay to the imbalance -- which is found to be a good approximation over our accessible timescales of a hundred hopping times (Fig.~\ref{fig:imbalance}) -- we observe that the decay coefficient $\gamma$ decreases with increasing potential strength $W$, until it vanishes for $W/J \gtrsim W_{\rm C}/J \approx 30$ in both S and NS models. This is consistent with a transition from an ergodic to an MBL phase. 

While we can disregard the possibility of rare weakly disordered regions due to the quasiperiodic, deterministic nature of the potential, our systems are subject to weak potential lines (WPLs). Within the parameter range studied in this work, WPLs do not appear to lead to global thermalization even at the long times studied in this work, as seen for example in Figs.~\ref{fig:NS_imbalance_WPL}(b) and~\ref{fig:S_imbalance_WPL}(b). To test this, we consider the influence of system size and indeed we observe that $W_{\rm C}$ remains constant within our accessible system sizes, Fig.~\ref{fig:decay_coefs}(a) and (b), in contrast to similar simulations of random 2D systems~\cite{Doggen2020} and also to the additional numerics in this paper (Fig.~\ref{fig:decay_coefs}(c)). 

Since our study is based on numerical simulations, it is difficult to make rigorous claims about the stability of MBL in infinite quasiperiodic systems and at infinite times. There could exist, in principle, extremely slow processes that are able to thermalize the system only at length and time scales much larger than the ones we probe. For example, resonances between distant sites in large systems could lead to slow dynamics that activates at extremely long timescale~\cite{Gopalakrishnan2020,Leonard2020,Morningstar2021}. 
Another caveat of the numerical analysis is that for finite WPLs the many-body level spacing could be larger than the decay rate of the localized parts coupled to the WPL, which is given by the Fermi golden rule, and thus the avalanche would not start. On the other hand, for infinite systems, ergodic WPLs have infinite length with the level spacing equal to zero, which means that they act as perfect baths. According to Refs.~\onlinecite{Chandran2016,DeRoeck2017a,DeRoeck2017b} this would lead to an unbounded growth of the ergodic regions until the whole system is eventually thermalized. 
Nevertheless, due to the small decay rate of the localized parts coupled to the WPL, the avalanche would be an extremely slow process with a timescale much larger than the ones accessible by the state-of-art experiments.

Interestingly, however, WPLs do affect the transport properties even on short timescales. This manifests in a different way for the S and NS models, as can be seen from the analysis of individual samples before averaging, see Fig.~\ref{fig:histograms}. 
While the distribution of $\gamma$ in the S model shrinks towards a delta function when tuning $W$ to larger values, in the NS model it remains relatively broad even inside the localized phase. 
This is attributed to the WPLs that spread out along the $x$- and $y$-direction in the NS model, which allow for the long range transport of particles, and therefore cause the imbalance to partially decay. As discussed in Sec.~\ref{subsec:WPL_nonseparable}, transport through such WPLs is not affected by the choice of the initial state, and exists even when the rest of the system is deeply inside the localized phase, see Figs.~\ref{fig:NS_density_evolution_WPL} and~\ref{fig:NS_imbalance_WPL}. Such a coexistence of finite particle transport within an MBL phase has, to the best of our knowledge, never been discussed before. 

WPLs also appear in the S model, but contrary to the NS model, they lie on diagonals. Consequently, the transport over them strongly depends on the choice of an initial state. 
The checkerboard initial state either completely fills the sites of the WPL or leaves them empty -- hence, transport is strongly suppressed when $W>W_{\rm C}$. In Sec.~\ref{subsec:WPL_separable} we considered a different initial state, namely a columnar density wave, that fills every second site of the diagonal WPL. In this case, long range transport is indeed recovered, Fig.~\ref{fig:S_imbalance_WPL}(a) and (b), with particles hopping alternatively on sites above and below the WPL in order to move along the diagonal, forming a zigzag pattern. The behavior is equivalent to a 1D mosaic lattice, which contains a mobility edge in the spectrum and extended single-particle states exist at large $W$ (which can go to infinity in a case of a WPL with zero potential), see App.~\ref{app:details_separable_model_mosaic}. Such extended states support transport both in single-particle and many-body cases, Fig.~\ref{fig:diamond_chain}(b). Since in an infinite 2D system it is always possible to find a WPL with infinitesimally low potential, transport over such WPL will not be suppressed no matter how strong the potential is, as long as the system evolves from a columnar density wave. Note, however, that a conducting WPL does not affect the rest of the 2D system, which remains localized.
On the other hand, if the localized 2D system is initialized in another state, e.g., where particles are randomly distributed throughout the system, the dynamics over a WPL can be mapped onto a diamond chain, which has an MBL transition at finite $W_{\rm C}$. As a consequence, no transport is possible anywhere in the localized system described by the S model, see App.~\ref{app:details_separable_model} and Fig.~\ref{fig:MB_diamond_chain}.

The results presented in this work are of direct relevance to experimental realizations, for instance using ultracold atoms in optical lattices. 
A particularly promising experiment is discussed in Refs.~\cite{Viebahn2019,Sbroscia2020,Sanchez-Palencia2005,Gautier2021}, where two square optical lattices are created with pairs of perpendicular light beams. In such a setup, by setting one optical lattice to be very deep and the other one shallow, one can create a tight-binding square lattice with an external 2D cosine potential, respectively. Quasiperiodicity of the external potential then comes from the incommensurate wavelengths of the shallow and deep lattices. To create a separable model, two square lattices should be aligned, while for the nonseparable model the lattices should form an angle of $45^{\circ}$. Interactions in ultracold atomic systems can be easily tuned. Such an experiment could go a step further and, along with testing the predictions of this paper, it could also shed light in what is happening inside the ergodic phase of the models ($W \ll W_{\rm C}$). 

Our work opens up a wide range of important avenues for investigation of MBL in 2D systems. One direction is to study MBL in more exotic models like Bose-Hubbard or Fermi-Hubbard models with the presence of a quasiperiodic AA potential. Another interesting question is related to the stability of the MBL in 2D systems that are subjected to long range interactions, which are of particular interest for experiments involving trapped ions~\cite{Smith2016}. WPLs may, in this case, be sufficient to globally thermalize the system on timescales much shorter than in our case of contact interactions. Furthermore, using our approach it is possible to investigate the effects of coupling the localized system to various types of thermal baths (e.g., one- and two-dimensional baths).
Our approach can also be deployed to study MBL in 2D nonperiodic lattices like the Penrose or Ammann-Beenker tiling, which are relevant to ongoing experiments in ultracold atomic systems~\cite{Viebahn2019,Sbroscia2020}. Such lattices with quasiperiodic arrangements of both onsite potential and hopping should not host any regions of weak potential, meaning that transport of particles should be fully suppressed inside the localized phase.

Lastly, the properties of the ergodic phase, i.e., the parameter regime with $W \ll W_{\rm C}$, of the models studied in our work remain unexplored. 
The entanglement entropy grows much faster inside the ergodic phase, compared to the MBL phase, and thus the TDVP method quickly becomes impractical due to the large bond-dimensions needed for such simulations, which make the numerical calculations exceedingly long. One could tackle this problem by mapping it to the analogous problem of Anderson localization on appropriate Bethe lattices~\cite{DeLuca2014, Bera2018, DeTomasi2020}. This simplified approach could give some insight into the transport properties inside the ergodic phase, for instance whether there is a subdiffusive regime present when approaching $W_{\rm C}$, like in the case of random 2D interacting systems, or it is absent like in a 1D Aubry-Andr\'{e} model~\cite{Znidaric2018,Znidaric2021}.

\textit{Note added:} While preparing this manuscript, we became aware of Refs.~\onlinecite{Agrawal2022,Crowley2022}, which use different approaches to argue the stability of MBL in 2D quasiperiodic systems. Ref.~\onlinecite{Crowley2022} also argues that large enough ergodic inclusions -- such as the WPLs -- will eventually destabilize the MBL phase and thermalize the whole system. While we do not observe it in our simulations, as mentioned in Sec.~\ref{sec:discussion_and_conclusion} this could be because such thermalizing process occur on exceptionally large length and timescales, beyond the ones studied in our work.

%
%%%%%%%%%%%%%%%%%%%%%%%%%%%%%%%%%%%%%%%%%%%%%%%%%%%
\section*{Acknowledgements}
%%%%%%%%%%%%%%%%%%%%%%%%%%%%%%%%%%%%%%%%%%%%%%%%%%%
%
We are grateful to P.~Crowley, I.~V.~Gornyi, A.~D.~Mirlin, D.~G.~Polyakov, and U.~Schneider for insightful discussions. 
A.Š. acknowledges financial support from the Swiss National Science Foundation (Grant No.~199969).
C.C. and A.Š. acknowledge financial support from the Engineering and Physical Sciences Research Council (EPSRC) (Grants No.~EP/M007065/1,~EP/P034616/1,~EP/T028580/1, and~EP/V062654/1).
Numerical simulations were performed using the TeNPy library~\cite{tenpy}. All data that support the plots within this paper are available from the corresponding author upon request.

%
%%%%%%%%%%%%%%%%%%%%%%%%%%%%%%%%%%%%%%%%%%%%%%%%%%%
\appendix
%%%%%%%%%%%%%%%%%%%%%%%%%%%%%%%%%%%%%%%%%%%%%%%%%%%
%

%
%%%
\section{Details of the separable model  \label{app:details_separable_model}}
%%%
%
\subsection{Single-particle properties of the diamond chain}

In the first part of this appendix, we discuss the single-particle properties of the diamond chain shown in Fig.~\ref{fig:diamond_chain}, cf.~Ref.~\cite{Danieli2015}.
Denoting the projection of the wave function onto the upper, middle, and lower site of the $j$-th unit cell (see Fig.~\ref{fig:diamond_chain}(a)) with $\psi^u_j$, $\psi^m_j$ and $\psi^d_j$, respectively, the (discrete) Schr\"{o}dinger equation of the system decomposes into the following set of coupled equations:
\begin{align}
    (E-U^u_j) \psi^u_j &= t (\psi^m_j + \psi^m_{j+1}) \nonumber\\
    (E-U^m_j) \psi^m_j &= t (\psi^u_j + \psi^d_j) + t (\psi^u_{j-1} + \psi^d_{j-1}) \nonumber\\
    (E-U^d_j) \psi^d_j &= t (\psi^m_j + \psi^m_{j+1}) \, ,
    \label{eq:Schroedinger_eqs_diamond_chain}
\end{align}
with the quasiperiodic potentials $U^{u,m,d}_j$ given by Eq.~\eqref{eq:WPL_potential}.

In the case where a WPL lies in the middle chain, i.e., when $\phi_{y'}=\pi/2$, the set of Eqs.~\eqref{eq:Schroedinger_eqs_diamond_chain} simplifies to
\begin{align}
    (E-\epsilon_j) \psi^u_j &= t (\psi^m_j + \psi^m_{j+1}) \nonumber\\
    E \psi^m_j &= t (\psi^u_j + \psi^d_j) + t (\psi^u_{j-1} + \psi^d_{j-1}) \nonumber\\
    (E+\epsilon_j) \psi^d_j &= t (\psi^m_j + \psi^m_{j+1}) \, ,
    \label{eq:WPL_eqs_diamond_chain}
\end{align}
where we defined
\begin{align}
  \epsilon_j &\equiv 2 W |\cos(\pi b + \pi/2)| \cos(2 \pi b j + \phi_{x'}) \nonumber\\ &\equiv \widetilde{W}\cos(2 \pi b j + \phi_{x'}) \, .
  \label{eq:epsilon_j_diamond_chain}
\end{align}
Note that the phase shift $\pi b$ in the second cosine in $U_j^{u,d}$, see Eqs.~\eqref{eq:WPL_potential}, can be trivially incorporated into $\phi_{x'}$, therefore, we omit it from Eq.~\eqref{eq:epsilon_j_diamond_chain}.

\begin{figure}[t!]
	\centering
    \includegraphics[scale=1]{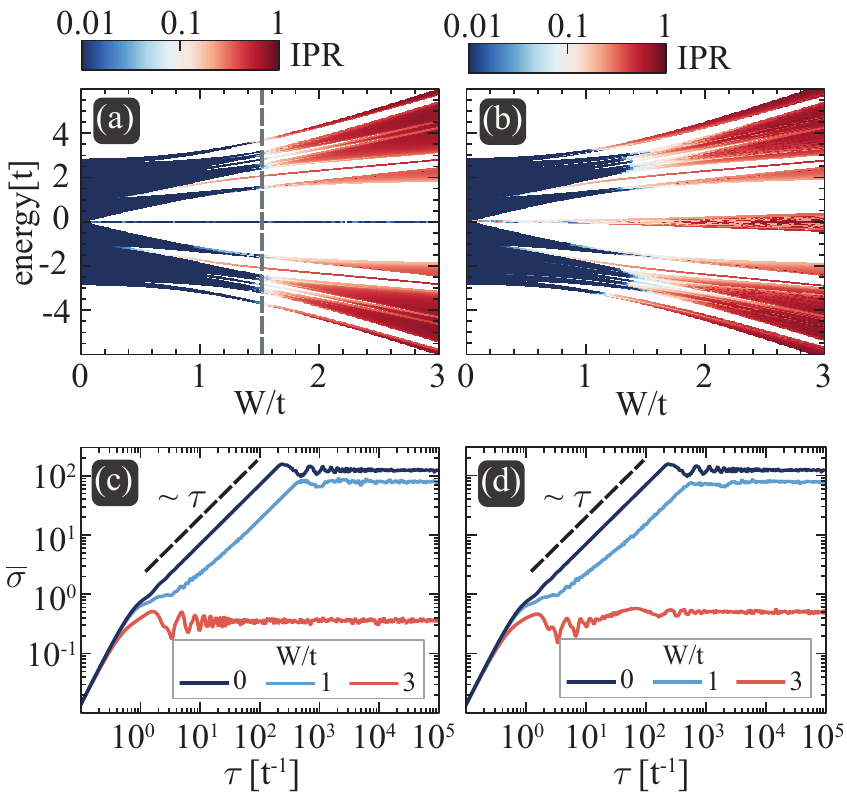}
    \caption{Single-particle localization properties of the diamond chain. The IPR for all the eigenstates is shown as a function of the potential strength $W/t$ for (a) $\phi_{y'}=\pi/2$ and (b) $\phi_{y'}=\pi/2+0.1$, with $\phi_{x'}=0$ in both. In (a) a sharp transition, marked by the gray dashed line, Eq.~\eqref{eq:diamond_chain_Wc}, can be seen for states with $E \neq 0$. In (b) there is no such transition, and a mobility edge appears. Panels (c) and (d) show the average displacement defined in Eq.~\eqref{eq:averaged_displacement} for $\phi_{y'}=\pi/2$ and $\phi_{y'}=\pi/2+0.1$, respectively. 
    The system in (a) and (b) has $L=233$ unit cells, while in (c) and (d) we used $L=610$ unit cells. Averaging in (c) and (d) was performed over $100$ realizations of the potential with randomly chosen $\phi_{x'}$. 
    }
    \label{fig:single_particle_diamond_chain}
\end{figure}
To analyze the spectrum of the diamond chain and its localization properties, we map the original lattice onto a Fano lattice~\cite{Flach2014, Danieli2015} by applying local transformations of the amplitudes 
\begin{align}
    \begin{pmatrix} 
        p_j \\
        c_j \\
        f_j
    \end{pmatrix} = \frac{1}{\sqrt{2}}
    \begin{pmatrix}
        1 & 0 & 1 \\
        0 & \sqrt{2} & 0 \\
        1 & 0 & -1
    \end{pmatrix}
    \begin{pmatrix}
        \psi^u_j \\
        \psi^m_j \\
        \psi^d_j
    \end{pmatrix} \, .
\end{align}
This also locally rotates the values of the onsite potential to $\epsilon_{\pm} = (U_j^u \pm U_j^d)/2$, which in the case of Eq.~\eqref{eq:WPL_eqs_diamond_chain} leads to $\epsilon_+ = 0$ and $\epsilon_- = \epsilon_j$. The Schr\"{o}dinger equation in the new basis reads
\begin{align}
    E p_j &= \epsilon_j f_j + \sqrt{2} t (c_j + c_{j+1}) \nonumber\\
    E f_j &= \epsilon_j p_j \nonumber\\
    E c_j &= \sqrt{2} t (p_{j-1} + p_j) \, .
    \label{eq:Fano_lattice_Sch}
\end{align}
Combining the three equations from above, and using a trigonometric identity
$ \epsilon_j^2 = \frac{\widetilde{W}^2}{2} + \frac{\widetilde{W}^2}{2} \cos(4 \pi b j + 2 \phi_{x'})$,
we arrive at the single equation for $p_j$
\begin{align}
  (E - 4t^2 - \frac{\widetilde{W}^2}{2}) p_j = &\frac{\widetilde{W}^2}{2} \cos(4 \pi b j + 2 \phi_{x'}) p_j \nonumber\\
  & + 2t^2 (p_{j-1} + p_{j+1}) \, .
  \label{eq:pj_equation_diamond_chain}
\end{align}
The form of the equation above is the same as in the 1D Aubry-Andr\'{e} model~\cite{Aubry1980}, which is known to have a localization-delocalization transition at the point where the model is self-dual, i.e., when the strength of the onsite potential is two times larger than the hopping. 
Therefore, we can use the same self-duality argument, which includes transforming $p_j = \exp(i \theta j) \sum_{k=-\infty}^{k=\infty} p_k \exp[i k (4 \pi b j + 2\phi_{x'})]$, where $\theta$ is an arbitrary phase, and comparing the equations for $p_j$ and $p_k$,
to conclude that the localization transition for all $p_j$ states occurs at 
\begin{align}
    \frac{\widetilde{W}_{\rm C}}{t} = 2 \sqrt{2} \quad \longrightarrow \quad \frac{W_{\rm C}}{t} = \frac{\sqrt{2}}{|\cos(\pi b + \frac{\pi}{2})|}\, .
    \label{eq:diamond_chain_Wc}
\end{align}

From the numerical results in Fig.~\ref{fig:single_particle_diamond_chain}(a), it is clear that all states with $E \neq 0$ indeed undergo a transition at $\widetilde{W}_{\rm C}/t$, as discussed above.
However, there are some states at zero energy that do not follow the same trend, and stay delocalized even well above $\widetilde{W}_{\rm C}$, as one can see from Eq.~\eqref{eq:Fano_lattice_Sch}. For $E=0$, the second line gives $p_j=0$, $\forall j$, and the third line is then automatically satisfied; the first line gives a wave function that is non-vanishing only on the middle chain, $c_j$, and on the antisymmetric combination of the upper and the lower chains, $f_j$.
There are precisely $L$ zero-energy states and they spread throughout the whole chain, with their weights on $u,m,d$ sites depending on the local configuration of the potential, $\epsilon_j$. 

If instead we had chosen the phase $\phi_{y'} = \pi/2 + \delta$, with $\delta \neq 0$, we would have obtained a WPL with finite onsite quasiperiodic potential. The localization properties are then radically different, as it can be seen from Fig.~\ref{fig:single_particle_diamond_chain}(b). Firstly, there is no longer a uniform localization transition for states at $E \neq 0$, as the spectrum develops a nontrivial mobility edge~\cite{Danieli2015}. Secondly, zero energy states that are present in the $\phi_{y'} = \pi/2$ case are now dispersed in a finite window that depends on $W$, and they localize for a strong enough potential.

Lastly, we investigate how the localization properties of the spectrum affect the transport through the chain when a particle is initially placed in the middle chain, i.e. $\psi_{j}^{\alpha} (\tau=0) = \delta_{\alpha, m} \delta_{j,j_0}$ at the position $j_0=L/2$. We calculate the average displacement of the initial wave packet 
\begin{align}
    \overline{\sigma} = \frac{1}{\mathcal{N}} \sum_{\phi_{x'}} \sum_{j=0}^{L} \Big( \expval{\psi_{\phi_{x'}}(\tau)| (j-j_0)^2 |\psi_{\phi_{x'}}(\tau)} \Big)^{1/2} \, ,
    \label{eq:averaged_displacement}
\end{align}
where the first sum is over $\mathcal{N}$ systems with different randomly chosen $\phi_{x'}$. The result is shown in Fig.~\ref{fig:single_particle_diamond_chain}(c) and (d) for cases where $\phi_{y'}=\pi/2$ and $\phi_{y'}=\pi/2+0.1$, respectively. In both cases, $\overline{\sigma}$ grows linearly when the spectrum contains only extended states (see curves for $W/t=0$ and $W/t=1$), which is expected for a ballistic expansion, until they saturate due to the finite system size. For $W/t=3$, on the other hand, the average displacement saturates within a few hopping times in both Fig.~\ref{fig:single_particle_diamond_chain}(c) and (d). Therefore we conclude that there is no particle transport when the potential is strong enough, namely when $W>W_{\rm C}$ for a WPL with $\phi_{y'}=\pi/2$, and when all states are eventually localized for a WPL with $\phi_{y'}=\pi/2+\delta$.

%
%%%%%
\subsection{Transport and localization in the interacting many-body diamond chain}
%%%%%
%
\begin{figure}[h!]
	\centering
    \includegraphics[scale=1]{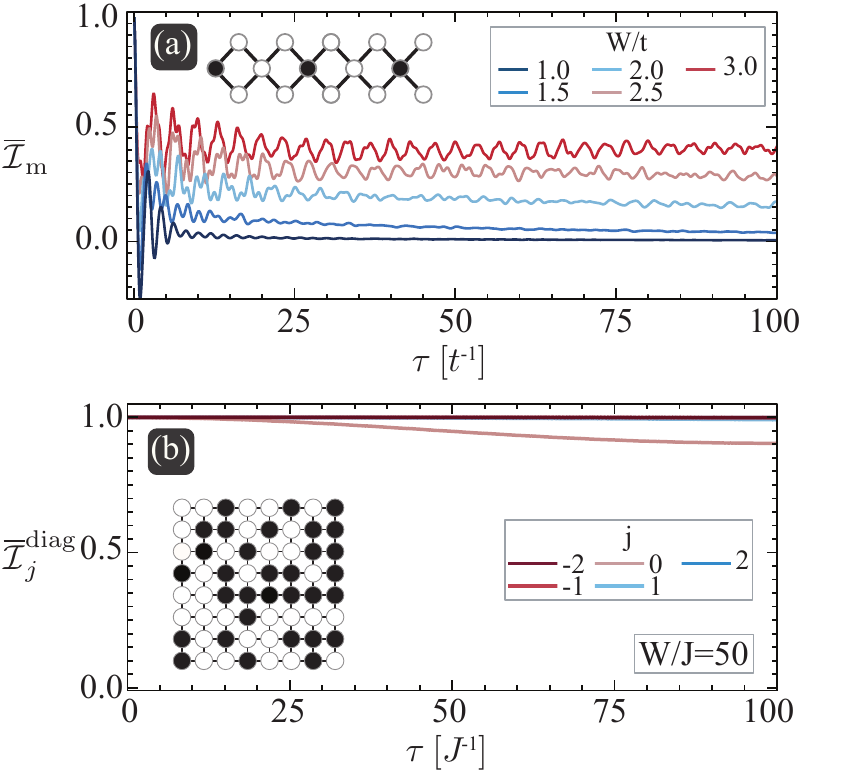}
    \caption{Panel (a): The average imbalance calculated over the middle chain defined in the main text, and with the initial state resembling a density wave only on the middle chain, as illustrated in the inset. The decay of the imbalance slows down with increasing $W/t$ until it stops for $W/t \gtrapprox 2.5$.
    We used systems with $L=20$ unit cells and the averaging was performed over $10$ different samples. We used a time step $\delta \tau = 0.1$ and bond dimension $\chi=128$.
    Panel: (b) Diagonal imbalance averaged over $10$ random initial states (the inset shows one such random state) and for the same samples used in Fig.~\ref{fig:S_imbalance_WPL}(b) that contain a WPL on the main diagonal. In this case, all diagonal imbalances saturate to a value that is close to $1$. We use the same parameters as in Fig.~\ref{fig:S_imbalance_WPL}(b).
    }
    \label{fig:MB_diamond_chain}
\end{figure}
Now we turn to the many-body version of the diamond chain analyzed in the previous section in presence of nearest-neighbor interactions. 
The question we address here is whether the interactions destabilize the localized phase and introduce some finite transport over the chain. We employ the same approach as in the main text and calculate the dynamics of the system 
after preparing it in a given initial state, using the TDVP method. We use a N\'{e}el state on the middle sites, illustrated in the inset of Fig.~\ref{fig:MB_diamond_chain}(a), and we calculate the particle imbalance as a function of time. 
The results averaged over samples with different $\phi_{x'}$, see Eq.~\eqref{eq:epsilon_j_diamond_chain}, is shown in Fig.~\ref{fig:MB_diamond_chain}(a). 
The imbalance decays to $0$ for low values of the potential $W/t$, but the decay slows down with increasing $W/t$. For $W/t \gtrapprox 2.5$, the averaged imbalance saturates at a finite value, indicating the presence of localization. The only effect of interactions in this case appears to be a shift of the single-particle transition point to a larger value of $W_{\rm C}/t \approx 2.5$.
The presence of the many-body localized phase is expected since all $E \neq 0$ single-particle states localize at $W_{\rm C}/t$, given by Eq.~\eqref{eq:diamond_chain_Wc}, and extended states at $E=0$ are extremely fragile, meaning that any additional perturbation leads to their localization.
Therefore, we conclude that for $W/t \gtrapprox 2.5$ there is no transport of particles across the diamond chain when the upper and lower sites are initially empty.  

The results in this appendix tell us that transport through diagonal WPLs in the 2D separable AA model is suppressed if one considers random initial states. To confirm this, we calculate the diagonal imbalance averaged over the same samples as in Fig.~\ref{fig:S_imbalance_WPL}(b), but now starting from different random product states (see the inset of Fig.~\ref{fig:MB_diamond_chain}(b) for an example of a random state). Contrary to the results for the initial columnar density wave state considered earlier, Fig.~\ref{fig:S_imbalance_WPL}(b), all diagonal imbalances saturate in Fig.~\ref{fig:MB_diamond_chain}(b). A small deviation of $\overline{\mathcal{I}}_{j=0}$ (where $j=0$ is the position of the WPL) from $1$ is due to some of the random initial configurations that locally, i.e., close to the WPL, look similar to a columnar density wave. Such local regions can be described as a mosaic lattice and particles can spread out across that part of the WPL. Note, however, that such local regions are statistically limited in size, and therefore particles remain unable to travel over the whole WPL.

%
%%%
\subsection{Mosaic lattice limit \label{app:details_separable_model_mosaic}}
%%%
%
\begin{figure}[h!]
	\centering
    \includegraphics[scale=1]{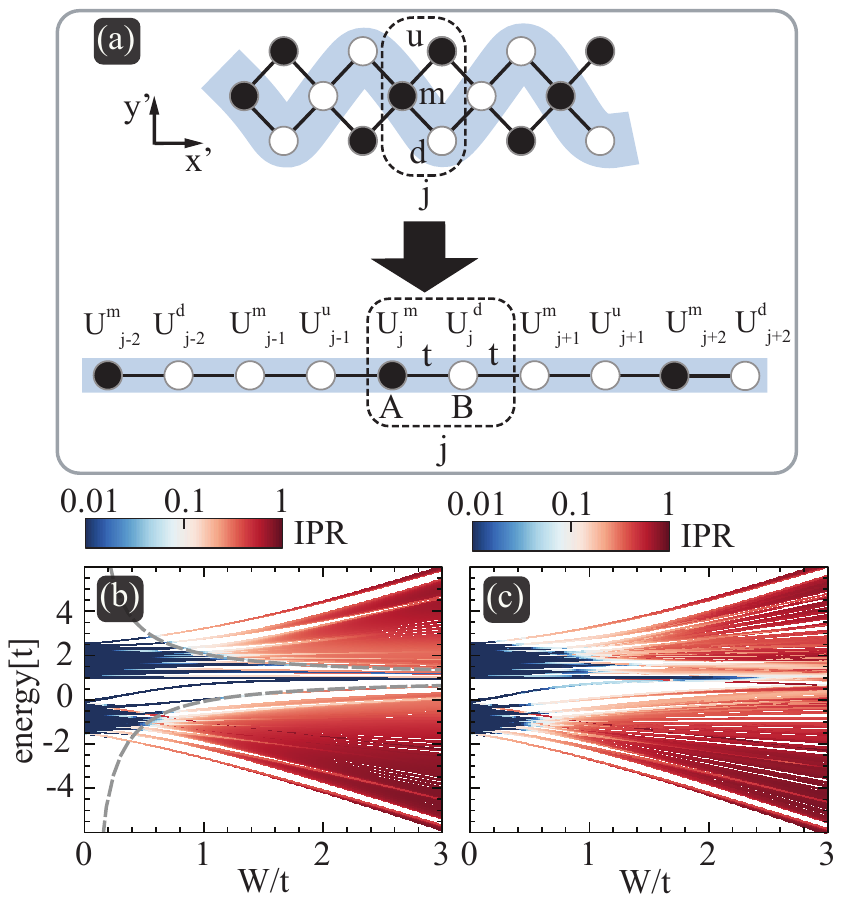}
    \caption{ Mosaic lattice and its localization properties.
    (a) A transformation from a half-filled diamond chain (up) with a density wave discussed in Fig.~\ref{fig:diamond_chain}(c) to a simpler mosaic lattice (down). Black and white circles denote occupied and unoccupied sites of a diamond chain. 
    The transformation consists of removing the occupied sites of upper and lower chains in a diamond chain. Blue wavy curve marks the sites in a diamond chain that are described by a mosaic lattice. For visualization purposes, we leave the black sites, which were a part of the blue wavy line in a diamond chain, in the mosaic lattice after the transformation. However, we concentrate only on the single-particle properties of the mosaic lattice, and not on a quarter-filling as shown here.
    In (b) and (c), IPR of each eigenstate is shown as a function of the potential strength W/t for $\delta=0$ and $\delta=0.1$, respectively. We used a system with $L=233$ unit cells, and without loss of generality set $\phi_{x'}=0$.
    }
        \label{fig:single_particle_mosaic_chain}
\end{figure}
When the many-body diamond chain is initialized in the state shown in Fig.~\ref{fig:single_particle_mosaic_chain}(a), see also the inset of Fig.~\ref{fig:diamond_chain}(b), particles initially located on the upper and lower sites are localized and impede the motion of particles that are on the adjacent middle sites. The latter can therefore hop only along a zigzag pattern, namely the sites highlighted in blue in Fig.~\ref{fig:single_particle_mosaic_chain}(a). Such zigzag pattern, together with the potentials on the upper and lower sites, can be mapped onto a quasiperiodic \textit{mosaic lattice}.
A mosaic lattice has two orbitals per site $A$ and $B$, each subject to a quasiperiodic onsite potential $U_j^A$ and $U_j^B$, respectively. In Ref.~\cite{Wang2020}, a quasiperiodic mosaic lattice was studied with $U_j^A=0$ and with $U_j^B$ given by a cosine modulation that is incommensurate with the underlying lattice. Such a model was shown to have analytical mobility edges that separate localized from extended states. 
In this appendix, we show that by removing the occupied sites from the upper and lower chains of the diamond chain, one obtains a quasiperiodic mosaic lattice similar to the one discussed in Ref.~\cite{Wang2020}. We first study a $\delta=0$ case which can be analytically solved.

Let us start by writing the Hamiltonian for the simplified chain shown in the bottom of Fig.~\ref{fig:single_particle_mosaic_chain}(a): 
\begin{align}
    H_{\delta=0} =& t \, \sum_{j=1}^L \left( b^{A \dagger}_j b^B_j + b^{B \dagger}_j b^A_{j+1} + {\rm h.c.} \right) \nonumber\\
    &+ \sum_{j=1}^L \left( U^A_j b^{A \dagger}_j b^{A}_j + U^B_j b^{B \dagger}_j b^{B}_j \right) \, .
\end{align}
After the transformation shown in Fig.~\ref{fig:single_particle_mosaic_chain}(a), it follows that
\begin{align}
    U_j^A &=  2 V \nonumber\\
    U_j^B &= 2 W (-1)^j \cos(-\pi b + \frac{\pi}{2}) \, \cos(2 \pi b j + \phi_{x'}) \nonumber \\
          &= (-1)^j \epsilon_j \, ,
\end{align}
where $U_j^A$ is a background potential imposed by the localized particles on $u$/$d$ sites, and $\epsilon_j$ is defined in Eq.~\eqref{eq:epsilon_j_diamond_chain}. The Hamiltonian then reads 
\begin{align}
    H_{\delta=0} =& t \, \sum_{j=1}^L \left( b^{A \dagger}_j b^B_j + b^{B \dagger}_j b^A_{j+1} + {\rm h.c.} \right) \nonumber\\
    &+ \sum_{j=1}^L (-1)^j \epsilon_j b^{B \dagger}_j b^{B}_j + 2 V\sum_{j=1}^L b^{A \dagger}_j b^{A}_j  \, ,
\end{align}
which under the transformation $b^B_j \rightarrow (-1)^j b^B_j$ becomes the model discussed in Ref.~\cite{Wang2020}: 
\begin{align}
    H_{\delta=0} =& t \, \sum_{j=1}^L \left( b^{A \dagger}_j b^B_j + b^{B \dagger}_j b^A_{j+1} + {\rm h.c.} \right) \nonumber\\
    &+ \sum_{j=1}^L \epsilon_j b^{B \dagger}_j b^{B}_j + 2 V\sum_{j=1}^L b^{A \dagger}_j b^{A}_j  \, .
\end{align}

To show that the model above has a mobility edge, we write the Schr\"{o}dinger equations for the wavefunctions $\psi^A_j$ and $\psi^B_j$ on the $A$ and $B$ sites, respectively,
\begin{align}
    (E-\epsilon_j) \psi^B_j &= t (\psi^A_j + \psi^A_{j+1}) \nonumber\\
    (E - 2V) \psi^A_j &= t (\psi^B_j  + \psi^B_{j-1} ) \, .
    \label{eq:Sch_equations_mosaic}
\end{align}
By combining the two equations above, we obtain a single equation for $\psi_j^B$ 
\begin{align}
    (E - \frac{2 t^2}{E-2V} - \epsilon_j ) \psi^B_j &= \frac{t^2}{E-2V} (\psi^B_{j-1} + \psi^B_{j+1}) \, ,
\end{align}
which has the same form as Eq.~\eqref{eq:pj_equation_diamond_chain}. Therefore, it follows that the model is self-dual when $\widetilde{W} = 2 t^2/(E-2V)$, which leads to the mobility edge
\begin{align}
    E_{\rm C} = 2V \pm \frac{t^2}{W |\cos(\pi b + \frac{\pi}{2})|} \, .
    \label{eq:E_c_mosaic_lattice}
\end{align}
This analytically obtained mobility edge is in excellent agreement with the
numerical results for the IPR, defined in Sec.~\ref{sec:single_particle} and shown in  Fig.~\ref{fig:single_particle_mosaic_chain}(b). The spectrum at small $W$ contains only extended states, while for larger $W$ extended and localized states coexist and are separated by $E_{\rm C}$.

For the case $\delta \neq 0$, which involves a finite quasiperiodic onsite potential on the  A sites, the mobility edge no longer follows from Eq.~\eqref{eq:E_c_mosaic_lattice}. From the numerically obtained IPR in Fig.~\ref{fig:single_particle_mosaic_chain}(c), we observe that the spectrum is divided into three regions depending on $W$: (i) for small $W$ all states are extended, (ii) for intermediate $W$ the spectrum contains both extended and localized states, and (iii) for large $W$ all states are localized.

%
%%%
\section{Details of the nonseparable model} \label{app:details_nonseparable_model}
%%%
%
\begin{figure}[t!]
	\centering
    \includegraphics[scale=1]{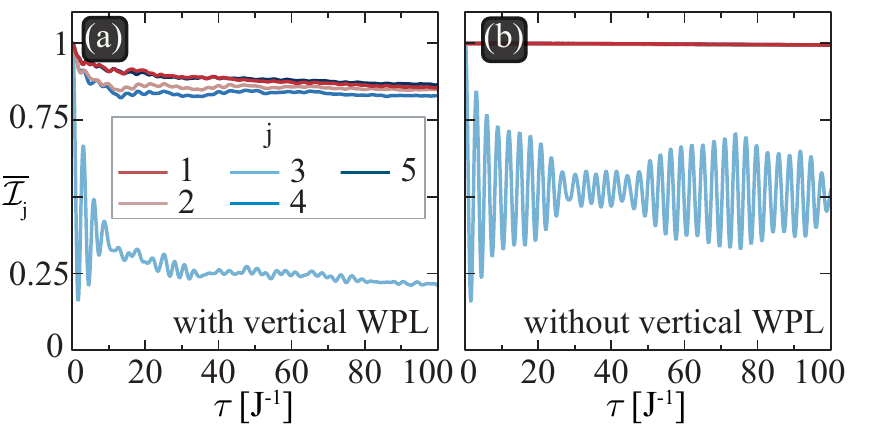}
    \caption{The $y$-dependent imbalance defined in the main text for a single sample that contains a horizontal WPL with zero potential, located in the middle of the sample ($j=3$). In (a) the sample has also a vertical WPL, while in (b) no vertical WPLs are present. The index $j$ labels the rows of the system starting from the bottom. We used the same parameters as in Fig.~\ref{fig:NS_imbalance_WPL}(b).
    }
    \label{fig:vertical_WPLs_comparison}
\end{figure}
In this appendix we further investigate the effects of vertical WPLs in the nonseparable AA model, when a horizontal WPL is simultaneously present in the system, as briefly discussed in the main text. 
Fig.~\ref{fig:vertical_WPLs_comparison} shows the corresponding behavior of the $y$-dependent imbalance. 
When a vertical WPL is present, we see that the imbalance decays for all rows, labeled by $j$, since particles are able to travel both horizontally and vertically. Without any vertical WPLs instead, the imbalance decays only for $j=3$, which coincides with the position of the horizontal WPL.

%
%%%
\section{Convergence of numerical simulations} \label{app:numdetails}
%%%
%
\begin{figure}[ht!]
	\centering
    \includegraphics[scale=1]{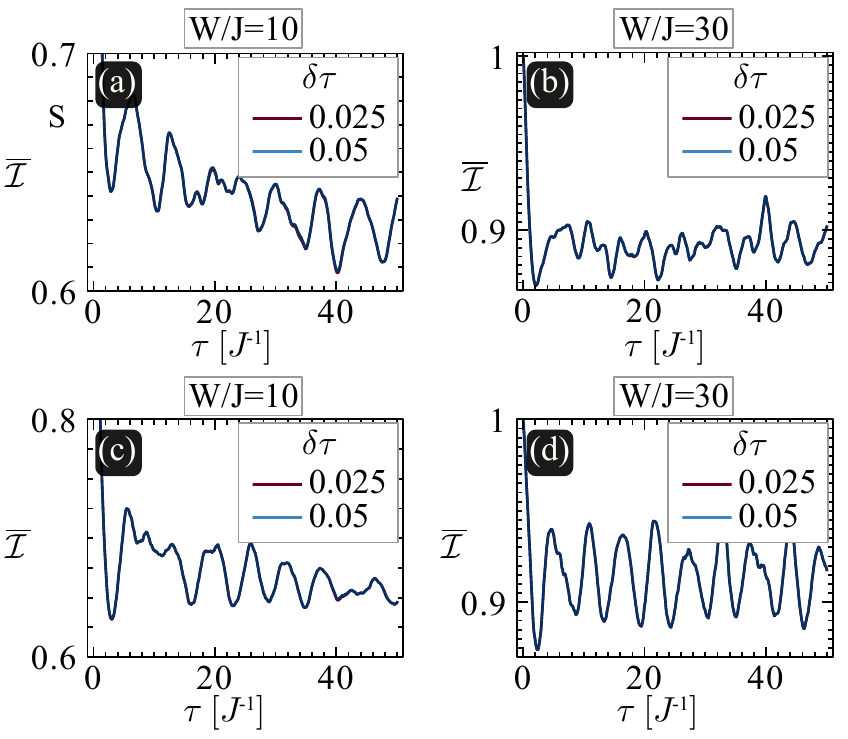}
    \caption{Convergence of the simulated dynamics of the averaged imbalance with the time step $\delta \tau$. Panels (a) and (b) show the case of a 2D nonseparable AA model for two different potential strengths $W/J$ and for two values of the time step $\delta \tau$. Panels (c) and (d) show the same for a 2D separable AA model. The imbalances are averaged over $5$ different samples of size $L_x \times L_y = 16 \times 5$ and the bond dimensions is set to $\chi = 128$ in all plots.
    }
    \label{fig:convergence_dt}
\end{figure}
In this appendix, we check that our numerical simulations, and in particular the imbalance, have indeed converged. The two relevant parameters are the time step $\delta \tau$ and the bond dimension $\chi$. A benefit of the TDVP algorithm is its stability with the respect to the choice of the time step~\cite{Doggen2020}. We find that $\delta \tau = 0.05$ provides converged results for all system sizes in the 2D models studied in this paper, as illustrated in Fig.~\ref{fig:convergence_dt}, and $\delta \tau=0.1$ is enough for our diamond chain studies in Fig.~\ref{fig:diamond_chain}.

\begin{figure}[ht!]
	\centering
    \includegraphics[scale=1]{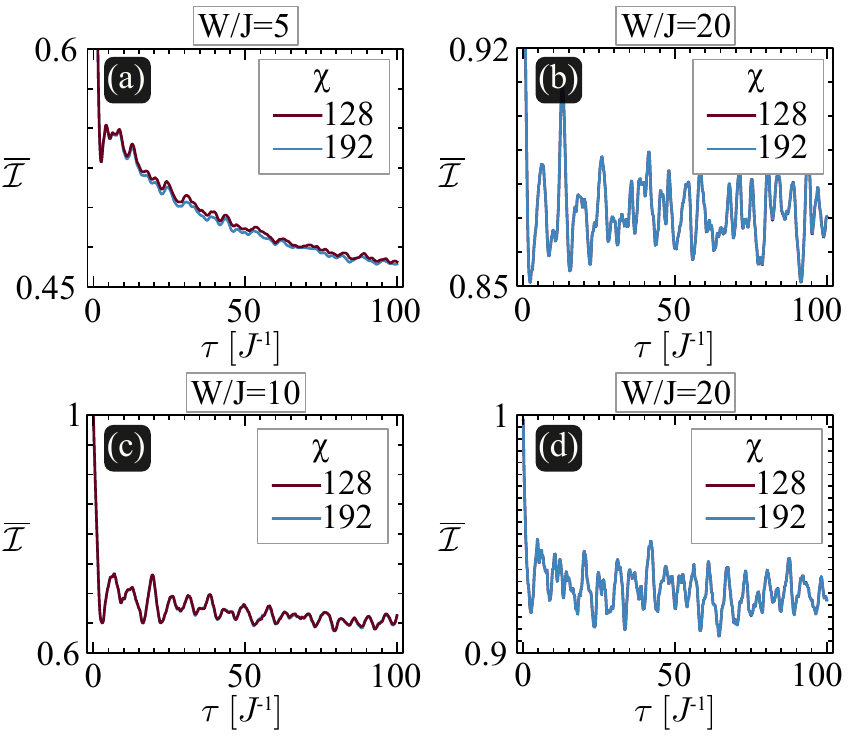}
    \caption{Convergence of the simulated dynamics of the averaged imbalance. Panels (a) and (b) show the case of the 2D nonseparable AA model for two different potential strengths $W/J$ and for two values of the bond dimension $\chi$. Panels (c) and (d) show the same for the 2D separable AA model. The imbalances are averaged over $5$ different samples of size $L_x \times L_y = 16 \times 5$ and the time step of $\delta \tau=0.05$ is used in all plots.
    }
    \label{fig:convergence_chi}
\end{figure}
Besides the time step, the dynamics of the imbalance also depends on the bond dimension $\chi$, which is crucial for determining the long-time precision of the simulations. The error caused by the truncation of the bond dimension $\chi$ tends to grow with time; therefore $\chi$ determines the rate at which the simulated imbalance departs from the exact value as time progresses. In the ergodic phase, where the entanglement in the system grows rapidly in time, a larger bond dimension is needed compared to the less entangled localized phase; for more details see Refs.~\cite{Haegeman2011, Haegeman2016a, Doggen2018,Doggen2019,Doggen2020,Strkalj2021,Doggen2021a}.
Since our choice of maximum simulation time decreases upon increasing the size of the system, and upon decreasing the potential strength, then if for a certain value of $\chi$ convergence is reached for some system size and $W'/J$, we can safely use the same $\chi$ for all smaller systems and/or all systems with $W>W'$.
In Fig.~\ref{fig:convergence_chi}, we show the dynamics of the averaged imbalance for a system of size $L_x \times L_y = 16 \times 5$ sites, for a few different potential strengths and for two values of the bond dimension. 
We observe that $\chi=128$ is large enough to reach convergence on a timescale of one hundred hopping times for systems with dimensions smaller or equal to $L_x \times L_y = 16 \times 5$, and for potential strengths $W/J \geq 5$ for the nonseparable AA model and $W/J \geq 10$ for the separable AA model.

%
%%%%%%%%%%%%%%%%%%%%%%%%%%%%%%%%%%%%%%%%%%%%
%apsrev4-2.bst 2019-01-14 (MD) hand-edited version of apsrev4-1.bst
%Control: key (0)
%Control: author (8) initials jnrlst
%Control: editor formatted (1) identically to author
%Control: production of article title (0) allowed
%Control: page (0) single
%Control: year (1) truncated
%Control: production of eprint (0) enabled
%


\begin{thebibliography}{85}%
	\makeatletter
	\providecommand \@ifxundefined [1]{%
		\@ifx{#1\undefined}
	}%
	\providecommand \@ifnum [1]{%
		\ifnum #1\expandafter \@firstoftwo
		\else \expandafter \@secondoftwo
		\fi
	}%
	\providecommand \@ifx [1]{%
		\ifx #1\expandafter \@firstoftwo
		\else \expandafter \@secondoftwo
		\fi
	}%
	\providecommand \natexlab [1]{#1}%
	\providecommand \enquote  [1]{``#1''}%
	\providecommand \bibnamefont  [1]{#1}%
	\providecommand \bibfnamefont [1]{#1}%
	\providecommand \citenamefont [1]{#1}%
	\providecommand \href@noop [0]{\@secondoftwo}%
	\providecommand \href [0]{\begingroup \@sanitize@url \@href}%
	\providecommand \@href[1]{\@@startlink{#1}\@@href}%
	\providecommand \@@href[1]{\endgroup#1\@@endlink}%
	\providecommand \@sanitize@url [0]{\catcode `\\12\catcode `\$12\catcode
		`\&12\catcode `\#12\catcode `\^12\catcode `\_12\catcode `\%12\relax}%
	\providecommand \@@startlink[1]{}%
	\providecommand \@@endlink[0]{}%
	\providecommand \url  [0]{\begingroup\@sanitize@url \@url }%
	\providecommand \@url [1]{\endgroup\@href {#1}{\urlprefix }}%
	\providecommand \urlprefix  [0]{URL }%
	\providecommand \Eprint [0]{\href }%
	\providecommand \doibase [0]{https://doi.org/}%
	\providecommand \selectlanguage [0]{\@gobble}%
	\providecommand \bibinfo  [0]{\@secondoftwo}%
	\providecommand \bibfield  [0]{\@secondoftwo}%
	\providecommand \translation [1]{[#1]}%
	\providecommand \BibitemOpen [0]{}%
	\providecommand \bibitemStop [0]{}%
	\providecommand \bibitemNoStop [0]{.\EOS\space}%
	\providecommand \EOS [0]{\spacefactor3000\relax}%
	\providecommand \BibitemShut  [1]{\csname bibitem#1\endcsname}%
	\let\auto@bib@innerbib\@empty
	%</preamble>
	\bibitem [{\citenamefont {Deutsch}(1991)}]{Deutsch1991}%
	\BibitemOpen
	\bibfield  {author} {\bibinfo {author} {\bibfnamefont {J.~M.}\ \bibnamefont
			{Deutsch}},\ }\bibfield  {title} {\bibinfo {title} {Quantum statistical
			mechanics in a closed system},\ }\href
	{https://doi.org/10.1103/PhysRevA.43.2046} {\bibfield  {journal} {\bibinfo
			{journal} {Phys. Rev. A}\ }\textbf {\bibinfo {volume} {43}},\ \bibinfo
		{pages} {2046} (\bibinfo {year} {1991})}\BibitemShut {NoStop}%
	\bibitem [{\citenamefont {Srednicki}(1994)}]{Srednicki1994}%
	\BibitemOpen
	\bibfield  {author} {\bibinfo {author} {\bibfnamefont {M.}~\bibnamefont
			{Srednicki}},\ }\bibfield  {title} {\bibinfo {title} {Chaos and quantum
			thermalization},\ }\href {https://doi.org/10.1103/PhysRevE.50.888} {\bibfield
		{journal} {\bibinfo  {journal} {Phys. Rev. E}\ }\textbf {\bibinfo {volume}
			{50}},\ \bibinfo {pages} {888} (\bibinfo {year} {1994})}\BibitemShut
	{NoStop}%
	\bibitem [{\citenamefont {Srednicki}(1999)}]{Srednicki1999}%
	\BibitemOpen
	\bibfield  {author} {\bibinfo {author} {\bibfnamefont {M.}~\bibnamefont
			{Srednicki}},\ }\bibfield  {title} {\bibinfo {title} {The approach to thermal
			equilibrium in quantized chaotic systems},\ }\href
	{https://doi.org/10.1088/0305-4470/32/7/007} {\bibfield  {journal} {\bibinfo
			{journal} {Journal of Physics A: Mathematical and General}\ }\textbf
		{\bibinfo {volume} {32}},\ \bibinfo {pages} {1163} (\bibinfo {year}
		{1999})}\BibitemShut {NoStop}%
	\bibitem [{\citenamefont {Polkovnikov}\ \emph {et~al.}(2011)\citenamefont
		{Polkovnikov}, \citenamefont {Sengupta}, \citenamefont {Silva},\ and\
		\citenamefont {Vengalattore}}]{Polkovnikov2011}%
	\BibitemOpen
	\bibfield  {author} {\bibinfo {author} {\bibfnamefont {A.}~\bibnamefont
			{Polkovnikov}}, \bibinfo {author} {\bibfnamefont {K.}~\bibnamefont
			{Sengupta}}, \bibinfo {author} {\bibfnamefont {A.}~\bibnamefont {Silva}},\
		and\ \bibinfo {author} {\bibfnamefont {M.}~\bibnamefont {Vengalattore}},\
	}\bibfield  {title} {\bibinfo {title} {Colloquium: Nonequilibrium dynamics of
			closed interacting quantum systems},\ }\href
	{https://doi.org/10.1103/RevModPhys.83.863} {\bibfield  {journal} {\bibinfo
			{journal} {Rev. Mod. Phys.}\ }\textbf {\bibinfo {volume} {83}},\ \bibinfo
		{pages} {863} (\bibinfo {year} {2011})}\BibitemShut {NoStop}%
	\bibitem [{\citenamefont {Nandkishore}\ and\ \citenamefont
		{Huse}(2015)}]{Nandkishore2015}%
	\BibitemOpen
	\bibfield  {author} {\bibinfo {author} {\bibfnamefont {R.}~\bibnamefont
			{Nandkishore}}\ and\ \bibinfo {author} {\bibfnamefont {D.~A.}\ \bibnamefont
			{Huse}},\ }\bibfield  {title} {\bibinfo {title} {Many-body localization and
			thermalization in quantum statistical mechanics},\ }\href
	{https://doi.org/10.1146/annurev-conmatphys-031214-014726} {\bibfield
		{journal} {\bibinfo  {journal} {Annu. Rev. Condens. Matter Phys}\ }\textbf
		{\bibinfo {volume} {6}},\ \bibinfo {pages} {15} (\bibinfo {year}
		{2015})}\BibitemShut {NoStop}%
	\bibitem [{\citenamefont {Abanin}\ \emph {et~al.}(2019)\citenamefont {Abanin},
		\citenamefont {Altman}, \citenamefont {Bloch},\ and\ \citenamefont
		{Serbyn}}]{Abanin2019}%
	\BibitemOpen
	\bibfield  {author} {\bibinfo {author} {\bibfnamefont {D.~A.}\ \bibnamefont
			{Abanin}}, \bibinfo {author} {\bibfnamefont {E.}~\bibnamefont {Altman}},
		\bibinfo {author} {\bibfnamefont {I.}~\bibnamefont {Bloch}},\ and\ \bibinfo
		{author} {\bibfnamefont {M.}~\bibnamefont {Serbyn}},\ }\bibfield  {title}
	{\bibinfo {title} {Colloquium: {M}any-body localization, thermalization, and
			entanglement},\ }\href {https://doi.org/10.1103/RevModPhys.91.021001}
	{\bibfield  {journal} {\bibinfo  {journal} {Rev. Mod. Phys.}\ }\textbf
		{\bibinfo {volume} {91}},\ \bibinfo {pages} {021001} (\bibinfo {year}
		{2019})}\BibitemShut {NoStop}%
	\bibitem [{\citenamefont {Anderson}(1958)}]{Anderson1958}%
	\BibitemOpen
	\bibfield  {author} {\bibinfo {author} {\bibfnamefont {P.~W.}\ \bibnamefont
			{Anderson}},\ }\bibfield  {title} {\bibinfo {title} {Absence of diffusion in
			certain random lattices},\ }\href {https://doi.org/10.1103/PhysRev.109.1492}
	{\bibfield  {journal} {\bibinfo  {journal} {Phys. Rev.}\ }\textbf {\bibinfo
			{volume} {109}},\ \bibinfo {pages} {1492} (\bibinfo {year}
		{1958})}\BibitemShut {NoStop}%
	\bibitem [{\citenamefont {Abrahams}(2010)}]{Abrahams2010}%
	\BibitemOpen
	\bibfield  {author} {\bibinfo {author} {\bibfnamefont {E.}~\bibnamefont
			{Abrahams}},\ }\href {https://doi.org/10.1142/9789814299084} {\emph {\bibinfo
			{title} {50 years of Anderson localization}}}\ (\bibinfo  {publisher} {World
		Scientific Publishing Co. Pte. Ltd.},\ \bibinfo {year} {2010})\BibitemShut
	{NoStop}%
	\bibitem [{\citenamefont {Gornyi}\ \emph {et~al.}(2005)\citenamefont {Gornyi},
		\citenamefont {Mirlin},\ and\ \citenamefont {Polyakov}}]{Gornyi2005}%
	\BibitemOpen
	\bibfield  {author} {\bibinfo {author} {\bibfnamefont {I.~V.}\ \bibnamefont
			{Gornyi}}, \bibinfo {author} {\bibfnamefont {A.~D.}\ \bibnamefont {Mirlin}},\
		and\ \bibinfo {author} {\bibfnamefont {D.~G.}\ \bibnamefont {Polyakov}},\
	}\bibfield  {title} {\bibinfo {title} {Interacting electrons in disordered
			wires: Anderson localization and low-$t$ transport},\ }\href
	{https://doi.org/10.1103/PhysRevLett.95.206603} {\bibfield  {journal}
		{\bibinfo  {journal} {Phys. Rev. Lett.}\ }\textbf {\bibinfo {volume} {95}},\
		\bibinfo {pages} {206603} (\bibinfo {year} {2005})}\BibitemShut {NoStop}%
	\bibitem [{\citenamefont {Basko}\ \emph {et~al.}(2006)\citenamefont {Basko},
		\citenamefont {Aleiner},\ and\ \citenamefont {Altshuler}}]{Basko2006}%
	\BibitemOpen
	\bibfield  {author} {\bibinfo {author} {\bibfnamefont {D.~M.}\ \bibnamefont
			{Basko}}, \bibinfo {author} {\bibfnamefont {I.~L.}\ \bibnamefont {Aleiner}},\
		and\ \bibinfo {author} {\bibfnamefont {B.~L.}\ \bibnamefont {Altshuler}},\
	}\bibfield  {title} {\bibinfo {title} {Metal–insulator transition in a
			weakly interacting many-electron system with localized single-particle
			states},\ }\href {https://doi.org/10.1016/j.aop.2005.11.014} {\bibfield
		{journal} {\bibinfo  {journal} {Ann. Phys. (N.Y.)}\ }\textbf {\bibinfo
			{volume} {321}},\ \bibinfo {pages} {1126} (\bibinfo {year}
		{2006})}\BibitemShut {NoStop}%
	\bibitem [{\citenamefont {Altshuler}\ \emph {et~al.}(1997)\citenamefont
		{Altshuler}, \citenamefont {Gefen}, \citenamefont {Kamenev},\ and\
		\citenamefont {Levitov}}]{Altshuler1997}%
	\BibitemOpen
	\bibfield  {author} {\bibinfo {author} {\bibfnamefont {B.~L.}\ \bibnamefont
			{Altshuler}}, \bibinfo {author} {\bibfnamefont {Y.}~\bibnamefont {Gefen}},
		\bibinfo {author} {\bibfnamefont {A.}~\bibnamefont {Kamenev}},\ and\ \bibinfo
		{author} {\bibfnamefont {L.~S.}\ \bibnamefont {Levitov}},\ }\bibfield
	{title} {\bibinfo {title} {Quasiparticle lifetime in a finite system: A
			nonperturbative approach},\ }\href
	{https://doi.org/10.1103/PhysRevLett.78.2803} {\bibfield  {journal} {\bibinfo
			{journal} {Phys. Rev. Lett.}\ }\textbf {\bibinfo {volume} {78}},\ \bibinfo
		{pages} {2803} (\bibinfo {year} {1997})}\BibitemShut {NoStop}%
	\bibitem [{\citenamefont {Imbrie}(2016)}]{Imbrie2016}%
	\BibitemOpen
	\bibfield  {author} {\bibinfo {author} {\bibfnamefont {J.~Z.}\ \bibnamefont
			{Imbrie}},\ }\bibfield  {title} {\bibinfo {title} {On many-body localization
			for quantum spin chains},\ }\href {https://doi.org/10.1007/s10955-016-1508-x}
	{\bibfield  {journal} {\bibinfo  {journal} {Journal of Statistical Physics}\
		}\textbf {\bibinfo {volume} {163}},\ \bibinfo {pages} {998} (\bibinfo {year}
		{2016})}\BibitemShut {NoStop}%
	\bibitem [{\citenamefont {Oganesyan}\ and\ \citenamefont
		{Huse}(2007)}]{Oganesyan2007}%
	\BibitemOpen
	\bibfield  {author} {\bibinfo {author} {\bibfnamefont {V.}~\bibnamefont
			{Oganesyan}}\ and\ \bibinfo {author} {\bibfnamefont {D.~A.}\ \bibnamefont
			{Huse}},\ }\bibfield  {title} {\bibinfo {title} {Localization of interacting
			fermions at high temperature},\ }\href
	{https://doi.org/10.1103/PhysRevB.75.155111} {\bibfield  {journal} {\bibinfo
			{journal} {Phys. Rev. B}\ }\textbf {\bibinfo {volume} {75}},\ \bibinfo
		{pages} {155111} (\bibinfo {year} {2007})}\BibitemShut {NoStop}%
	\bibitem [{\citenamefont {\ifmmode \check{Z}\else
			\v{Z}\fi{}nidari\ifmmode~\check{c}\else \v{c}\fi{}}\ \emph
		{et~al.}(2008)\citenamefont {\ifmmode \check{Z}\else
			\v{Z}\fi{}nidari\ifmmode~\check{c}\else \v{c}\fi{}}, \citenamefont {Prosen},\
		and\ \citenamefont {Prelov\ifmmode~\check{s}\else
			\v{s}\fi{}ek}}]{Znidaric2008}%
	\BibitemOpen
	\bibfield  {author} {\bibinfo {author} {\bibfnamefont {M.}~\bibnamefont
			{\ifmmode \check{Z}\else \v{Z}\fi{}nidari\ifmmode~\check{c}\else
				\v{c}\fi{}}}, \bibinfo {author} {\bibfnamefont {T.}~\bibnamefont {Prosen}},\
		and\ \bibinfo {author} {\bibfnamefont {P.}~\bibnamefont
			{Prelov\ifmmode~\check{s}\else \v{s}\fi{}ek}},\ }\bibfield  {title} {\bibinfo
		{title} {Many-body localization in the {H}eisenberg $xxz$ magnet in a random
			field},\ }\href {https://doi.org/10.1103/PhysRevB.77.064426} {\bibfield
		{journal} {\bibinfo  {journal} {Phys. Rev. B}\ }\textbf {\bibinfo {volume}
			{77}},\ \bibinfo {pages} {064426} (\bibinfo {year} {2008})}\BibitemShut
	{NoStop}%
	\bibitem [{\citenamefont {Pal}\ and\ \citenamefont {Huse}(2010)}]{Pal2010}%
	\BibitemOpen
	\bibfield  {author} {\bibinfo {author} {\bibfnamefont {A.}~\bibnamefont
			{Pal}}\ and\ \bibinfo {author} {\bibfnamefont {D.~A.}\ \bibnamefont {Huse}},\
	}\bibfield  {title} {\bibinfo {title} {Many-body localization phase
			transition},\ }\href {https://doi.org/10.1103/PhysRevB.82.174411} {\bibfield
		{journal} {\bibinfo  {journal} {Phys. Rev. B}\ }\textbf {\bibinfo {volume}
			{82}},\ \bibinfo {pages} {174411} (\bibinfo {year} {2010})}\BibitemShut
	{NoStop}%
	\bibitem [{\citenamefont {Gopalakrishnan}\ and\ \citenamefont
		{Parameswaran}(2020)}]{Gopalakrishnan2020}%
	\BibitemOpen
	\bibfield  {author} {\bibinfo {author} {\bibfnamefont {S.}~\bibnamefont
			{Gopalakrishnan}}\ and\ \bibinfo {author} {\bibfnamefont {S.}~\bibnamefont
			{Parameswaran}},\ }\bibfield  {title} {\bibinfo {title} {Dynamics and
			transport at the threshold of many-body localization},\ }\href
	{https://doi.org/10.1016/j.physrep.2020.03.003} {\bibfield  {journal}
		{\bibinfo  {journal} {Physics Reports}\ }\textbf {\bibinfo {volume} {862}},\
		\bibinfo {pages} {1} (\bibinfo {year} {2020})}\BibitemShut {NoStop}%
	\bibitem [{\citenamefont {Schreiber}\ \emph {et~al.}(2015)\citenamefont
		{Schreiber}, \citenamefont {Hodgman}, \citenamefont {Bordia}, \citenamefont
		{Lüschen}, \citenamefont {Fischer}, \citenamefont {Vosk}, \citenamefont
		{Altman}, \citenamefont {Schneider},\ and\ \citenamefont
		{Bloch}}]{Schreiber2015}%
	\BibitemOpen
	\bibfield  {author} {\bibinfo {author} {\bibfnamefont {M.}~\bibnamefont
			{Schreiber}}, \bibinfo {author} {\bibfnamefont {S.~S.}\ \bibnamefont
			{Hodgman}}, \bibinfo {author} {\bibfnamefont {P.}~\bibnamefont {Bordia}},
		\bibinfo {author} {\bibfnamefont {H.~P.}\ \bibnamefont {Lüschen}}, \bibinfo
		{author} {\bibfnamefont {M.~H.}\ \bibnamefont {Fischer}}, \bibinfo {author}
		{\bibfnamefont {R.}~\bibnamefont {Vosk}}, \bibinfo {author} {\bibfnamefont
			{E.}~\bibnamefont {Altman}}, \bibinfo {author} {\bibfnamefont
			{U.}~\bibnamefont {Schneider}},\ and\ \bibinfo {author} {\bibfnamefont
			{I.}~\bibnamefont {Bloch}},\ }\bibfield  {title} {\bibinfo {title}
		{Observation of many-body localization of interacting fermions in a
			quasirandom optical lattice},\ }\href
	{https://doi.org/10.1126/science.aaa7432} {\bibfield  {journal} {\bibinfo
			{journal} {Science}\ }\textbf {\bibinfo {volume} {349}},\ \bibinfo {pages}
		{842} (\bibinfo {year} {2015})}\BibitemShut {NoStop}%
	\bibitem [{\citenamefont {Alet}\ and\ \citenamefont
		{Laflorencie}(2018)}]{Alet2018}%
	\BibitemOpen
	\bibfield  {author} {\bibinfo {author} {\bibfnamefont {F.}~\bibnamefont
			{Alet}}\ and\ \bibinfo {author} {\bibfnamefont {N.}~\bibnamefont
			{Laflorencie}},\ }\bibfield  {title} {\bibinfo {title} {Many-body
			localization: An introduction and selected topics},\ }\href
	{https://doi.org/https://doi.org/10.1016/j.crhy.2018.03.003} {\bibfield
		{journal} {\bibinfo  {journal} {Comptes Rendus Physique}\ }\textbf {\bibinfo
			{volume} {19}},\ \bibinfo {pages} {498} (\bibinfo {year} {2018})},\ \bibinfo
	{note} {quantum simulation / Simulation quantique}\BibitemShut {NoStop}%
	\bibitem [{\citenamefont {Luitz}\ \emph {et~al.}(2015)\citenamefont {Luitz},
		\citenamefont {Laflorencie},\ and\ \citenamefont {Alet}}]{Luitz2015}%
	\BibitemOpen
	\bibfield  {author} {\bibinfo {author} {\bibfnamefont {D.~J.}\ \bibnamefont
			{Luitz}}, \bibinfo {author} {\bibfnamefont {N.}~\bibnamefont {Laflorencie}},\
		and\ \bibinfo {author} {\bibfnamefont {F.}~\bibnamefont {Alet}},\ }\bibfield
	{title} {\bibinfo {title} {Many-body localization edge in the random-field
			heisenberg chain},\ }\href {https://doi.org/10.1103/PhysRevB.91.081103}
	{\bibfield  {journal} {\bibinfo  {journal} {Phys. Rev. B}\ }\textbf {\bibinfo
			{volume} {91}},\ \bibinfo {pages} {081103} (\bibinfo {year}
		{2015})}\BibitemShut {NoStop}%
	\bibitem [{\citenamefont {Mondaini}\ and\ \citenamefont
		{Rigol}(2015)}]{Mondaini2015}%
	\BibitemOpen
	\bibfield  {author} {\bibinfo {author} {\bibfnamefont {R.}~\bibnamefont
			{Mondaini}}\ and\ \bibinfo {author} {\bibfnamefont {M.}~\bibnamefont
			{Rigol}},\ }\bibfield  {title} {\bibinfo {title} {Many-body localization and
			thermalization in disordered hubbard chains},\ }\href
	{https://doi.org/10.1103/PhysRevA.92.041601} {\bibfield  {journal} {\bibinfo
			{journal} {Phys. Rev. A}\ }\textbf {\bibinfo {volume} {92}},\ \bibinfo
		{pages} {041601} (\bibinfo {year} {2015})}\BibitemShut {NoStop}%
	\bibitem [{\citenamefont {Prelov\ifmmode~\check{s}\else \v{s}\fi{}ek}\ \emph
		{et~al.}(2016)\citenamefont {Prelov\ifmmode~\check{s}\else \v{s}\fi{}ek},
		\citenamefont {Bari\ifmmode \check{s}\else \v{s}\fi{}i\ifmmode~\acute{c}\else
			\'{c}\fi{}},\ and\ \citenamefont {\ifmmode \check{Z}\else
			\v{Z}\fi{}nidari\ifmmode~\check{c}\else \v{c}\fi{}}}]{Prelovsek2016}%
	\BibitemOpen
	\bibfield  {author} {\bibinfo {author} {\bibfnamefont {P.}~\bibnamefont
			{Prelov\ifmmode~\check{s}\else \v{s}\fi{}ek}}, \bibinfo {author}
		{\bibfnamefont {O.~S.}\ \bibnamefont {Bari\ifmmode \check{s}\else
				\v{s}\fi{}i\ifmmode~\acute{c}\else \'{c}\fi{}}},\ and\ \bibinfo {author}
		{\bibfnamefont {M.}~\bibnamefont {\ifmmode \check{Z}\else
				\v{Z}\fi{}nidari\ifmmode~\check{c}\else \v{c}\fi{}}},\ }\bibfield  {title}
	{\bibinfo {title} {Absence of full many-body localization in the disordered
			hubbard chain},\ }\href {https://doi.org/10.1103/PhysRevB.94.241104}
	{\bibfield  {journal} {\bibinfo  {journal} {Phys. Rev. B}\ }\textbf {\bibinfo
			{volume} {94}},\ \bibinfo {pages} {241104} (\bibinfo {year}
		{2016})}\BibitemShut {NoStop}%
	\bibitem [{\citenamefont {Zakrzewski}\ and\ \citenamefont
		{Delande}(2018)}]{Zakrzewski2018}%
	\BibitemOpen
	\bibfield  {author} {\bibinfo {author} {\bibfnamefont {J.}~\bibnamefont
			{Zakrzewski}}\ and\ \bibinfo {author} {\bibfnamefont {D.}~\bibnamefont
			{Delande}},\ }\bibfield  {title} {\bibinfo {title} {Spin-charge separation
			and many-body localization},\ }\href
	{https://doi.org/10.1103/PhysRevB.98.014203} {\bibfield  {journal} {\bibinfo
			{journal} {Phys. Rev. B}\ }\textbf {\bibinfo {volume} {98}},\ \bibinfo
		{pages} {014203} (\bibinfo {year} {2018})}\BibitemShut {NoStop}%
	\bibitem [{\citenamefont {Sierant}\ and\ \citenamefont
		{Zakrzewski}(2018)}]{Sierant2018}%
	\BibitemOpen
	\bibfield  {author} {\bibinfo {author} {\bibfnamefont {P.}~\bibnamefont
			{Sierant}}\ and\ \bibinfo {author} {\bibfnamefont {J.}~\bibnamefont
			{Zakrzewski}},\ }\bibfield  {title} {\bibinfo {title} {Many-body localization
			of bosons in optical lattices},\ }\href
	{https://doi.org/10.1088/1367-2630/aabb17} {\bibfield  {journal} {\bibinfo
			{journal} {New Journal of Physics}\ }\textbf {\bibinfo {volume} {20}},\
		\bibinfo {pages} {043032} (\bibinfo {year} {2018})}\BibitemShut {NoStop}%
	\bibitem [{\citenamefont {Orell}\ \emph {et~al.}(2019)\citenamefont {Orell},
		\citenamefont {Michailidis}, \citenamefont {Serbyn},\ and\ \citenamefont
		{Silveri}}]{Orell2019}%
	\BibitemOpen
	\bibfield  {author} {\bibinfo {author} {\bibfnamefont {T.}~\bibnamefont
			{Orell}}, \bibinfo {author} {\bibfnamefont {A.~A.}\ \bibnamefont
			{Michailidis}}, \bibinfo {author} {\bibfnamefont {M.}~\bibnamefont
			{Serbyn}},\ and\ \bibinfo {author} {\bibfnamefont {M.}~\bibnamefont
			{Silveri}},\ }\bibfield  {title} {\bibinfo {title} {Probing the many-body
			localization phase transition with superconducting circuits},\ }\href
	{https://doi.org/10.1103/PhysRevB.100.134504} {\bibfield  {journal} {\bibinfo
			{journal} {Phys. Rev. B}\ }\textbf {\bibinfo {volume} {100}},\ \bibinfo
		{pages} {134504} (\bibinfo {year} {2019})}\BibitemShut {NoStop}%
	\bibitem [{\citenamefont {Hopjan}\ and\ \citenamefont
		{Heidrich-Meisner}(2020)}]{Hopjan2020}%
	\BibitemOpen
	\bibfield  {author} {\bibinfo {author} {\bibfnamefont {M.}~\bibnamefont
			{Hopjan}}\ and\ \bibinfo {author} {\bibfnamefont {F.}~\bibnamefont
			{Heidrich-Meisner}},\ }\bibfield  {title} {\bibinfo {title} {Many-body
			localization from a one-particle perspective in the disordered
			one-dimensional bose-hubbard model},\ }\href
	{https://doi.org/10.1103/PhysRevA.101.063617} {\bibfield  {journal} {\bibinfo
			{journal} {Phys. Rev. A}\ }\textbf {\bibinfo {volume} {101}},\ \bibinfo
		{pages} {063617} (\bibinfo {year} {2020})}\BibitemShut {NoStop}%
	\bibitem [{\citenamefont {y.~Choi}\ \emph {et~al.}(2016)\citenamefont
		{y.~Choi}, \citenamefont {Hild}, \citenamefont {Zeiher}, \citenamefont
		{Schauss}, \citenamefont {Rubio-Abadal}, \citenamefont {Yefsah},
		\citenamefont {Khemani}, \citenamefont {Huse}, \citenamefont {Bloch},\ and\
		\citenamefont {Gross}}]{Choi2016}%
	\BibitemOpen
	\bibfield  {author} {\bibinfo {author} {\bibfnamefont {J.}~\bibnamefont
			{y.~Choi}}, \bibinfo {author} {\bibfnamefont {S.}~\bibnamefont {Hild}},
		\bibinfo {author} {\bibfnamefont {J.}~\bibnamefont {Zeiher}}, \bibinfo
		{author} {\bibfnamefont {P.}~\bibnamefont {Schauss}}, \bibinfo {author}
		{\bibfnamefont {A.}~\bibnamefont {Rubio-Abadal}}, \bibinfo {author}
		{\bibfnamefont {T.}~\bibnamefont {Yefsah}}, \bibinfo {author} {\bibfnamefont
			{V.}~\bibnamefont {Khemani}}, \bibinfo {author} {\bibfnamefont {D.~A.}\
			\bibnamefont {Huse}}, \bibinfo {author} {\bibfnamefont {I.}~\bibnamefont
			{Bloch}},\ and\ \bibinfo {author} {\bibfnamefont {C.}~\bibnamefont {Gross}},\
	}\bibfield  {title} {\bibinfo {title} {Exploring the many-body localization
			transition in two dimensions},\ }\href
	{https://doi.org/10.1126/science.aaf8834} {\bibfield  {journal} {\bibinfo
			{journal} {Science}\ }\textbf {\bibinfo {volume} {352}},\ \bibinfo {pages}
		{1547} (\bibinfo {year} {2016})}\BibitemShut {NoStop}%
	\bibitem [{\citenamefont {Wahl}\ \emph {et~al.}(2018)\citenamefont {Wahl},
		\citenamefont {Pal},\ and\ \citenamefont {Simon}}]{Wahl2018}%
	\BibitemOpen
	\bibfield  {author} {\bibinfo {author} {\bibfnamefont {T.~B.}\ \bibnamefont
			{Wahl}}, \bibinfo {author} {\bibfnamefont {A.}~\bibnamefont {Pal}},\ and\
		\bibinfo {author} {\bibfnamefont {S.~H.}\ \bibnamefont {Simon}},\ }\bibfield
	{title} {\bibinfo {title} {Signatures of the many-body localized regime in
			two dimensions},\ }\href {https://doi.org/10.1038/s41567-018-0339-x}
	{\bibfield  {journal} {\bibinfo  {journal} {Nature Physics}\ }\textbf
		{\bibinfo {volume} {15}},\ \bibinfo {pages} {164} (\bibinfo {year}
		{2018})}\BibitemShut {NoStop}%
	\bibitem [{\citenamefont {Kennes}(2018)}]{Kennes2018}%
	\BibitemOpen
	\bibfield  {author} {\bibinfo {author} {\bibfnamefont {D.~M.}\ \bibnamefont
			{Kennes}},\ }\href@noop {} {\bibinfo {title} {Many-body localization in two
			dimensions from projected entangled-pair states}} (\bibinfo {year} {2018}),\
	\Eprint {https://arxiv.org/abs/1811.04126} {arXiv:1811.04126
		[cond-mat.dis-nn]} \BibitemShut {NoStop}%
	\bibitem [{\citenamefont {Th\'eveniaut}\ \emph {et~al.}(2020)\citenamefont
		{Th\'eveniaut}, \citenamefont {Lan}, \citenamefont {Meyer},\ and\
		\citenamefont {Alet}}]{Theveniaut2020}%
	\BibitemOpen
	\bibfield  {author} {\bibinfo {author} {\bibfnamefont {H.}~\bibnamefont
			{Th\'eveniaut}}, \bibinfo {author} {\bibfnamefont {Z.}~\bibnamefont {Lan}},
		\bibinfo {author} {\bibfnamefont {G.}~\bibnamefont {Meyer}},\ and\ \bibinfo
		{author} {\bibfnamefont {F.}~\bibnamefont {Alet}},\ }\bibfield  {title}
	{\bibinfo {title} {Transition to a many-body localized regime in a
			two-dimensional disordered quantum dimer model},\ }\href
	{https://doi.org/10.1103/PhysRevResearch.2.033154} {\bibfield  {journal}
		{\bibinfo  {journal} {Phys. Rev. Research}\ }\textbf {\bibinfo {volume}
			{2}},\ \bibinfo {pages} {033154} (\bibinfo {year} {2020})}\BibitemShut
	{NoStop}%
	\bibitem [{\citenamefont {Kshetrimayum}\ \emph {et~al.}(2020)\citenamefont
		{Kshetrimayum}, \citenamefont {Goihl},\ and\ \citenamefont
		{Eisert}}]{Kshetrimayum2020}%
	\BibitemOpen
	\bibfield  {author} {\bibinfo {author} {\bibfnamefont {A.}~\bibnamefont
			{Kshetrimayum}}, \bibinfo {author} {\bibfnamefont {M.}~\bibnamefont
			{Goihl}},\ and\ \bibinfo {author} {\bibfnamefont {J.}~\bibnamefont
			{Eisert}},\ }\bibfield  {title} {\bibinfo {title} {Time evolution of
			many-body localized systems in two spatial dimensions},\ }\href
	{https://doi.org/10.1103/PhysRevB.102.235132} {\bibfield  {journal} {\bibinfo
			{journal} {Phys. Rev. B}\ }\textbf {\bibinfo {volume} {102}},\ \bibinfo
		{pages} {235132} (\bibinfo {year} {2020})}\BibitemShut {NoStop}%
	\bibitem [{\citenamefont {Li}\ \emph {et~al.}(2021)\citenamefont {Li},
		\citenamefont {Chan},\ and\ \citenamefont {Wahl}}]{Li2021}%
	\BibitemOpen
	\bibfield  {author} {\bibinfo {author} {\bibfnamefont {J.}~\bibnamefont
			{Li}}, \bibinfo {author} {\bibfnamefont {A.}~\bibnamefont {Chan}},\ and\
		\bibinfo {author} {\bibfnamefont {T.~B.}\ \bibnamefont {Wahl}},\ }\href@noop
	{} {\bibinfo {title} {Fermionic quantum circuits reproduce experimental
			two-dimensional many-body localization transition point}} (\bibinfo {year}
	{2021}),\ \Eprint {https://arxiv.org/abs/2108.08268} {arXiv:2108.08268
		[cond-mat.dis-nn]} \BibitemShut {NoStop}%
	\bibitem [{\citenamefont {Foo}\ \emph {et~al.}(2022)\citenamefont {Foo},
		\citenamefont {Swain}, \citenamefont {Sengupta}, \citenamefont {Lemarié},\
		and\ \citenamefont {Adam}}]{Foo2022}%
	\BibitemOpen
	\bibfield  {author} {\bibinfo {author} {\bibfnamefont {D.~C.~W.}\
			\bibnamefont {Foo}}, \bibinfo {author} {\bibfnamefont {N.}~\bibnamefont
			{Swain}}, \bibinfo {author} {\bibfnamefont {P.}~\bibnamefont {Sengupta}},
		\bibinfo {author} {\bibfnamefont {G.}~\bibnamefont {Lemarié}},\ and\
		\bibinfo {author} {\bibfnamefont {S.}~\bibnamefont {Adam}},\ }\href@noop {}
	{\bibinfo {title} {A stabilization mechanism for many-body localization in
			two dimensions}} (\bibinfo {year} {2022}),\ \Eprint
	{https://arxiv.org/abs/2202.09072} {arXiv:2202.09072 [cond-mat.dis-nn]}
	\BibitemShut {NoStop}%
	\bibitem [{\citenamefont {De~Roeck}\ and\ \citenamefont
		{Huveneers}(2017)}]{DeRoeck2017a}%
	\BibitemOpen
	\bibfield  {author} {\bibinfo {author} {\bibfnamefont {W.}~\bibnamefont
			{De~Roeck}}\ and\ \bibinfo {author} {\bibfnamefont {F.}~\bibnamefont
			{Huveneers}},\ }\bibfield  {title} {\bibinfo {title} {Stability and
			instability towards delocalization in many-body localization systems},\
	}\href {https://doi.org/10.1103/PhysRevB.95.155129} {\bibfield  {journal}
		{\bibinfo  {journal} {Phys. Rev. B}\ }\textbf {\bibinfo {volume} {95}},\
		\bibinfo {pages} {155129} (\bibinfo {year} {2017})}\BibitemShut {NoStop}%
	\bibitem [{\citenamefont {Roeck}\ and\ \citenamefont
		{Imbrie}(2017)}]{DeRoeck2017b}%
	\BibitemOpen
	\bibfield  {author} {\bibinfo {author} {\bibfnamefont {W.~D.}\ \bibnamefont
			{Roeck}}\ and\ \bibinfo {author} {\bibfnamefont {J.~Z.}\ \bibnamefont
			{Imbrie}},\ }\bibfield  {title} {\bibinfo {title} {Many-body localization:
			stability and instability},\ }\href {https://doi.org/10.1098/rsta.2016.0422}
	{\bibfield  {journal} {\bibinfo  {journal} {Philosophical Transactions of the
				Royal Society A: Mathematical, Physical and Engineering Sciences}\ }\textbf
		{\bibinfo {volume} {375}},\ \bibinfo {pages} {20160422} (\bibinfo {year}
		{2017})}\BibitemShut {NoStop}%
	\bibitem [{\citenamefont {Gopalakrishnan}\ and\ \citenamefont
		{Huse}(2019)}]{Gopalakrishnan2019}%
	\BibitemOpen
	\bibfield  {author} {\bibinfo {author} {\bibfnamefont {S.}~\bibnamefont
			{Gopalakrishnan}}\ and\ \bibinfo {author} {\bibfnamefont {D.~A.}\
			\bibnamefont {Huse}},\ }\bibfield  {title} {\bibinfo {title} {Instability of
			many-body localized systems as a phase transition in a nonstandard
			thermodynamic limit},\ }\href {https://doi.org/10.1103/PhysRevB.99.134305}
	{\bibfield  {journal} {\bibinfo  {journal} {Phys. Rev. B}\ }\textbf {\bibinfo
			{volume} {99}},\ \bibinfo {pages} {134305} (\bibinfo {year}
		{2019})}\BibitemShut {NoStop}%
	\bibitem [{\citenamefont {Potirniche}\ \emph {et~al.}(2019)\citenamefont
		{Potirniche}, \citenamefont {Banerjee},\ and\ \citenamefont
		{Altman}}]{Potirniche2019}%
	\BibitemOpen
	\bibfield  {author} {\bibinfo {author} {\bibfnamefont {I.-D.}\ \bibnamefont
			{Potirniche}}, \bibinfo {author} {\bibfnamefont {S.}~\bibnamefont
			{Banerjee}},\ and\ \bibinfo {author} {\bibfnamefont {E.}~\bibnamefont
			{Altman}},\ }\bibfield  {title} {\bibinfo {title} {Exploration of the
			stability of many-body localization in $d > 1$},\ }\href
	{https://doi.org/10.1103/PhysRevB.99.205149} {\bibfield  {journal} {\bibinfo
			{journal} {Phys. Rev. B}\ }\textbf {\bibinfo {volume} {99}},\ \bibinfo
		{pages} {205149} (\bibinfo {year} {2019})}\BibitemShut {NoStop}%
	\bibitem [{\citenamefont {Doggen}\ \emph {et~al.}(2020)\citenamefont {Doggen},
		\citenamefont {Gornyi}, \citenamefont {Mirlin},\ and\ \citenamefont
		{Polyakov}}]{Doggen2020}%
	\BibitemOpen
	\bibfield  {author} {\bibinfo {author} {\bibfnamefont {E.~V.~H.}\
			\bibnamefont {Doggen}}, \bibinfo {author} {\bibfnamefont {I.~V.}\
			\bibnamefont {Gornyi}}, \bibinfo {author} {\bibfnamefont {A.~D.}\
			\bibnamefont {Mirlin}},\ and\ \bibinfo {author} {\bibfnamefont {D.~G.}\
			\bibnamefont {Polyakov}},\ }\bibfield  {title} {\bibinfo {title} {Slow
			many-body delocalization beyond one dimension},\ }\href
	{https://doi.org/10.1103/PhysRevLett.125.155701} {\bibfield  {journal}
		{\bibinfo  {journal} {Phys. Rev. Lett.}\ }\textbf {\bibinfo {volume} {125}},\
		\bibinfo {pages} {155701} (\bibinfo {year} {2020})}\BibitemShut {NoStop}%
	\bibitem [{\citenamefont {\ifmmode~\check{S}\else \v{S}\fi{}untajs}\ \emph
		{et~al.}(2020)\citenamefont {\ifmmode~\check{S}\else \v{S}\fi{}untajs},
		\citenamefont {Bon\ifmmode~\check{c}\else \v{c}\fi{}a}, \citenamefont
		{Prosen},\ and\ \citenamefont {Vidmar}}]{Suntajs2020}%
	\BibitemOpen
	\bibfield  {author} {\bibinfo {author} {\bibfnamefont {J.}~\bibnamefont
			{\ifmmode~\check{S}\else \v{S}\fi{}untajs}}, \bibinfo {author} {\bibfnamefont
			{J.}~\bibnamefont {Bon\ifmmode~\check{c}\else \v{c}\fi{}a}}, \bibinfo
		{author} {\bibfnamefont {T.}~\bibnamefont {Prosen}},\ and\ \bibinfo {author}
		{\bibfnamefont {L.}~\bibnamefont {Vidmar}},\ }\bibfield  {title} {\bibinfo
		{title} {Quantum chaos challenges many-body localization},\ }\href
	{https://doi.org/10.1103/PhysRevE.102.062144} {\bibfield  {journal} {\bibinfo
			{journal} {Phys. Rev. E}\ }\textbf {\bibinfo {volume} {102}},\ \bibinfo
		{pages} {062144} (\bibinfo {year} {2020})}\BibitemShut {NoStop}%
	\bibitem [{\citenamefont {Thiery}\ \emph {et~al.}(2018)\citenamefont {Thiery},
		\citenamefont {Huveneers}, \citenamefont {M\"uller},\ and\ \citenamefont
		{De~Roeck}}]{Thiery2018}%
	\BibitemOpen
	\bibfield  {author} {\bibinfo {author} {\bibfnamefont {T.}~\bibnamefont
			{Thiery}}, \bibinfo {author} {\bibfnamefont {F.}~\bibnamefont {Huveneers}},
		\bibinfo {author} {\bibfnamefont {M.}~\bibnamefont {M\"uller}},\ and\
		\bibinfo {author} {\bibfnamefont {W.}~\bibnamefont {De~Roeck}},\ }\bibfield
	{title} {\bibinfo {title} {Many-body delocalization as a quantum avalanche},\
	}\href {https://doi.org/10.1103/PhysRevLett.121.140601} {\bibfield  {journal}
		{\bibinfo  {journal} {Phys. Rev. Lett.}\ }\textbf {\bibinfo {volume} {121}},\
		\bibinfo {pages} {140601} (\bibinfo {year} {2018})}\BibitemShut {NoStop}%
	\bibitem [{\citenamefont {Léonard}\ \emph {et~al.}(2020)\citenamefont
		{Léonard}, \citenamefont {Rispoli}, \citenamefont {Lukin}, \citenamefont
		{Schittko}, \citenamefont {Kim}, \citenamefont {Kwan}, \citenamefont {Sels},
		\citenamefont {Demler},\ and\ \citenamefont {Greiner}}]{Leonard2020}%
	\BibitemOpen
	\bibfield  {author} {\bibinfo {author} {\bibfnamefont {J.}~\bibnamefont
			{Léonard}}, \bibinfo {author} {\bibfnamefont {M.}~\bibnamefont {Rispoli}},
		\bibinfo {author} {\bibfnamefont {A.}~\bibnamefont {Lukin}}, \bibinfo
		{author} {\bibfnamefont {R.}~\bibnamefont {Schittko}}, \bibinfo {author}
		{\bibfnamefont {S.}~\bibnamefont {Kim}}, \bibinfo {author} {\bibfnamefont
			{J.}~\bibnamefont {Kwan}}, \bibinfo {author} {\bibfnamefont {D.}~\bibnamefont
			{Sels}}, \bibinfo {author} {\bibfnamefont {E.}~\bibnamefont {Demler}},\ and\
		\bibinfo {author} {\bibfnamefont {M.}~\bibnamefont {Greiner}},\ }\href@noop
	{} {\bibinfo {title} {Signatures of bath-induced quantum avalanches in a
			many-body--localized system}} (\bibinfo {year} {2020}),\ \Eprint
	{https://arxiv.org/abs/2012.15270} {arXiv:2012.15270 [cond-mat.quant-gas]}
	\BibitemShut {NoStop}%
	\bibitem [{\citenamefont {Morningstar}\ \emph {et~al.}(2022)\citenamefont
		{Morningstar}, \citenamefont {Colmenarez}, \citenamefont {Khemani},
		\citenamefont {Luitz},\ and\ \citenamefont {Huse}}]{Morningstar2021}%
	\BibitemOpen
	\bibfield  {author} {\bibinfo {author} {\bibfnamefont {A.}~\bibnamefont
			{Morningstar}}, \bibinfo {author} {\bibfnamefont {L.}~\bibnamefont
			{Colmenarez}}, \bibinfo {author} {\bibfnamefont {V.}~\bibnamefont {Khemani}},
		\bibinfo {author} {\bibfnamefont {D.~J.}\ \bibnamefont {Luitz}},\ and\
		\bibinfo {author} {\bibfnamefont {D.~A.}\ \bibnamefont {Huse}},\ }\bibfield
	{title} {\bibinfo {title} {Avalanches and many-body resonances in many-body
			localized systems},\ }\href {https://doi.org/10.1103/PhysRevB.105.174205}
	{\bibfield  {journal} {\bibinfo  {journal} {Phys. Rev. B}\ }\textbf {\bibinfo
			{volume} {105}},\ \bibinfo {pages} {174205} (\bibinfo {year}
		{2022})}\BibitemShut {NoStop}%
	\bibitem [{\citenamefont {Sels}(2022)}]{Sels2022}%
	\BibitemOpen
	\bibfield  {author} {\bibinfo {author} {\bibfnamefont {D.}~\bibnamefont
			{Sels}},\ }\bibfield  {title} {\bibinfo {title} {Bath-induced delocalization
			in interacting disordered spin chains},\ }\href
	{https://doi.org/10.1103/PhysRevB.106.L020202} {\bibfield  {journal}
		{\bibinfo  {journal} {Phys. Rev. B}\ }\textbf {\bibinfo {volume} {106}},\
		\bibinfo {pages} {L020202} (\bibinfo {year} {2022})}\BibitemShut {NoStop}%
	\bibitem [{\citenamefont {Aubry}\ and\ \citenamefont
		{Andr{\'e}}(1980)}]{Aubry1980}%
	\BibitemOpen
	\bibfield  {author} {\bibinfo {author} {\bibfnamefont {S.}~\bibnamefont
			{Aubry}}\ and\ \bibinfo {author} {\bibfnamefont {G.}~\bibnamefont
			{Andr{\'e}}},\ }\bibfield  {title} {\bibinfo {title} {Analyticity breaking
			and {A}nderson localization in incommensurate lattices},\ }\href@noop {}
	{\bibfield  {journal} {\bibinfo  {journal} {Ann. Israel Phys. Soc}\ }\textbf
		{\bibinfo {volume} {3}},\ \bibinfo {pages} {18} (\bibinfo {year}
		{1980})}\BibitemShut {NoStop}%
	\bibitem [{\citenamefont {Senechal}(1995)}]{Senechal1995}%
	\BibitemOpen
	\bibfield  {author} {\bibinfo {author} {\bibfnamefont {M.}~\bibnamefont
			{Senechal}},\ }\href {https://books.google.ch/books?id=Ojh5zQEACAAJ} {\emph
		{\bibinfo {title} {Quasicrystals and Geometry}}}\ (\bibinfo  {publisher}
	{Cambridge University Press},\ \bibinfo {year} {1995})\BibitemShut {NoStop}%
	\bibitem [{\citenamefont {Devakul}\ and\ \citenamefont
		{Huse}(2017)}]{Devakul2017}%
	\BibitemOpen
	\bibfield  {author} {\bibinfo {author} {\bibfnamefont {T.}~\bibnamefont
			{Devakul}}\ and\ \bibinfo {author} {\bibfnamefont {D.~A.}\ \bibnamefont
			{Huse}},\ }\bibfield  {title} {\bibinfo {title} {Anderson localization
			transitions with and without random potentials},\ }\href
	{https://doi.org/10.1103/PhysRevB.96.214201} {\bibfield  {journal} {\bibinfo
			{journal} {Phys. Rev. B}\ }\textbf {\bibinfo {volume} {96}},\ \bibinfo
		{pages} {214201} (\bibinfo {year} {2017})}\BibitemShut {NoStop}%
	\bibitem [{\citenamefont {Bordia}\ \emph {et~al.}(2017)\citenamefont {Bordia},
		\citenamefont {L\"uschen}, \citenamefont {Scherg}, \citenamefont
		{Gopalakrishnan}, \citenamefont {Knap}, \citenamefont {Schneider},\ and\
		\citenamefont {Bloch}}]{Bordia2017}%
	\BibitemOpen
	\bibfield  {author} {\bibinfo {author} {\bibfnamefont {P.}~\bibnamefont
			{Bordia}}, \bibinfo {author} {\bibfnamefont {H.}~\bibnamefont {L\"uschen}},
		\bibinfo {author} {\bibfnamefont {S.}~\bibnamefont {Scherg}}, \bibinfo
		{author} {\bibfnamefont {S.}~\bibnamefont {Gopalakrishnan}}, \bibinfo
		{author} {\bibfnamefont {M.}~\bibnamefont {Knap}}, \bibinfo {author}
		{\bibfnamefont {U.}~\bibnamefont {Schneider}},\ and\ \bibinfo {author}
		{\bibfnamefont {I.}~\bibnamefont {Bloch}},\ }\bibfield  {title} {\bibinfo
		{title} {Probing slow relaxation and many-body localization in
			two-dimensional quasiperiodic systems},\ }\href
	{https://doi.org/10.1103/PhysRevX.7.041047} {\bibfield  {journal} {\bibinfo
			{journal} {Phys. Rev. X}\ }\textbf {\bibinfo {volume} {7}},\ \bibinfo {pages}
		{041047} (\bibinfo {year} {2017})}\BibitemShut {NoStop}%
	\bibitem [{\citenamefont {Szab\'o}\ and\ \citenamefont
		{Schneider}(2020)}]{Szabo2020}%
	\BibitemOpen
	\bibfield  {author} {\bibinfo {author} {\bibfnamefont {A.}~\bibnamefont
			{Szab\'o}}\ and\ \bibinfo {author} {\bibfnamefont {U.}~\bibnamefont
			{Schneider}},\ }\bibfield  {title} {\bibinfo {title} {Mixed spectra and
			partially extended states in a two-dimensional quasiperiodic model},\ }\href
	{https://doi.org/10.1103/PhysRevB.101.014205} {\bibfield  {journal} {\bibinfo
			{journal} {Phys. Rev. B}\ }\textbf {\bibinfo {volume} {101}},\ \bibinfo
		{pages} {014205} (\bibinfo {year} {2020})}\BibitemShut {NoStop}%
	\bibitem [{\citenamefont {Johnstone}\ \emph {et~al.}(2021)\citenamefont
		{Johnstone}, \citenamefont {Öhberg},\ and\ \citenamefont
		{Duncan}}]{Johnstone2021}%
	\BibitemOpen
	\bibfield  {author} {\bibinfo {author} {\bibfnamefont {D.}~\bibnamefont
			{Johnstone}}, \bibinfo {author} {\bibfnamefont {P.}~\bibnamefont {Öhberg}},\
		and\ \bibinfo {author} {\bibfnamefont {C.~W.}\ \bibnamefont {Duncan}},\
	}\bibfield  {title} {\bibinfo {title} {The mean-field bose glass in
			quasicrystalline systems},\ }\href {https://doi.org/10.1088/1751-8121/ac1dc0}
	{\bibfield  {journal} {\bibinfo  {journal} {Journal of Physics A:
				Mathematical and Theoretical}\ }\textbf {\bibinfo {volume} {54}},\ \bibinfo
		{pages} {395001} (\bibinfo {year} {2021})}\BibitemShut {NoStop}%
	\bibitem [{\citenamefont {Johnstone}\ \emph {et~al.}(2022)\citenamefont
		{Johnstone}, \citenamefont {Öhberg},\ and\ \citenamefont
		{Duncan}}]{Johnstone2022}%
	\BibitemOpen
	\bibfield  {author} {\bibinfo {author} {\bibfnamefont {D.}~\bibnamefont
			{Johnstone}}, \bibinfo {author} {\bibfnamefont {P.}~\bibnamefont {Öhberg}},\
		and\ \bibinfo {author} {\bibfnamefont {C.~W.}\ \bibnamefont {Duncan}},\
	}\bibfield  {title} {\bibinfo {title} {Barriers to macroscopic superfluidity
			and insulation in a {2D} {A}ubry–{A}ndr{é} model},\ }\href
	{https://doi.org/10.1088/1361-6455/ac6d34} {\bibfield  {journal} {\bibinfo
			{journal} {Journal of Physics B: Atomic, Molecular and Optical Physics}\
		}\textbf {\bibinfo {volume} {55}},\ \bibinfo {pages} {125302} (\bibinfo
		{year} {2022})}\BibitemShut {NoStop}%
	\bibitem [{footnote_phases()}]{footnote_phases}%
	\BibitemOpen
	\bibinfo {note} {The fact that changing the phases of the cosines does not
		change the localization properties of a system in the thermodynamic limit can
		be seen from the following argument. Let us imagine a unit circle where the
		projections of its points onto the $x$-axis determine the values of the
		potential $\cos(2 \pi b i + \phi)$, where $i$ can take only integer values
		and $b$ is irrational -- as it is the case in Eq.~\eqref{eq:QP_potentials}.
		Since the frequency $b$ is irrational, by changing $i$ one can cover the
		whole unit circle. In other words, each point in the circle can be attributed
		to some $i$. The phase $\phi$ only rotates the coordinate system along the
		unit circle, but changing $i$ from $0$ to $\infty$ will in any case cover the
		whole circle. Therefore, one can always account for a phase change from
		$\phi$ to $\phi'$ by relabelling the sites from $i$ to $i'\in \mathcal{N}$
		such that $\cos(2 \pi b i + \phi) = \cos(2 \pi b i' + \phi')$.}\BibitemShut
	{Stop}%
	\bibitem [{\citenamefont {Rossignolo}\ and\ \citenamefont
		{Dell'Anna}(2019)}]{Rossignolo2019}%
	\BibitemOpen
	\bibfield  {author} {\bibinfo {author} {\bibfnamefont {M.}~\bibnamefont
			{Rossignolo}}\ and\ \bibinfo {author} {\bibfnamefont {L.}~\bibnamefont
			{Dell'Anna}},\ }\bibfield  {title} {\bibinfo {title} {Localization
			transitions and mobility edges in coupled {A}ubry-{A}ndr\'e chains},\ }\href
	{https://doi.org/10.1103/PhysRevB.99.054211} {\bibfield  {journal} {\bibinfo
			{journal} {Phys. Rev. B}\ }\textbf {\bibinfo {volume} {99}},\ \bibinfo
		{pages} {054211} (\bibinfo {year} {2019})}\BibitemShut {NoStop}%
	\bibitem [{\citenamefont {Huang}\ and\ \citenamefont {Liu}(2019)}]{Huang2019}%
	\BibitemOpen
	\bibfield  {author} {\bibinfo {author} {\bibfnamefont {B.}~\bibnamefont
			{Huang}}\ and\ \bibinfo {author} {\bibfnamefont {W.~V.}\ \bibnamefont
			{Liu}},\ }\bibfield  {title} {\bibinfo {title} {Moir\'e localization in
			two-dimensional quasiperiodic systems},\ }\href
	{https://doi.org/10.1103/PhysRevB.100.144202} {\bibfield  {journal} {\bibinfo
			{journal} {Phys. Rev. B}\ }\textbf {\bibinfo {volume} {100}},\ \bibinfo
		{pages} {144202} (\bibinfo {year} {2019})}\BibitemShut {NoStop}%
	\bibitem [{\citenamefont {Goblot}\ \emph {et~al.}(2020)\citenamefont {Goblot},
		\citenamefont {{\v{S}}trkalj}, \citenamefont {Pernet}, \citenamefont {Lado},
		\citenamefont {Dorow}, \citenamefont {Lema{\^{\i}}tre}, \citenamefont
		{Gratiet}, \citenamefont {Harouri}, \citenamefont {Sagnes}, \citenamefont
		{Ravets}, \citenamefont {Amo}, \citenamefont {Bloch},\ and\ \citenamefont
		{Zilberberg}}]{Strkalj2020}%
	\BibitemOpen
	\bibfield  {author} {\bibinfo {author} {\bibfnamefont {V.}~\bibnamefont
			{Goblot}}, \bibinfo {author} {\bibfnamefont {A.}~\bibnamefont
			{{\v{S}}trkalj}}, \bibinfo {author} {\bibfnamefont {N.}~\bibnamefont
			{Pernet}}, \bibinfo {author} {\bibfnamefont {J.~L.}\ \bibnamefont {Lado}},
		\bibinfo {author} {\bibfnamefont {C.}~\bibnamefont {Dorow}}, \bibinfo
		{author} {\bibfnamefont {A.}~\bibnamefont {Lema{\^{\i}}tre}}, \bibinfo
		{author} {\bibfnamefont {L.~L.}\ \bibnamefont {Gratiet}}, \bibinfo {author}
		{\bibfnamefont {A.}~\bibnamefont {Harouri}}, \bibinfo {author} {\bibfnamefont
			{I.}~\bibnamefont {Sagnes}}, \bibinfo {author} {\bibfnamefont
			{S.}~\bibnamefont {Ravets}}, \bibinfo {author} {\bibfnamefont
			{A.}~\bibnamefont {Amo}}, \bibinfo {author} {\bibfnamefont {J.}~\bibnamefont
			{Bloch}},\ and\ \bibinfo {author} {\bibfnamefont {O.}~\bibnamefont
			{Zilberberg}},\ }\bibfield  {title} {\bibinfo {title} {Emergence of
			criticality through a cascade of delocalization transitions in quasiperiodic
			chains},\ }\href {https://doi.org/10.1038/s41567-020-0908-7} {\bibfield
		{journal} {\bibinfo  {journal} {Nature Physics}\ }\textbf {\bibinfo {volume}
			{16}},\ \bibinfo {pages} {832} (\bibinfo {year} {2020})}\BibitemShut
	{NoStop}%
	\bibitem [{\citenamefont {Bell}\ and\ \citenamefont {Dean}(1970)}]{Bell1970}%
	\BibitemOpen
	\bibfield  {author} {\bibinfo {author} {\bibfnamefont {R.~J.}\ \bibnamefont
			{Bell}}\ and\ \bibinfo {author} {\bibfnamefont {P.}~\bibnamefont {Dean}},\
	}\bibfield  {title} {\bibinfo {title} {Atomic vibrations in vitreous
			silica},\ }\href {https://doi.org/10.1039/DF9705000055} {\bibfield  {journal}
		{\bibinfo  {journal} {Discuss. Faraday Soc.}\ }\textbf {\bibinfo {volume}
			{50}},\ \bibinfo {pages} {55} (\bibinfo {year} {1970})}\BibitemShut {NoStop}%
	\bibitem [{\citenamefont {Edwards}\ and\ \citenamefont
		{Thouless}(1972)}]{Edwards1972}%
	\BibitemOpen
	\bibfield  {author} {\bibinfo {author} {\bibfnamefont {J.~T.}\ \bibnamefont
			{Edwards}}\ and\ \bibinfo {author} {\bibfnamefont {D.~J.}\ \bibnamefont
			{Thouless}},\ }\bibfield  {title} {\bibinfo {title} {Numerical studies of
			localization in disordered systems},\ }\href
	{https://doi.org/10.1088/0022-3719/5/8/007} {\bibfield  {journal} {\bibinfo
			{journal} {Journal of Physics C: Solid State Physics}\ }\textbf {\bibinfo
			{volume} {5}},\ \bibinfo {pages} {807} (\bibinfo {year} {1972})}\BibitemShut
	{NoStop}%
	\bibitem [{\citenamefont {Doggen}\ and\ \citenamefont
		{Mirlin}(2019)}]{Doggen2019}%
	\BibitemOpen
	\bibfield  {author} {\bibinfo {author} {\bibfnamefont {E.~V.~H.}\
			\bibnamefont {Doggen}}\ and\ \bibinfo {author} {\bibfnamefont {A.~D.}\
			\bibnamefont {Mirlin}},\ }\bibfield  {title} {\bibinfo {title} {Many-body
			delocalization dynamics in long {A}ubry-{A}ndr\'e quasiperiodic chains},\
	}\href {https://doi.org/10.1103/PhysRevB.100.104203} {\bibfield  {journal}
		{\bibinfo  {journal} {Phys. Rev. B}\ }\textbf {\bibinfo {volume} {100}},\
		\bibinfo {pages} {104203} (\bibinfo {year} {2019})}\BibitemShut {NoStop}%
	\bibitem [{\citenamefont {Haegeman}\ \emph {et~al.}(2016)\citenamefont
		{Haegeman}, \citenamefont {Lubich}, \citenamefont {Oseledets}, \citenamefont
		{Vandereycken},\ and\ \citenamefont {Verstraete}}]{Haegeman2016a}%
	\BibitemOpen
	\bibfield  {author} {\bibinfo {author} {\bibfnamefont {J.}~\bibnamefont
			{Haegeman}}, \bibinfo {author} {\bibfnamefont {C.}~\bibnamefont {Lubich}},
		\bibinfo {author} {\bibfnamefont {I.}~\bibnamefont {Oseledets}}, \bibinfo
		{author} {\bibfnamefont {B.}~\bibnamefont {Vandereycken}},\ and\ \bibinfo
		{author} {\bibfnamefont {F.}~\bibnamefont {Verstraete}},\ }\bibfield  {title}
	{\bibinfo {title} {Unifying time evolution and optimization with matrix
			product states},\ }\href {https://doi.org/10.1103/PhysRevB.94.165116}
	{\bibfield  {journal} {\bibinfo  {journal} {Phys. Rev. B}\ }\textbf {\bibinfo
			{volume} {94}},\ \bibinfo {pages} {165116} (\bibinfo {year}
		{2016})}\BibitemShut {NoStop}%
	\bibitem [{\citenamefont {Schollw\"ock}(2011)}]{Schollwock2011a}%
	\BibitemOpen
	\bibfield  {author} {\bibinfo {author} {\bibfnamefont {U.}~\bibnamefont
			{Schollw\"ock}},\ }\bibfield  {title} {\bibinfo {title} {The density-matrix
			renormalization group in the age of matrix product states},\ }\href
	{https://doi.org/10.1016/j.aop.2010.09.012} {\bibfield  {journal} {\bibinfo
			{journal} {Ann. Phys. (N. Y.)}\ }\textbf {\bibinfo {volume} {326}},\ \bibinfo
		{pages} {96 } (\bibinfo {year} {2011})}\BibitemShut {NoStop}%
	\bibitem [{\citenamefont {Doggen}\ \emph {et~al.}(2021)\citenamefont {Doggen},
		\citenamefont {Gornyi}, \citenamefont {Mirlin},\ and\ \citenamefont
		{Polyakov}}]{Doggen2021a}%
	\BibitemOpen
	\bibfield  {author} {\bibinfo {author} {\bibfnamefont {E.~V.~H.}\
			\bibnamefont {Doggen}}, \bibinfo {author} {\bibfnamefont {I.~V.}\
			\bibnamefont {Gornyi}}, \bibinfo {author} {\bibfnamefont {A.~D.}\
			\bibnamefont {Mirlin}},\ and\ \bibinfo {author} {\bibfnamefont {D.~G.}\
			\bibnamefont {Polyakov}},\ }\bibfield  {title} {\bibinfo {title} {Many-body
			localization in large systems: Matrix-product-state approach},\ }\href
	{https://doi.org/https://doi.org/10.1016/j.aop.2021.168437} {\bibfield
		{journal} {\bibinfo  {journal} {Ann. Phys. (N.Y.)}\ }\textbf {\bibinfo
			{volume} {435}},\ \bibinfo {pages} {168437} (\bibinfo {year} {2021})},\
	\bibinfo {note} {special Issue on Localisation 2020}\BibitemShut {NoStop}%
	\bibitem [{\citenamefont {Paeckel}\ \emph {et~al.}(2019)\citenamefont
		{Paeckel}, \citenamefont {Köhler}, \citenamefont {Swoboda}, \citenamefont
		{Manmana}, \citenamefont {Schollwöck},\ and\ \citenamefont
		{Hubig}}]{Paeckel2019a}%
	\BibitemOpen
	\bibfield  {author} {\bibinfo {author} {\bibfnamefont {S.}~\bibnamefont
			{Paeckel}}, \bibinfo {author} {\bibfnamefont {T.}~\bibnamefont {Köhler}},
		\bibinfo {author} {\bibfnamefont {A.}~\bibnamefont {Swoboda}}, \bibinfo
		{author} {\bibfnamefont {S.~R.}\ \bibnamefont {Manmana}}, \bibinfo {author}
		{\bibfnamefont {U.}~\bibnamefont {Schollwöck}},\ and\ \bibinfo {author}
		{\bibfnamefont {C.}~\bibnamefont {Hubig}},\ }\bibfield  {title} {\bibinfo
		{title} {Time-evolution methods for matrix-product states},\ }\href
	{https://doi.org/https://doi.org/10.1016/j.aop.2019.167998} {\bibfield
		{journal} {\bibinfo  {journal} {Ann. Phys. (N.Y.)}\ }\textbf {\bibinfo
			{volume} {411}},\ \bibinfo {pages} {167998} (\bibinfo {year}
		{2019})}\BibitemShut {NoStop}%
	\bibitem [{\citenamefont {Bordia}\ \emph {et~al.}(2016)\citenamefont {Bordia},
		\citenamefont {L\"uschen}, \citenamefont {Hodgman}, \citenamefont
		{Schreiber}, \citenamefont {Bloch},\ and\ \citenamefont
		{Schneider}}]{Bordia2016}%
	\BibitemOpen
	\bibfield  {author} {\bibinfo {author} {\bibfnamefont {P.}~\bibnamefont
			{Bordia}}, \bibinfo {author} {\bibfnamefont {H.~P.}\ \bibnamefont
			{L\"uschen}}, \bibinfo {author} {\bibfnamefont {S.~S.}\ \bibnamefont
			{Hodgman}}, \bibinfo {author} {\bibfnamefont {M.}~\bibnamefont {Schreiber}},
		\bibinfo {author} {\bibfnamefont {I.}~\bibnamefont {Bloch}},\ and\ \bibinfo
		{author} {\bibfnamefont {U.}~\bibnamefont {Schneider}},\ }\bibfield  {title}
	{\bibinfo {title} {Coupling identical one-dimensional many-body localized
			systems},\ }\href {https://doi.org/10.1103/PhysRevLett.116.140401} {\bibfield
		{journal} {\bibinfo  {journal} {Phys. Rev. Lett.}\ }\textbf {\bibinfo
			{volume} {116}},\ \bibinfo {pages} {140401} (\bibinfo {year}
		{2016})}\BibitemShut {NoStop}%
	\bibitem [{\citenamefont {\ifmmode~\check{S}\else \v{S}\fi{}trkalj}\ \emph
		{et~al.}(2021)\citenamefont {\ifmmode~\check{S}\else \v{S}\fi{}trkalj},
		\citenamefont {Doggen}, \citenamefont {Gornyi},\ and\ \citenamefont
		{Zilberberg}}]{Strkalj2021}%
	\BibitemOpen
	\bibfield  {author} {\bibinfo {author} {\bibfnamefont {A.}~\bibnamefont
			{\ifmmode~\check{S}\else \v{S}\fi{}trkalj}}, \bibinfo {author} {\bibfnamefont
			{E.~V.~H.}\ \bibnamefont {Doggen}}, \bibinfo {author} {\bibfnamefont {I.~V.}\
			\bibnamefont {Gornyi}},\ and\ \bibinfo {author} {\bibfnamefont
			{O.}~\bibnamefont {Zilberberg}},\ }\bibfield  {title} {\bibinfo {title}
		{Many-body localization in the interpolating {A}ubry-{A}ndr\'e-{F}ibonacci
			model},\ }\href {https://doi.org/10.1103/PhysRevResearch.3.033257} {\bibfield
		{journal} {\bibinfo  {journal} {Phys. Rev. Research}\ }\textbf {\bibinfo
			{volume} {3}},\ \bibinfo {pages} {033257} (\bibinfo {year}
		{2021})}\BibitemShut {NoStop}%
	\bibitem [{\citenamefont {Efron}(1979)}]{Efron1979}%
	\BibitemOpen
	\bibfield  {author} {\bibinfo {author} {\bibfnamefont {B.}~\bibnamefont
			{Efron}},\ }\bibfield  {title} {\bibinfo {title} {{Bootstrap Methods:
				{A}nother Look at the Jackknife}},\ }\href
	{https://doi.org/10.1214/aos/1176344552} {\bibfield  {journal} {\bibinfo
			{journal} {The Annals of Statistics}\ }\textbf {\bibinfo {volume} {7}},\
		\bibinfo {pages} {1 } (\bibinfo {year} {1979})}\BibitemShut {NoStop}%
	\bibitem [{footnote:weakWPL()}]{footnote:weakWPL}%
	\BibitemOpen
	\bibinfo {note} {Here we chose for convenience the extreme case of a WPL with
		exactly zero potential. If this is not able to thermalize the system, it
		stands to reason that neither will generic WPLs that have weak but
		nonvanishing potential.}\BibitemShut {Stop}%
	\bibitem [{\citenamefont {Vidal}\ \emph {et~al.}(2000)\citenamefont {Vidal},
		\citenamefont {Dou\ifmmode~\mbox{\c{c}}\else \c{c}\fi{}ot}, \citenamefont
		{Mosseri},\ and\ \citenamefont {Butaud}}]{Vidal_2000}%
	\BibitemOpen
	\bibfield  {author} {\bibinfo {author} {\bibfnamefont {J.}~\bibnamefont
			{Vidal}}, \bibinfo {author} {\bibfnamefont {B.}~\bibnamefont
			{Dou\ifmmode~\mbox{\c{c}}\else \c{c}\fi{}ot}}, \bibinfo {author}
		{\bibfnamefont {R.}~\bibnamefont {Mosseri}},\ and\ \bibinfo {author}
		{\bibfnamefont {P.}~\bibnamefont {Butaud}},\ }\bibfield  {title} {\bibinfo
		{title} {Interaction induced delocalization for two particles in a periodic
			potential},\ }\href {https://doi.org/10.1103/PhysRevLett.85.3906} {\bibfield
		{journal} {\bibinfo  {journal} {Phys. Rev. Lett.}\ }\textbf {\bibinfo
			{volume} {85}},\ \bibinfo {pages} {3906} (\bibinfo {year}
		{2000})}\BibitemShut {NoStop}%
	\bibitem [{\citenamefont {Danieli}\ \emph {et~al.}(2015)\citenamefont
		{Danieli}, \citenamefont {Bodyfelt},\ and\ \citenamefont
		{Flach}}]{Danieli2015}%
	\BibitemOpen
	\bibfield  {author} {\bibinfo {author} {\bibfnamefont {C.}~\bibnamefont
			{Danieli}}, \bibinfo {author} {\bibfnamefont {J.~D.}\ \bibnamefont
			{Bodyfelt}},\ and\ \bibinfo {author} {\bibfnamefont {S.}~\bibnamefont
			{Flach}},\ }\bibfield  {title} {\bibinfo {title} {Flat-band engineering of
			mobility edges},\ }\href {https://doi.org/10.1103/PhysRevB.91.235134}
	{\bibfield  {journal} {\bibinfo  {journal} {Phys. Rev. B}\ }\textbf {\bibinfo
			{volume} {91}},\ \bibinfo {pages} {235134} (\bibinfo {year}
		{2015})}\BibitemShut {NoStop}%
	\bibitem [{\citenamefont {Roy}\ \emph {et~al.}(2020)\citenamefont {Roy},
		\citenamefont {Ramachandran},\ and\ \citenamefont {Sharma}}]{Roy2020}%
	\BibitemOpen
	\bibfield  {author} {\bibinfo {author} {\bibfnamefont {N.}~\bibnamefont
			{Roy}}, \bibinfo {author} {\bibfnamefont {A.}~\bibnamefont {Ramachandran}},\
		and\ \bibinfo {author} {\bibfnamefont {A.}~\bibnamefont {Sharma}},\
	}\bibfield  {title} {\bibinfo {title} {Interplay of disorder and interactions
			in a flat-band supporting diamond chain},\ }\href
	{https://doi.org/10.1103/PhysRevResearch.2.043395} {\bibfield  {journal}
		{\bibinfo  {journal} {Phys. Rev. Research}\ }\textbf {\bibinfo {volume}
			{2}},\ \bibinfo {pages} {043395} (\bibinfo {year} {2020})}\BibitemShut
	{NoStop}%
	\bibitem [{\citenamefont {Wang}\ \emph {et~al.}(2020)\citenamefont {Wang},
		\citenamefont {Xia}, \citenamefont {Zhang}, \citenamefont {Yao},
		\citenamefont {Chen}, \citenamefont {You}, \citenamefont {Zhou},\ and\
		\citenamefont {Liu}}]{Wang2020}%
	\BibitemOpen
	\bibfield  {author} {\bibinfo {author} {\bibfnamefont {Y.}~\bibnamefont
			{Wang}}, \bibinfo {author} {\bibfnamefont {X.}~\bibnamefont {Xia}}, \bibinfo
		{author} {\bibfnamefont {L.}~\bibnamefont {Zhang}}, \bibinfo {author}
		{\bibfnamefont {H.}~\bibnamefont {Yao}}, \bibinfo {author} {\bibfnamefont
			{S.}~\bibnamefont {Chen}}, \bibinfo {author} {\bibfnamefont {J.}~\bibnamefont
			{You}}, \bibinfo {author} {\bibfnamefont {Q.}~\bibnamefont {Zhou}},\ and\
		\bibinfo {author} {\bibfnamefont {X.-J.}\ \bibnamefont {Liu}},\ }\bibfield
	{title} {\bibinfo {title} {One-dimensional quasiperiodic mosaic lattice with
			exact mobility edges},\ }\href
	{https://doi.org/10.1103/PhysRevLett.125.196604} {\bibfield  {journal}
		{\bibinfo  {journal} {Phys. Rev. Lett.}\ }\textbf {\bibinfo {volume} {125}},\
		\bibinfo {pages} {196604} (\bibinfo {year} {2020})}\BibitemShut {NoStop}%
	\bibitem [{\citenamefont {Chandran}\ \emph {et~al.}(2016)\citenamefont
		{Chandran}, \citenamefont {Pal}, \citenamefont {Laumann},\ and\ \citenamefont
		{Scardicchio}}]{Chandran2016}%
	\BibitemOpen
	\bibfield  {author} {\bibinfo {author} {\bibfnamefont {A.}~\bibnamefont
			{Chandran}}, \bibinfo {author} {\bibfnamefont {A.}~\bibnamefont {Pal}},
		\bibinfo {author} {\bibfnamefont {C.~R.}\ \bibnamefont {Laumann}},\ and\
		\bibinfo {author} {\bibfnamefont {A.}~\bibnamefont {Scardicchio}},\
	}\bibfield  {title} {\bibinfo {title} {Many-body localization beyond
			eigenstates in all dimensions},\ }\href
	{https://doi.org/10.1103/PhysRevB.94.144203} {\bibfield  {journal} {\bibinfo
			{journal} {Phys. Rev. B}\ }\textbf {\bibinfo {volume} {94}},\ \bibinfo
		{pages} {144203} (\bibinfo {year} {2016})}\BibitemShut {NoStop}%
	\bibitem [{\citenamefont {Viebahn}\ \emph {et~al.}(2019)\citenamefont
		{Viebahn}, \citenamefont {Sbroscia}, \citenamefont {Carter}, \citenamefont
		{Yu},\ and\ \citenamefont {Schneider}}]{Viebahn2019}%
	\BibitemOpen
	\bibfield  {author} {\bibinfo {author} {\bibfnamefont {K.}~\bibnamefont
			{Viebahn}}, \bibinfo {author} {\bibfnamefont {M.}~\bibnamefont {Sbroscia}},
		\bibinfo {author} {\bibfnamefont {E.}~\bibnamefont {Carter}}, \bibinfo
		{author} {\bibfnamefont {J.-C.}\ \bibnamefont {Yu}},\ and\ \bibinfo {author}
		{\bibfnamefont {U.}~\bibnamefont {Schneider}},\ }\bibfield  {title} {\bibinfo
		{title} {Matter-wave diffraction from a quasicrystalline optical lattice},\
	}\href {https://doi.org/10.1103/PhysRevLett.122.110404} {\bibfield  {journal}
		{\bibinfo  {journal} {Phys. Rev. Lett.}\ }\textbf {\bibinfo {volume} {122}},\
		\bibinfo {pages} {110404} (\bibinfo {year} {2019})}\BibitemShut {NoStop}%
	\bibitem [{\citenamefont {Sbroscia}\ \emph {et~al.}(2020)\citenamefont
		{Sbroscia}, \citenamefont {Viebahn}, \citenamefont {Carter}, \citenamefont
		{Yu}, \citenamefont {Gaunt},\ and\ \citenamefont {Schneider}}]{Sbroscia2020}%
	\BibitemOpen
	\bibfield  {author} {\bibinfo {author} {\bibfnamefont {M.}~\bibnamefont
			{Sbroscia}}, \bibinfo {author} {\bibfnamefont {K.}~\bibnamefont {Viebahn}},
		\bibinfo {author} {\bibfnamefont {E.}~\bibnamefont {Carter}}, \bibinfo
		{author} {\bibfnamefont {J.-C.}\ \bibnamefont {Yu}}, \bibinfo {author}
		{\bibfnamefont {A.}~\bibnamefont {Gaunt}},\ and\ \bibinfo {author}
		{\bibfnamefont {U.}~\bibnamefont {Schneider}},\ }\bibfield  {title} {\bibinfo
		{title} {Observing localization in a 2d quasicrystalline optical lattice},\
	}\href {https://doi.org/10.1103/PhysRevLett.125.200604} {\bibfield  {journal}
		{\bibinfo  {journal} {Phys. Rev. Lett.}\ }\textbf {\bibinfo {volume} {125}},\
		\bibinfo {pages} {200604} (\bibinfo {year} {2020})}\BibitemShut {NoStop}%
	\bibitem [{\citenamefont {Sanchez-Palencia}\ and\ \citenamefont
		{Santos}(2005)}]{Sanchez-Palencia2005}%
	\BibitemOpen
	\bibfield  {author} {\bibinfo {author} {\bibfnamefont {L.}~\bibnamefont
			{Sanchez-Palencia}}\ and\ \bibinfo {author} {\bibfnamefont {L.}~\bibnamefont
			{Santos}},\ }\bibfield  {title} {\bibinfo {title} {Bose-{E}instein
			condensates in optical quasicrystal lattices},\ }\href
	{https://doi.org/10.1103/PhysRevA.72.053607} {\bibfield  {journal} {\bibinfo
			{journal} {Phys. Rev. A}\ }\textbf {\bibinfo {volume} {72}},\ \bibinfo
		{pages} {053607} (\bibinfo {year} {2005})}\BibitemShut {NoStop}%
	\bibitem [{\citenamefont {Gautier}\ \emph {et~al.}(2021)\citenamefont
		{Gautier}, \citenamefont {Yao},\ and\ \citenamefont
		{Sanchez-Palencia}}]{Gautier2021}%
	\BibitemOpen
	\bibfield  {author} {\bibinfo {author} {\bibfnamefont {R.}~\bibnamefont
			{Gautier}}, \bibinfo {author} {\bibfnamefont {H.}~\bibnamefont {Yao}},\ and\
		\bibinfo {author} {\bibfnamefont {L.}~\bibnamefont {Sanchez-Palencia}},\
	}\bibfield  {title} {\bibinfo {title} {Strongly interacting bosons in a
			two-dimensional quasicrystal lattice},\ }\href
	{https://doi.org/10.1103/PhysRevLett.126.110401} {\bibfield  {journal}
		{\bibinfo  {journal} {Phys. Rev. Lett.}\ }\textbf {\bibinfo {volume} {126}},\
		\bibinfo {pages} {110401} (\bibinfo {year} {2021})}\BibitemShut {NoStop}%
	\bibitem [{\citenamefont {Smith}\ \emph {et~al.}(2016)\citenamefont {Smith},
		\citenamefont {Lee}, \citenamefont {Richerme}, \citenamefont {Neyenhuis},
		\citenamefont {Hess}, \citenamefont {Hauke}, \citenamefont {Heyl},
		\citenamefont {Huse},\ and\ \citenamefont {Monroe}}]{Smith2016}%
	\BibitemOpen
	\bibfield  {author} {\bibinfo {author} {\bibfnamefont {J.}~\bibnamefont
			{Smith}}, \bibinfo {author} {\bibfnamefont {A.}~\bibnamefont {Lee}}, \bibinfo
		{author} {\bibfnamefont {P.}~\bibnamefont {Richerme}}, \bibinfo {author}
		{\bibfnamefont {B.}~\bibnamefont {Neyenhuis}}, \bibinfo {author}
		{\bibfnamefont {P.~W.}\ \bibnamefont {Hess}}, \bibinfo {author}
		{\bibfnamefont {P.}~\bibnamefont {Hauke}}, \bibinfo {author} {\bibfnamefont
			{M.}~\bibnamefont {Heyl}}, \bibinfo {author} {\bibfnamefont {D.~A.}\
			\bibnamefont {Huse}},\ and\ \bibinfo {author} {\bibfnamefont
			{C.}~\bibnamefont {Monroe}},\ }\bibfield  {title} {\bibinfo {title}
		{Many-body localization in a quantum simulator with programmable random
			disorder},\ }\href {https://doi.org/10.1038/nphys3783} {\bibfield  {journal}
		{\bibinfo  {journal} {Nature Physics}\ }\textbf {\bibinfo {volume} {12}},\
		\bibinfo {pages} {907} (\bibinfo {year} {2016})}\BibitemShut {NoStop}%
	\bibitem [{\citenamefont {De~Luca}\ \emph {et~al.}(2014)\citenamefont
		{De~Luca}, \citenamefont {Altshuler}, \citenamefont {Kravtsov},\ and\
		\citenamefont {Scardicchio}}]{DeLuca2014}%
	\BibitemOpen
	\bibfield  {author} {\bibinfo {author} {\bibfnamefont {A.}~\bibnamefont
			{De~Luca}}, \bibinfo {author} {\bibfnamefont {B.~L.}\ \bibnamefont
			{Altshuler}}, \bibinfo {author} {\bibfnamefont {V.~E.}\ \bibnamefont
			{Kravtsov}},\ and\ \bibinfo {author} {\bibfnamefont {A.}~\bibnamefont
			{Scardicchio}},\ }\bibfield  {title} {\bibinfo {title} {Anderson localization
			on the {B}ethe lattice: Nonergodicity of extended states},\ }\href
	{https://doi.org/10.1103/PhysRevLett.113.046806} {\bibfield  {journal}
		{\bibinfo  {journal} {Phys. Rev. Lett.}\ }\textbf {\bibinfo {volume} {113}},\
		\bibinfo {pages} {046806} (\bibinfo {year} {2014})}\BibitemShut {NoStop}%
	\bibitem [{\citenamefont {Bera}\ \emph {et~al.}(2018)\citenamefont {Bera},
		\citenamefont {De~Tomasi}, \citenamefont {Khaymovich},\ and\ \citenamefont
		{Scardicchio}}]{Bera2018}%
	\BibitemOpen
	\bibfield  {author} {\bibinfo {author} {\bibfnamefont {S.}~\bibnamefont
			{Bera}}, \bibinfo {author} {\bibfnamefont {G.}~\bibnamefont {De~Tomasi}},
		\bibinfo {author} {\bibfnamefont {I.~M.}\ \bibnamefont {Khaymovich}},\ and\
		\bibinfo {author} {\bibfnamefont {A.}~\bibnamefont {Scardicchio}},\
	}\bibfield  {title} {\bibinfo {title} {Return probability for the {A}nderson
			model on the random regular graph},\ }\href
	{https://doi.org/10.1103/PhysRevB.98.134205} {\bibfield  {journal} {\bibinfo
			{journal} {Phys. Rev. B}\ }\textbf {\bibinfo {volume} {98}},\ \bibinfo
		{pages} {134205} (\bibinfo {year} {2018})}\BibitemShut {NoStop}%
	\bibitem [{\citenamefont {De~Tomasi}\ \emph {et~al.}(2020)\citenamefont
		{De~Tomasi}, \citenamefont {Bera}, \citenamefont {Scardicchio},\ and\
		\citenamefont {Khaymovich}}]{DeTomasi2020}%
	\BibitemOpen
	\bibfield  {author} {\bibinfo {author} {\bibfnamefont {G.}~\bibnamefont
			{De~Tomasi}}, \bibinfo {author} {\bibfnamefont {S.}~\bibnamefont {Bera}},
		\bibinfo {author} {\bibfnamefont {A.}~\bibnamefont {Scardicchio}},\ and\
		\bibinfo {author} {\bibfnamefont {I.~M.}\ \bibnamefont {Khaymovich}},\
	}\bibfield  {title} {\bibinfo {title} {Subdiffusion in the {A}nderson model
			on the random regular graph},\ }\href
	{https://doi.org/10.1103/PhysRevB.101.100201} {\bibfield  {journal} {\bibinfo
			{journal} {Phys. Rev. B}\ }\textbf {\bibinfo {volume} {101}},\ \bibinfo
		{pages} {100201} (\bibinfo {year} {2020})}\BibitemShut {NoStop}%
	\bibitem [{\citenamefont {{\v Z}nidari{\v c}}\ and\ \citenamefont
		{Ljubotina}(2018)}]{Znidaric2018}%
	\BibitemOpen
	\bibfield  {author} {\bibinfo {author} {\bibfnamefont {M.}~\bibnamefont {{\v
					Z}nidari{\v c}}}\ and\ \bibinfo {author} {\bibfnamefont {M.}~\bibnamefont
			{Ljubotina}},\ }\bibfield  {title} {\bibinfo {title} {Interaction instability
			of localization in quasiperiodic systems},\ }\href
	{https://doi.org/10.1073/pnas.1800589115} {\bibfield  {journal} {\bibinfo
			{journal} {Proceedings of the National Academy of Sciences}\ }\textbf
		{\bibinfo {volume} {115}},\ \bibinfo {pages} {4595} (\bibinfo {year}
		{2018})}\BibitemShut {NoStop}%
	\bibitem [{\citenamefont {\ifmmode \check{Z}\else
			\v{Z}\fi{}nidari\ifmmode~\check{c}\else \v{c}\fi{}}(2021)}]{Znidaric2021}%
	\BibitemOpen
	\bibfield  {author} {\bibinfo {author} {\bibfnamefont {M.}~\bibnamefont
			{\ifmmode \check{Z}\else \v{Z}\fi{}nidari\ifmmode~\check{c}\else
				\v{c}\fi{}}},\ }\bibfield  {title} {\bibinfo {title} {Comment on
			``nonequilibrium steady state phases of the interacting
			{A}ubry-{A}ndr\'e-{H}arper model''},\ }\href
	{https://doi.org/10.1103/PhysRevB.103.237101} {\bibfield  {journal} {\bibinfo
			{journal} {Phys. Rev. B}\ }\textbf {\bibinfo {volume} {103}},\ \bibinfo
		{pages} {237101} (\bibinfo {year} {2021})}\BibitemShut {NoStop}%
	\bibitem [{\citenamefont {Agrawal}\ \emph {et~al.}(2022)\citenamefont
		{Agrawal}, \citenamefont {Vasseur},\ and\ \citenamefont
		{Gopalakrishnan}}]{Agrawal2022}%
	\BibitemOpen
	\bibfield  {author} {\bibinfo {author} {\bibfnamefont {U.}~\bibnamefont
			{Agrawal}}, \bibinfo {author} {\bibfnamefont {R.}~\bibnamefont {Vasseur}},\
		and\ \bibinfo {author} {\bibfnamefont {S.}~\bibnamefont {Gopalakrishnan}},\
	}\bibfield  {title} {\bibinfo {title} {Quasiperiodic many-body localization
			transition in dimension d$>$1},\ }\href
	{https://doi.org/10.1103/PhysRevB.106.094206} {\bibfield  {journal} {\bibinfo
			{journal} {Phys. Rev. B}\ }\textbf {\bibinfo {volume} {106}},\ \bibinfo
		{pages} {094206} (\bibinfo {year} {2022})}\BibitemShut {NoStop}%
	\bibitem [{\citenamefont {Crowley}\ and\ \citenamefont
		{Chandran}(2022)}]{Crowley2022}%
	\BibitemOpen
	\bibfield  {author} {\bibinfo {author} {\bibfnamefont {P.~J.~D.}\
			\bibnamefont {Crowley}}\ and\ \bibinfo {author} {\bibfnamefont
			{A.}~\bibnamefont {Chandran}},\ }\href@noop {} {\bibinfo {title} {Mean field
			theory of failed thermalizing avalanches}} (\bibinfo {year} {2022}),\ \Eprint
	{https://arxiv.org/abs/2204.09688} {arXiv:2204.09688 [cond-mat.dis-nn]}
	\BibitemShut {NoStop}%
	\bibitem [{\citenamefont {Hauschild}\ and\ \citenamefont
		{Pollmann}(2018)}]{tenpy}%
	\BibitemOpen
	\bibfield  {author} {\bibinfo {author} {\bibfnamefont {J.}~\bibnamefont
			{Hauschild}}\ and\ \bibinfo {author} {\bibfnamefont {F.}~\bibnamefont
			{Pollmann}},\ }\bibfield  {title} {\bibinfo {title} {{Efficient numerical
				simulations with Tensor Networks: Tensor Network Python (TeNPy)}},\ }\href
	{https://doi.org/10.21468/SciPostPhysLectNotes.5} {\bibfield  {journal}
		{\bibinfo  {journal} {SciPost Phys. Lect. Notes}\ ,\ \bibinfo {pages} {5}}
		(\bibinfo {year} {2018})},\ \bibinfo {note} {code available from
		\url{https://github.com/tenpy/tenpy}}\BibitemShut {NoStop}%
	\bibitem [{\citenamefont {Flach}\ \emph {et~al.}(2014)\citenamefont {Flach},
		\citenamefont {Leykam}, \citenamefont {Bodyfelt}, \citenamefont {Matthies},\
		and\ \citenamefont {Desyatnikov}}]{Flach2014}%
	\BibitemOpen
	\bibfield  {author} {\bibinfo {author} {\bibfnamefont {S.}~\bibnamefont
			{Flach}}, \bibinfo {author} {\bibfnamefont {D.}~\bibnamefont {Leykam}},
		\bibinfo {author} {\bibfnamefont {J.~D.}\ \bibnamefont {Bodyfelt}}, \bibinfo
		{author} {\bibfnamefont {P.}~\bibnamefont {Matthies}},\ and\ \bibinfo
		{author} {\bibfnamefont {A.~S.}\ \bibnamefont {Desyatnikov}},\ }\bibfield
	{title} {\bibinfo {title} {Detangling flat bands into fano lattices},\ }\href
	{https://doi.org/10.1209/0295-5075/105/30001} {\bibfield  {journal} {\bibinfo
			{journal} {{EPL} (Europhysics Letters)}\ }\textbf {\bibinfo {volume}
			{105}},\ \bibinfo {pages} {30001} (\bibinfo {year} {2014})}\BibitemShut
	{NoStop}%
	\bibitem [{\citenamefont {Haegeman}\ \emph {et~al.}(2011)\citenamefont
		{Haegeman}, \citenamefont {Cirac}, \citenamefont {Osborne}, \citenamefont
		{Pi\ifmmode~\check{z}\else \v{z}\fi{}orn}, \citenamefont {Verschelde},\ and\
		\citenamefont {Verstraete}}]{Haegeman2011}%
	\BibitemOpen
	\bibfield  {author} {\bibinfo {author} {\bibfnamefont {J.}~\bibnamefont
			{Haegeman}}, \bibinfo {author} {\bibfnamefont {J.~I.}\ \bibnamefont {Cirac}},
		\bibinfo {author} {\bibfnamefont {T.~J.}\ \bibnamefont {Osborne}}, \bibinfo
		{author} {\bibfnamefont {I.}~\bibnamefont {Pi\ifmmode~\check{z}\else
				\v{z}\fi{}orn}}, \bibinfo {author} {\bibfnamefont {H.}~\bibnamefont
			{Verschelde}},\ and\ \bibinfo {author} {\bibfnamefont {F.}~\bibnamefont
			{Verstraete}},\ }\bibfield  {title} {\bibinfo {title} {Time-dependent
			variational principle for quantum lattices},\ }\href
	{https://doi.org/10.1103/PhysRevLett.107.070601} {\bibfield  {journal}
		{\bibinfo  {journal} {Phys. Rev. Lett.}\ }\textbf {\bibinfo {volume} {107}},\
		\bibinfo {pages} {070601} (\bibinfo {year} {2011})}\BibitemShut {NoStop}%
	\bibitem [{\citenamefont {Doggen}\ \emph {et~al.}(2018)\citenamefont {Doggen},
		\citenamefont {Schindler}, \citenamefont {Tikhonov}, \citenamefont {Mirlin},
		\citenamefont {Neupert}, \citenamefont {Polyakov},\ and\ \citenamefont
		{Gornyi}}]{Doggen2018}%
	\BibitemOpen
	\bibfield  {author} {\bibinfo {author} {\bibfnamefont {E.~V.~H.}\
			\bibnamefont {Doggen}}, \bibinfo {author} {\bibfnamefont {F.}~\bibnamefont
			{Schindler}}, \bibinfo {author} {\bibfnamefont {K.~S.}\ \bibnamefont
			{Tikhonov}}, \bibinfo {author} {\bibfnamefont {A.~D.}\ \bibnamefont
			{Mirlin}}, \bibinfo {author} {\bibfnamefont {T.}~\bibnamefont {Neupert}},
		\bibinfo {author} {\bibfnamefont {D.~G.}\ \bibnamefont {Polyakov}},\ and\
		\bibinfo {author} {\bibfnamefont {I.~V.}\ \bibnamefont {Gornyi}},\ }\bibfield
	{title} {\bibinfo {title} {Many-body localization and delocalization in
			large quantum chains},\ }\href {https://doi.org/10.1103/PhysRevB.98.174202}
	{\bibfield  {journal} {\bibinfo  {journal} {Phys. Rev. B}\ }\textbf {\bibinfo
			{volume} {98}},\ \bibinfo {pages} {174202} (\bibinfo {year}
		{2018})}\BibitemShut {NoStop}%
\end{thebibliography}
\end{document}